\documentclass[fleqn,usenatbib,usedcolumn]{mnras}
\pdfoutput=1
\usepackage[british]{babel}
\usepackage{graphicx}
\usepackage{color,soul}
\usepackage{mathtools}
\usepackage{txfonts} 
\usepackage[T1]{fontenc}
\usepackage{bm}
\usepackage{amsmath}
\usepackage{hyperref}
\usepackage{booktabs}
\usepackage[inline]{enumitem}

\addto\extrasbritish{
  }

\let\mr\mathrm
\DeclarePairedDelimiter\abs{\lvert}{\rvert}%
\newcommand{\HII}{\mbox{H\,{\sc ii}} }
\newcommand{\lan}{\langle }
\newcommand{\ran}{\rangle }
\newcommand{\sTheta}{\bm{\Theta}} 
\newcommand{\sPsi}{\bm{\Psi}}     
\newcommand{\sSigma}{\bm{\Sigma}} 
\newcommand{\sdelta}{\delta}      
\newcommand{\sDelta}{\Delta}      
\newcommand{\sLambda}{\Lambda}    
\graphicspath{{Figures/}}

\begin{document}
\title[Contamination of the Epoch of Reionization power spectrum in the presence of foregrounds]{Contamination of the Epoch of Reionization power spectrum in the presence of foregrounds}
\author[Sims et al.]{Peter H. Sims,$^1$\thanks{E-mail: ps550@cam.ac.uk} Lindley Lentati,$^1$ Paul Alexander,$^1$ Chris L. Carilli$^{1,2}$\\
$^1$Cavendish Laboratory, 19 J. J. Thomson Ave., Cambridge, CB3 0HE \\
$^2$National Radio Astronomy Observatory, Socorro, NM, USA \\}
\date{8th August 2015}
\maketitle
\label{firstpage}

\begin{abstract}
We construct foreground simulations comprising spatially correlated extragalactic and diffuse Galactic emission components and calculate the `intrinsic' (instrument-free) two-dimensional spatial power spectrum and the cylindrically and spherically averaged three-dimensional $k$-space power spectra of the Epoch of Reionization (EoR) and our foreground simulations using a Bayesian power spectral estimation framework. This leads us to identify a model dependent region of optimal signal estimation for our foreground and EoR models, within which the spatial power in the EoR signal relative to foregrounds is maximised. We identify a target field dependent region, in $k$-space, of intrinsic foreground power spectral contamination at low $k_{\perp}$ and $k_{\parallel}$ and a transition to a relatively foreground-free intrinsic EoR window in the complement to this region. The contaminated region of $k$-space demonstrates that simultaneous estimation of the EoR and foregrounds is important for obtaining statistically robust estimates of the EoR power spectrum; biased results will be obtained from methodologies that ignore their covariance. Using simulated observations with frequency dependent $uv$-coverage and primary beam, with the former derived for HERA in 37-antenna and 331-antenna configuration, we recover instrumental power spectra consistent with their intrinsic counterparts. We discuss the implications of these results for optimal strategies for unbiased estimation of the EoR power spectrum.
\end{abstract}


\section{Introduction}
\label{Sec:Introduction}

The redshifted 21-cm hyperfine line emission from the neutral hydrogen that pervades the intergalactic medium (IGM) at high redshifts ($z \gtrsim 6$) provides a unique probe of the Cosmic Dawn and the Epoch of Reionization (EoR) when the first stars, galaxies and quasars formed, effecting a transition of the IGM from a neutral to ionized state. 

The redshift dependent hydrogen spin-temperature distribution during reionization (see e.g. \citealt{1958PIRE...46..240F, 1959ApJ...129..525F, 1990MNRAS.247..510S}) encodes a wealth of information which, if extracted, can be used to constrain cosmological parameters (e.g. \citealt{2006ApJ...653..815M, 2008PhRvD..78b3529M, 2009MNRAS.394.1667F, 2014ApJ...782...66P, 2015aska.confE..12P}), probe directly the initial stages of structure formation and deduce the nature of the first ionizing sources (e.g. \citealt{2012MNRAS.424..762D, 2013MNRAS.431..621M, 2014MNRAS.439.3262M, 2015MNRAS.449.4246G}).

Over the last few years, the first generation of low-frequency interferometers designed to detect the highly redshifted 21-cm signal from the EoR have begun taking measurements. These include: the Giant Metrewave Radio Telescope (GMRT; \citealt{2013MNRAS.433..639P})\footnote{http://www.gmrt.ncra.tifr.res.in}, the LOw Frequency ARray (LOFAR; \citealt{2013A&A...556A...2V})\footnote{http://www.lofar.org/}, the Murchison Widefield Array (MWA; \citealt{2013PASA...30....7T})\footnote{http://www.mwatelescope.org/} and the Donald C. Backer Precision Array for Probing the Epoch of Reionization (PAPER; \citealt{2010AJ....139.1468P})\footnote{http://eor.berkeley.edu/}. Over the next few years, the second generation of instruments with larger collecting areas will begin construction and operation. These include the Hydrogen Epoch of Reionization Array (HERA; \citealt{2016arXiv160607473D})\footnote{http://reionization.org/} and the Square Kilometre Array (SKA; \citealt{2013ExA....36..235M})\footnote{https://www.skatelescope.org/}. 

Statistical information about the expansion of ionized bubbles and the astrophysical sources that produced them is encoded in the 21-cm temperature power spectrum of the EoR.
For isotropic emission, the spatial power spectrum can be calculated through two-dimensional spatial averaging over circular annuli as a function of spatial scale.
For the redshifted 21-cm line emission, observation frequency maps to line of sight distance. Thus, for a sufficiently slowly evolving 21-cm signal averaging of the power spectrum can be extended to three dimensions for narrow frequency bands within which the evolution of the signal is minimal \citep{2012MNRAS.424.1877D, 2014MNRAS.442.1491D}. Performing three-dimensional averaging of this form using spherical shells or cylindrical annuli defines the spherical and cylindrical power spectrum respectively. As a result of the significant gains in signal to noise, facilitated by averaging of the data, detection of the power spectrum of the EoR signal and disentangling the cosmological and astrophysical information that it encodes is the primary goal of current EoR experiments (see e.g. \citealt{2015MNRAS.449.4246G}).

Theoretical models predict the 21-cm temperature signal from the EoR to be of the order of $10~\mr{mK}$ (e.g. \citealt{2011MNRAS.411..955M}). 
The largest challenge faced by both current and next generation instruments is the detection of this signal in the presence of astrophysical foreground contaminants that are up to 5 orders of magnitude brighter in intensity \citep{1999A&A...345..380S}. In order to detect the faint EoR power spectrum and disentangle it from the foregrounds in an unbiased manner, accurate characterisation of each in power spectral space can provide valuable insight. 

In this paper we develop EoR and foreground simulations and present the results of applying a Bayesian methodology to infer the `intrinsic' (instrument-free) two-dimensional spatial power spectrum, and cylindrically and spherically averaged three-dimensional $k$-space power spectra of these simulations. We also estimate the spherically averaged three-dimensional $k$-space power spectra from the simulated interferometric observation of the foregrounds. While the foregrounds do not possess the frequency-to-redshift mapping valid for the 21-cm emission, their $k$-space power spectra nevertheless describe foreground power at the spatial and spectral scales of interest for measuring the EoR. Their power spectra, therefore, define a fundamental measurement of contamination of the EoR power spectrum by foreground emission. 

In \autoref{Sec:ForegroundRemoval} we summarize common foreground avoidance / removal strategies used to date. In \autoref{Sec:PowerSpectralModelandAssumptions} we summarize the method of Bayesian power spectral estimation as presented in Lentati et al. 2016 (in prep.) with particular focus on the method of direct sampling from the spherical power spectrum coefficients of the EoR which we make use of in what follows. In Sections \ref{Sec:CosmologicalSignal}, \ref{Sec:GalacticForegroundEmission} and \ref{Sec:ExtragalacticForegroundEmission} we develop our EoR, Galactic and extragalactic emission simulations respectively. In \autoref{Sec:SimulatedHERAObservations} we describe our instrumental simulation modelled on HERA in 37 and 331-antenna configurations. We use a frequency dependent Gaussian approximation to the HERA primary beam and include realistic frequency dependent $uv$-coverage of the simulated observations. In \autoref{Sec:Analysis} we analyse the power spectra of our foreground and 21-cm simulations. Using the spatial power spectrum we calculate a model dependent region of optimal signal estimation within which the ratio of EoR to foreground power is maximised. We estimate the cylindrically and spherically averaged intrinsic power spectra of each of our simulation components and analyse contamination of the EoR power spectrum by foregrounds as a function of position in $k$-space. Finally, we derive the corresponding instrumental power spectra and show their detected coefficients are fully consistent with the intrinsic power spectra. We discuss the implications of these results for optimal strategies for unbiased estimation of the EoR power spectrum and in the context of current experimental approaches to EoR power spectral estimation. We offer some concluding remarks and discuss future work in \autoref{Sec:Conclusions}.

\section{Foreground Removal}
\label{Sec:ForegroundRemoval}

Experiments designed to measure statistical properties of the redshifted 21-cm emission from the EoR must extract statistics from noisy data dominated by strong astrophysical foregrounds, distorted by the ionosphere and instrumental response of the interferometer and with uneven sampling in frequency as a result of data flagging in the presence of radio frequency interference (RFI) and Galactic radio recombination lines.

The focus of statistical extraction techniques discussed in the literature to date has been the derivation of power spectral parameters and error estimates either in a frequentist manner, or, more recently in the analysis of \citet{2015MNRAS.452.1587G}, using a hybrid methodology applying foreground subtraction as an independent step to Bayesian power spectral estimation. In \autoref{Sec:ForegroundRemovalTechniques}, we summarise existing power spectral estimation methodologies. 

In this paper we estimate the 21-cm power spectrum in the presence of astrophysical foregrounds following a new, fully Bayesian analysis derived in Lentati et al. 2016 (in prep.; hereafter L16). In \autoref{Sec:Foregroundsimulations} we consider requirements for foreground simulations to provide a robust test of the effectiveness of this power spectral estimation framework and discuss emission features that can influence the power spectrum and that we incorporate into our simulations.

\subsection{Foreground removal techniques}
\label{Sec:ForegroundRemovalTechniques}

As discussed above, detection of the EoR power spectrum is dependent on adequate techniques to deal with the strong foregrounds. Proposed methods share a common initial step: the removal of bright contaminating sources (\citealt{2006ApJ...648..767M}). This includes both astrophysical sources, that can be excised, for example, through source peeling (e.g. \citealt{2009A&A...501.1185I}) and RFI that can be avoided by flagging affected channels not to be included in the analysis. 

Following the removal of bright sources, at best down to the confusion noise level of the instrument (see e.g. \citealt{2012ApJ...758...23C}), residuals due to imperfections in foreground subtraction and remaining foregrounds not subtracted in the previous step will remain. The challenge of estimating the EoR signal and avoiding contamination by these foregrounds has been considered by many authors. We can generally categorise the proposed approaches as belonging to one of two strategies: i) signal estimation following blind foreground removal \citep{2002ApJ...564..576D, 2003MNRAS.346..871O, 2004MNRAS.355.1053D, 2006ApJ...648..767M, 2008MNRAS.389.1319J, 2009ApJ...695..183B, 2009MNRAS.397.1138H, 2009MNRAS.398..401L, 2010MNRAS.409.1647J, 2011MNRAS.413.2103P, 2011PhRvD..83j3006L, 2012MNRAS.419.3491L, 2012MNRAS.423.2518C, 2013ApJ...763...90W, 2015ApJ...799...90B, 2015MNRAS.447.1973B, 2015MNRAS.447..400A, 2015MNRAS.452.1587G}; ii) foreground avoidance strategies -- estimation of the cosmological signal in a region of three-dimensional cosmological $k$-space with minimal contamination by foreground emission, often referred to as the `EoR window' and visualised by its two-dimensional projection in $k_{\perp}$ vs. $k_{\parallel}$-space \citep{2009AJ....138..219P, 2012ApJ...756..165P, 2013ApJ...768L..36P, 2013ApJ...769..154M, 2013PhRvD..87d3005D, 2013ApJ...770..156H, 2014PhRvD..90b3018L, 2014arXiv1408.4695C}. Both of these strategies rely on exploiting the spectral smoothness of the foregrounds relative to the EoR signal, which is expected to vary rapidly in all three spatial dimensions.

The blind foreground removal strategies that have been studied can all be considered part of a general framework of performing a least $\chi$-squared minimisation of a set of basis functions to the total emission. The goal is then to choose basis vectors that best differentiate between 21-cm signal that is characterised by rapid fluctuation in all dimensions and the foregrounds which can vary rapidly spatially but are dominated by a smoothly varying component of power-law-like emission spectrally. The chosen basis can then be used to attempt to isolate the foreground emission and subtract it so that what remains is the 21-cm signal plus thermal noise which can be averaged down with increasing observation time. Calculation of the power spectrum of the subtraction residuals both in a frequentist manner and more recently using a Bayesian framework \citep{2015MNRAS.452.1587G} is then performed. What differentiates the foreground subtraction approaches are the choice or method of derivation of the basis vectors and the number of basis vectors chosen to best fit the foregrounds. Two recent papers (\citealt{2015MNRAS.447..400A, 2015aska.confE...5C}) consider a number of foreground fitting methods including polynomial fitting, principle component analysis (PCA) and independent component analysis (ICA). This is done in the context of redshifted 21-cm intensity mapping 21-cm power spectral estimation respectively. They find that the different methods perform to a similar level but that each has the potential to produce a biased estimate of the power spectrum particularly when the foreground signal deviates from spectral smoothness. \citet{2006ApJ...648..767M} have provided a method for tackling this bias within the foreground subtraction framework by calculating the shape of residual foreground contamination in the power spectrum for a given choice of basis vectors. They demonstrate this method in the case of a quadratic model for foreground emission. Assuming foreground residual templates can be derived for more complex foreground models, this enables the amplitude of both an EoR and a foreground template to be estimated in power spectral space thus mitigating the risk of foreground bias expected from direct calculation of the power spectrum of the residuals following foreground subtraction. Additionally, \citet{2016ApJ...818..139T} have recently proposed a method of joint estimation of the EoR signal and foregrounds and demonstrated its use with MWA data. This allows for a full covariant understanding of the outputs which is a characteristic shared by the approach proposed in this paper. The notable difference is that \citet{2016ApJ...818..139T} do not forward model the instrument in their likelihood and consider joint estimation in the context of a maximum likelihood as opposed to Bayesian estimation of the EoR power spectrum.

Foreground avoidance methods rely on the relative spectral smoothness of the foregrounds by confining measurements of the EoR power spectrum to the `EoR window'. The `EoR window' can be understood through a number of routes but the `delay spectrum' approach is particularly intuitive. Performing a frequency Fourier transform along the visibility spectrum of a single baseline maps the received emission to `delay space' where delay, $\tau_{d}$, is the light travel time along the projected baseline length between two antennas (e.g. \citealt{2012ApJ...756..165P}). The delay spectrum at each baseline separation is therefore confined below the delay corresponding to the light travel time between the antennas. That is, the measured delay $\tau_{d} \le \abs{\mathbfit{u}}/c$ where $\abs{\mathbfit{u}}=(u^2+v^2)^{1/2}$ is the baseline separation and $c$ is the speed of light. Therefore, the centres of all measured delay spectra are confined to lie on or below the line defining the `horizon limit', $\tau_{d}=\abs{\mathbfit{u}}/c$, producing a wedge of emission in signal delay vs. $uv$-distance space. Additionally, since the Fourier transform of the broad, smooth function describing the foreground spectra is a narrow, highly peaked function in delay space, the emission from smooth spectrum foregrounds will be strongly localised in delay. On the contrary, the rapidly varying 21-cm emission will have a broad convolution kernel in delay space causing a larger proportion of the EoR signal to be convolved above the horizon line where it can be measured relatively free of foreground contamination. As a result, the area below $\tau_{d}=\abs{\mathbfit{u}}/c$ in signal delay vs. $uv$-distance space has been referred to as the `foreground wedge' and the area above, as the foreground-free `EoR window' \citep{2010ApJ...724..526D}. The impact of the instrument on the power spectrum is inextricably linked to these features. As such, in this paper we will refer to them as the `instrumental foreground wedge' and `instrumental EoR window'\footnote{In \autoref{Sec:Analysis} we will show, using the EoR and foreground models described in this paper, that the intrinsic (non-instrumental) $k$-space power spectrum of the EoR is characterised by regions of greater or lesser foreground contamination. These regions can be described as an intrinsic foreground wedge resulting solely from the astrophysical structure of the emission and a corresponding intrinsic EoR window in the complement to this region. We therefore refer to an `intrinsic' and an 'instrumental' foreground wedge and EoR window to differentiate between the distribution of power in $k$-space resulting from the astrophysical structure of the emission and that tied to the instrument respectively.}.

Each of the methods described above has limitations. The method of foreground avoidance restricts the measurement of the power spectrum to the EoR window, relegating a large portion of the signal to the contaminated foreground wedge below the horizon line. Any deviation from assumed smooth spectra, for example from synchrotron self-absorption of source spectra (see \autoref{Sec:OpticallyThick}), other astrophysical mechanisms, or instrumental effects (e.g. polarization leakage or other imperfect instrumental calibration \citealt{2011MNRAS.418..516G, 2010MNRAS.409.1647J, 2015MNRAS.451.3709A, 2016ApJ...823...88K}), have the potential to scatter foreground power into the EoR window biasing the extracted power spectrum parameters. 

Foreground subtraction prior to estimation of the EoR power spectrum requires a priori knowledge of the complexity and covariance of these two components. Blind foreground subtraction without this knowledge will bias the extracted signal, either through contamination by unmodelled foreground emission or absorption of the 21-cm signal within the foreground model. Additionally, if the two components are correlated in the data, as will be the case in interferometric observations of the foreground and EoR signal, fixing the foreground model, and then estimating the EoR signal, will not result in correct uncertainties on the signal parameters. Fitting for a foreground residual power spectral template and the EoR signal as outlined by \citet{2006ApJ...648..767M} could mitigate these issues if, for a given foreground model, an accurate power spectral template can be obtained.

\subsection{Foreground simulations}
\label{Sec:Foregroundsimulations}

In this paper we are interested in estimating the power in the foregrounds on scales relevant to the detection of the EoR. To achieve this, we make use of a new power spectral estimation strategy. We jointly estimate the foreground and EoR signal and do this within a unified Bayesian framework, allowing for robust estimation of both the signal parameters and their uncertainties (see \autoref{Sec:PowerSpectralModelandAssumptions} for details). 

Astrophysical foreground emission in the frequency range relevant to the detection of the redshifted EoR signal $(\sim 50$ -- $200~\mr{MHz})$ results from three key sources (see e.g. \citealt{1999A&A...345..380S}): Galactic diffuse synchrotron emission (GDSE; accounting for $\sim70\%$ of total intensity emission), extragalactic sources (EGS; $\sim30\%$) and free--free emission ($\sim1\%$). The challenge of simulating each, or a subset, of these foregrounds for testing EoR estimation frameworks has been considered by a number of authors (e.g. \citealt{2008MNRAS.389.1319J, 2009ApJ...695..183B, 2010MNRAS.409.1647J, 2012ApJ...744...29M, 2012MNRAS.424.2562Z, 2015ApJ...804...14T, 2015MNRAS.447.1973B}).

The foreground components have markedly different spatial and spectral structures. As a result, their power spectra will be similarly distinct. The general form of the foreground power spectra as a function of position in $k$-space can be predicted for a set of assumed spatial and spectral characteristics (e.g. \citealt{2004ApJ...615....7M, 2006ApJ...648..767M}). An important prediction is that, as a result of their relative spectral smoothness, the power in foregrounds will be concentrated at low values of $k_{\parallel}$ (corresponding to large spectral scales). However, detailed quantitative description of these features in the intrinsic power spectra of the foregrounds has not, before now, been provided. For example, how far does significant foreground power (relative to the EoR) extend in $k_{\parallel}$ and how does it vary as a function of $k_{\perp}$ and by foreground type. The power spectrum of a sum of signals is equal to the sum of their power spectra. Therefore, to disentangle these effects and to estimate the contribution of individual foreground components, in addition to the total foreground power spectrum, we construct foreground simulations for each component.

As described in \autoref{Sec:ForegroundRemovalTechniques}, there are a number of results in the literature which suggest that in ideal conditions -- assuming perfect instrumental calibration, and ignoring complications resulting from a complex spectral window function caused by flagging of RFI -- foregrounds can be well fit by sufficiently complex foreground models. However, there has been far less investigation of the covariance of these models with the EoR signal. The effects of this covariance can be seen, for example, in the results of \citet{2015MNRAS.452.1587G} where power spectral estimation of the residual signal following application of generalised morphological component analysis for foreground subtraction has been performed. In that case, foregrounds are found to be over-subtracted at low $k$. Assuming a more-complex-than-necessary model to fit the foreground simulations was not applied, this implies that the foreground simulations used there do indeed possess power on scales of interest for estimating the power in the EoR signal.

Similarly, in the absence of information on the covariance between specific foreground models and the EoR at the scales of interest for estimating the EoR power spectrum, model fitting of foreground features such as synchrotron self-absorption in EGS (see e.g. \citealt{2012MNRAS.424.2562Z}) or the similarity in subtraction residuals from correlated versus uncorrelated temperature--spectral index distributions in GDSE (see e.g. \citealt{2008MNRAS.389.1319J}), provides limited insight into the impact of those features on the contribution to the power spectrum of the foregrounds.

Further, L16 highlight the challenge to power spectral estimation frameworks posed by large scale Galactic structure. Methods that have been used to model diffuse Galactic emission broadly fall into two categories: those constructed as a random realisation of the sky with a prescribed spatial power spectrum and using statistical matching (such as the mean and standard deviation) to observational data (e.g. \citealt{2008MNRAS.389.1319J, 2009ApJ...695..183B, 2010MNRAS.409.1647J}) and those that are constructed from observational results directly (e.g. \citealt{2012ApJ...744...29M, 2012MNRAS.424.2562Z, 2015ApJ...804...14T, 2015MNRAS.447.1973B}). When calculating the power spectrum of emission using data from a subset of the sky, L16 show that, if not modelled when estimating the power spectrum, large-scale structure resulting from, for example, the full sky emission gradient in GDSE towards the plane of the Galaxy has the potential to bias power spectral estimation. Therefore foreground simulations constructed directly from observational data and reflecting large scale structure are preferred.

We aim to address these issues when constructing our foreground simulations. We approach this in the following ways:

\begin{itemize}
\item \textit{Modelling of large scale Galactic structure:}
The spatial structure model for our GDSE simulation is derived using a Bayesian power spectral decomposition of a $48\fdg0 \times 48\fdg0$ field of the Haslam all-sky survey \citep{1981A&A...100..209H, 1982A&AS...47....1H}. This has the twin benefit of enabling us to accurately model the synchrotron intensity distribution and large scale structure present in the data and to isolate in power spectral space the most likely synchrotron-only component of the map.
\item \textit{Spatial and temperature dependence of the GDSE spectral index distribution:} The spectral model we use for the GDSE simulation incorporates the spatial dependence and temperature correlation of the spectral index distribution of the emission such that the impact of the level of correlation on the measured power spectrum can be quantitatively analysed.
\item \textit{Physically motivated EGS spectral structure:} Our EGS simulation uses the Square Kilometre Array (SKA) Simulated Skies ($\mr{S^{3}}$) Simulation of Extragalactic Sources ($\mr{S^{3}}$-SEX) \citep{2008MNRAS.388.1335W}. We draw spectral indices for the sources from an experimentally derived distribution and model deviations from power law spectra in compact sources with a physically motivated spectral absorption model. The impact of deviations from a simple power law structure will thus be accounted for in the resulting power spectrum.
\end{itemize}

\section{Power-spectral model and assumptions}
\label{Sec:PowerSpectralModelandAssumptions}

Our methodology for estimating the intrinsic power spectra of the EoR and foregrounds relies on Bayesian inference to perform parameter estimation and is based on the method described in L16. We now present the key aspects of this approach.

Given a set of data $\bm{D}$, Bayesian inference provides a statistically robust approach to estimating a set of parameters, $\sTheta$, from a hypothesis or model, $M$, via Bayes' theorem,
\begin{equation}\label{Eq:BayesEqn}
\mr{Pr}(\sTheta\vert\bm{D},M) = \dfrac{\mr{Pr}(\bm{D}\vert\sTheta,M)\ \mr{Pr}(\sTheta\vert M)}{\mr{\mr{Pr}}(\bm{D}\vert M)} = \dfrac{\mathcal{L}(\sTheta)\pi(\sTheta)}{\mathcal{Z}} \ , 
\end{equation}
where $\mr{Pr}(\sTheta\vert\bm{D},M)$ is the posterior probability distribution of the parameters\footnote{In what follows, for notational simplicity, we leave the model dependence implicit, denoting the posterior probability distribution as $\mr{Pr}(\sTheta\vert\bm{D})$.
}, $\mr{Pr}(\bm{D}\vert\sTheta,M) \equiv \mathcal{L}(\sTheta)$ is the likelihood, $\mr{Pr}(\sTheta\vert M) \equiv \pi(\sTheta)$ is the prior probability distribution of the parameters and $\mr{Pr}(\bm{D}\vert M)\equiv\mathcal{Z}$ is the Bayesian evidence. 

Since the evidence is independent of the parameters $\sTheta$, to make inferences regarding the model parameters we sample from the unnormalised posterior,
\begin{equation}
\label{Eq:UnnoramisedPosterior}
\mr{Pr}(\sTheta\vert\bm{D}) \propto \mathcal{L}(\sTheta)\pi(\sTheta) \ .
\end{equation}
We choose to perform our model comparison in the $uv$-domain (the measurement domain of interferometric EoR experiments) to enable us to trivially restrict our analysis to specific spatial scales of interest when estimating the three-dimensional $k$-space power spectrum. 

We define a Gaussian likelihood function for our power spectral model as,
\begin{eqnarray}
\label{Eq:BasicVisLike}
\mathcal{L}(\sTheta) \propto \frac{1}{\sqrt{{\mr{det}(\mathbfss{N})}}} \exp\left[-\frac{1}{2}\left(\mathbfit{d} - \mathbfit{m}(\sTheta)\right)^{\dagger}\mathbfss{N}^{-1}\left(\mathbfit{d} - \mathbfit{m}(\sTheta)\right)\right] \ ,
\end{eqnarray}
where $\mathbfit{d} = \mathbfit{s} + \sdelta\mathbfit{n}$, is our data vector and is comprised of the signal $\mathbfit{s}$ and of noise $\sdelta\mathbfit{n}$. The signal (corrupted by noise) is in principle observed. In this paper it is obtained from simulated image cubes through the two-dimensional discrete Fourier transform (DFT) to $uv$-space of each channel. The noise in the $uv$-domain is modelled as an uncorrelated Gaussian random field, with covariance matrix $\mathbfss{N}$. The elements of the covariance matrix are given by, $\mathbfss{N}_{ij} = \left< n_in_j^*\right> = \delta_{ij}(\sigma_{j}^{2}+\alpha_{j}^{2})$. Here $\left< ... \right>$ represents the expectation value, $\sigma_{j}$ is the RMS value of the noise in visibility element $j$ and $\alpha_j$ is a small-scale-structure model parameter which accounts for the high-frequency structure on scales smaller than the channel width which manifests itself as an additional source of noise.

\subsection{Data model}
\label{Sec:DataModel}

We want to make inferences regarding the $k$-space power spectrum of the signal from the $uv$-domain representation of our EoR and foreground simulations. We construct our data model, $\mathbfit{m}$, via a transformation matrix $\mathbfss{T}$ for the model parameters, $\sTheta$, from a three-dimensional grid in $k$-space, $K_{m}(k_x, k_y, k_z)$, to their measurement-domain representation, $V_{m}({u, v, \nu_{i}})$.

Included in the definition of $\mathbfss{T}$ is a model for low-frequency structure in the data vector, both spatial and spectral, in addition to a model for power at well-sampled frequencies -- those frequencies fulfilling the Nyquist sampling criterion, (see e.g. \citealt{1928TAIEE..47..617N}). This, in combination with the high-frequency structure model in $\mathbfss{N}$, means that power on frequencies above and below those measurable with perfect fidelity, given the data sampling rate, will not leak into the well-sampled scales of interest, enabling unbiased parameter estimates and robust estimates on their uncertainties to be obtained.

A model for low-frequency spatial structure in the data vector (corresponding to spatial scales greater than or equal to the image size, $\theta \ge \theta_{\mr{im}}$ and $|\mathbfit{u}| \le 1/\theta_{\mr{im}}$ where $\theta_{\mr{im}}$ is the image size), is built directly into the $k$-space grid. This is achieved in a computationally efficient manner\footnote{Reducing our model cell width by a factor $N$ to ($\sDelta \mathbfit{u}=1/N\theta_{\mr{im}}$) would be adequate for modelling large scale spatial structure in the data but would increase the dimensionality and corresponding computation time by a factor $N^{4}$.}
by defining for each $k_z$ a joint model grid in $(k_x, k_y)$ comprised of two parts. The first is a `coarse' grid with spacing $\sDelta\mathbfit{u}\simeq 1/\theta_{\mr{im}}$. The second, embedded at the origin of the coarse grid and modelling power on large spatial scales, is a more finely spaced `sub-harmonic grid' containing a set of 10 log-uniformly spaced spatial scales between the size of the image and 10 times the size of the image. As will be shown in \autoref{Sec:GalacticForegroundEmission}, inclusion of a large spatial scale model will be particularly pertinent in the case of Galactic foregrounds where large scale structure, resulting from, for example, full-sky emission gradients towards the plane of the Galaxy, is present in the data.

The transformation matrix $\mathbfss{T}$, can be expressed as the multiplication of a sequence of intermediate transformations between $K_{m}(k_x, k_y, k_z)$ and $V_{m}({u, v, \nu_{i}})$. Firstly a redshift dependent coordinate re-definition of the parameters from $K_{m}(k_x, k_y, k_z)$ to $K_{m}(u^{\prime}, v^{\prime}, \eta_{i})$ is performed where primed coordinates signify the inclusion of the `sub-harmonic grid' in addition to the coarse grid. The relation between $k$-space and $uv\eta$-coordinates (both primed and non-primed) is given by (see e.g. \citealt{2004ApJ...615....7M}),
\begin{eqnarray}
\label{Eq:CoordinateConversion}
u &=& \dfrac{k_x D_M(z)}{2\pi}, \nonumber \\
v &=& \dfrac{k_y D_M(z)}{2\pi}, \\
\eta &\approx& \dfrac{c(1+z)^2 k_z}{2\pi H_\mr{0}f_{21}E(z)}. \nonumber 
\end{eqnarray} 
where $f_{21} \simeq 1420$ MHz is the rest frame frequency of the 21-cm line, $D_{M}$ is the transverse comoving distance from the observer to the redshift $z$ of the EoR observation (which for $\Omega_{\rm k}=0$, as assumed here, is simply equal to the comoving distance $D_{C}= (c/H_\mr{0})\int_{0}^{z} \mr{d}z^\prime / E(z^\prime)$, \citealt{1999astro.ph..5116H}) and $E(z) \equiv \sqrt{\Omega_{\rm M}(1+z)^{3} + \Omega_{\rm k}(1+z)^{2} + \Omega_{\sLambda}}$ is the dimensionless Hubble parameter. 

To move to the $(\mathbfit{u}, \nu)$ measurement space of the data vector, $K_{m}(u^{\prime}, v^{\prime}, \eta_{i})$ is transformed along $\eta$ to form $K_{m}(u^{\prime}, v^{\prime}, \nu_{i})$. At this point a large spectral scale model is incorporated. Well-sampled fluctuations in the data on scales smaller than the bandwidth ($1/B \ge \eta \ge N_{c}/2B$, with $N_{c}$ the number of channels and $B$ the simulation bandwidth) and reconstructible with perfect fidelity are modelled by Fourier modes encoded by a one-dimensional DFT matrix, $\mathbfss{F}_z$. These are fit for jointly with a set of quadratics\footnote{See L16 for a discussion of using quadratics vs. using lower frequency Fourier modes to model low-frequency structure in the data.}
which model the low-frequency fluctuations in the data, on scales longer than the bandwidth ($1/2B \ge \eta$), and are defined by the matrix $\mathbfss{Q}_z$. Since the foregrounds are expected to be dominated by spectrally smooth structure on scales larger than the bandwidth, this component will be modelled by the quadratics.

Next, a transform from $K_{m}(u^{\prime}, v^{\prime}, \nu_{i})$ to a uniformly gridded image $I_{m}(x_{p}, y_{q},\nu_{i})$ is performed using the two-dimensional DFT matrix $\mathbfss{F}^{\prime}$ where the prime denotes the transform from the $(u^{\prime}, v^{\prime}, \nu_{i})$ coordinates of our coarse-plus-sub-harmonic grid to the uniformly gridded coordinates of our image-space simulation cube. The final transformation performed takes one of two forms dependent upon whether the intrinsic power spectrum of the signal from the full gridded $uv$-plane or the power spectrum from a subset of the $uv$-plane given by an instrumental sampling function is to be calculated. In the first case, the transform is from $I_{m}(x_{p}, y_{q},\nu_{i})$ to the uniformly gridded visibilities $V_{m}({u, v, \nu_{i}})$ using a standard inverse discrete Fourier transform matrix $\mathbfss{F}^{-1}$. In the second case, the transformation has two components. Firstly from the gridded image $I_{m}(x_{p}, y_{q},\nu_{i})$ to a primary beam corrected model image via multiplication by the primary beam matrix $\mathbfss{P}$. Secondly to the instrumentally sampled $uv$-coordinates via a non-uniform DFT, $\mathbfss{F}_{\mr{n}}^{-1}$ (see L16 for further details). In what follows we assume the first case, however the instrumental equivalent can be obtained at any stage by substituting the operator $\mathbfss{F}^{-1}$ for $\mathbfss{F}_{\mr{n}}^{-1} \mathbfss{P}$.

By dividing the $k$-space model parameters into two sets, $\mathbfit{a}$ and $\mathbfit{q}$, which represent the amplitudes on well--sampled scales and large spectral scale fluctuations along the $\eta$-axis respectively, the final model vector is written as,
\begin{equation}
\label{Eq:FullModel}
\mathbfit{m} = \mathbfss{T}\sTheta = \mathbfss{F}^{-1}\mathbfss{F}^{\prime}\left(\mathbfss{F}_z\mathbfit{a} + \mathbfss{Q}_z\mathbfit{q}\right) \ .
\end{equation}
Here $\mathbfss{F}_z$ and $\mathbfss{Q}_z$ represent block diagonal matrices, where the $i$th block multiplies the vector of coefficients, $\mathbfit{a}_{i}$ and $\mathbfit{q}_{i}$, for model $uv$-cell $i$. 

In addition to the data model, the redshifted 21-cm signal is assumed to be spatially isotropic and, over the redshift interval under consideration, homogeneous (assuming the power spectrum of the 21-cm signal is approximately constant) and uncorrelated between spatial scales. The covariance matrix $\sPsi$ of the $k$-space coefficients $\mathbfit{a}$ is given by the outer product of a vector $\bm{\varphi}$ and the identity matrix,
\begin{equation}
\label{Eq:Prior}
\Psi_{ij} = \left< a(k_{i})a(k_{j})\right> = \varphi_{i}I_{ij} \ ,
\end{equation}
where $\bm{\varphi}$ represents the theoretical power spectrum of the EoR signal with units mK$^2$.

When constructing the spherically averaged power spectrum of the signal, the spatial homogeneity of the signal over a narrow redshift interval is made use of to average over spherical shells in $k$-space. The model for the spherically averaged power spectrum $\bm{\varphi}$ is hence given by a set of independent parameters $\varphi_{i}$, one for each $k$ bin $i$, where $k=\sqrt{k_x^2 + k_y^2 + k_z^2}$.

The final joint probability density of the model coefficients that define the power spectrum and the $k$-space signal coefficients is therefore,
\begin{equation}
\label{Eq:Prob}
\mr{Pr}(\bm{\varphi}, \mathbfit{a} , \mathbfit{q}\;|\; \mathbfit{d}) \; \propto \; \mr{Pr}(\mathbfit{d} | \mathbfit{q}, \mathbfit{a}) \; \mr{Pr}(\mathbfit{a} | \bm{\varphi}) \; \mr{Pr}(\bm{\varphi}) \; \mr{Pr}(\mathbfit{q}) \ .
\end{equation}

\subsection{Priors}
\label{Sec:Priors}

For this work, we assume a uniform prior on the amplitude of the quadratic parameters $\mathbfit{q}$ such that $\mr{Pr}(\mathbfit{q})=1$. In the high signal to noise regime we select the least informative prior for our choice of $\mr{Pr}(\bm{\varphi})$: a log-uniform prior on the amplitude of the power spectral signal coefficients. In the low signal to noise regime when the power spectral coefficients are undetected we apply a uniform prior in order to calculate upper limits.

To implement these priors we sample from the parameter $\rho_{i}$, which parametrises $\varphi_{i}$, such that,
\begin{equation}
\label{Eq:PowerSpectrumPrior}
\varphi_{i} = \gamma(k_{i}) 10^{2\rho_{i}} \ . 
\end{equation}
Here $\gamma=2\pi^{2}N_{\mr{im}}/ k^3V$ is a conversion factor\footnote{$10^{\bm{\rho}}$ is the power spectrum of the signal in units mK$^2$. The often used physical power spectrum is simply related by $P(k)=\dfrac{(2\pi)^{3}\bm{\varphi}}{4\pi~k^{3}}$, with $P(k)$ in units of $\mr{mK^2(h^{-1}Mpc)^3}$.} 
between the dimensional power spectrum $\bm{\varphi}$ and the dimensionless power spectral coefficients $10^{\bm{\rho}}$, where $N_{\mr{im}}$ is the number of voxels in our model image cube and $V$ the surveyed volume in $\mr{(h^{-1}Mpc)^3}$.

In \autoref{Eq:Prob} we can now write,
$\mr{Pr}(\mathbfit{a}\vert\bm{\varphi})\mr{Pr}(\bm{\varphi}) = \mr{Pr}(\mathbfit{a}\vert\bm{\rho})\mr{Pr}(\bm{\rho})$.
For our log-uniform prior on $\bm{\varphi}$ we have $\mr{Pr}(\bm{\rho})=1$ which gives us,
\begin{equation}
\label{Eq:LogUniformPrior}
\mr{Pr}(\mathbfit{a}\vert\bm{\rho})\mr{Pr}(\bm{\rho}) \propto \dfrac{1}{\sqrt{\mr{det}(\sPsi)}} \exp\left[ -\dfrac{1}{2}\mathbfit{a}^{\dagger}\sPsi^{-1}\mathbfit{a} \right] \ .
\end{equation}
In the case of a uniform prior on $\bm{\varphi}$ we have $\mr{Pr}(\bm{\rho}) = \prod_{s=1}^{N_{s}}10^{\rho_{s}}$ where $N_{s}$ is the number of spherical power spectrum bins used in the prior which gives us,
\begin{equation}
\label{Eq:UniformPrior}
\mr{Pr}(\mathbfit{a}\vert\bm{\rho})\mr{Pr}(\bm{\rho}) \propto \dfrac{1}{\sqrt{\mr{det}(\sPsi)}} \exp\left[ -\dfrac{1}{2}\mathbfit{a}^{\dagger}\sPsi^{-1}\mathbfit{a} \right] \prod_{s=1}^{N_{s}}10^{\rho_{s}} \ .
\end{equation}

\subsection{Marginalisation over the signal coefficients}
\label{Sec:MarginalisationOvertheSignalCoefficients}

Since we are interested specifically in the power spectrum, the model coefficients $\mathbfit{a}$ and $\mathbfit{q}$ in \autoref{Eq:Prob} can be marginalised over analytically, enabling us to sample from the far smaller dimensional space of the power spectral coefficients $\bm{\varphi}$ in our analysis. This gives our final posterior probability distribution, $\mr{Pr}(\bm{\varphi}|\mathbfit{d})$. For a detailed derivation of this distribution from Equations \ref{Eq:BasicVisLike} and \ref{Eq:Prob}, see L16. Here we quote the solution for the marginalised distribution for the case of log-uniform priors on the power spectral coefficients given in \autoref{Eq:LogUniformPrior},
\begin{eqnarray}
\label{Eq:Margin}
\mr{Pr}(\bm{\varphi}|\mathbfit{d}) &\propto& \frac{\mr{det} \left(\sSigma\right)^{-\frac{1}{2}}}{\sqrt{\mr{det} \left(\sPsi\right)~\mr{det}\left(\mathbfss{N}\right)}} \\
&\times&\exp\left[-\frac{1}{2}\left(\mathbfit{d}^T\mathbfss{N}^{-1} \mathbfit{d} - \bar{\mathbfit{d}}^T{\sSigma}^{-1}\bar{\mathbfit{d}}\right)\right], \nonumber
\end{eqnarray}
where $\overline{\mathbfit{d}}=(\mathbfss{F}^{-1}\mathbfss{F}^{\prime}(\mathbfss{F}_{z} + \mathbfss{Q}_{z}))^{\dagger}\mathbfss{N}^{-1}\mathbfit{d}$ represents the weighted, gridded visibilities, $\sSigma= (\mathbfss{F}^{-1}\mathbfss{F}^{\prime}(\mathbfss{F}_{z} + \mathbfss{Q}_{z}))^{\dagger}\mathbfss{N}^{-1}(\mathbfss{F}^{-1}\mathbfss{F}^{\prime}(\mathbfss{F}_{z} + \mathbfss{Q}_{z})) + \sPsi^{-1}$ is the covariance matrix of $\overline{\mathbfit{d}}$ and the elements of the matrix $\sPsi^{-1}$ that correspond to the coefficients $\mathbfit{q}$ are set to zero. It is simple to show that the maximum likelihood solution for the coefficients $\mathbfit{a}$ and $\mathbfit{q}$ is given by $\sSigma^{-1}\bar{\mathbfit{d}}$.

For all components of our power spectral analysis we perform this analytic marginalisation and make use of \autoref{Eq:Margin}, sampling directly from the marginalised posterior for the spherical power spectrum coefficients $\mr{Pr}(\bm{\varphi}|\mathbfit{d})$ using nested sampling as implemented by the {\sc{MultiNest}} algorithm \citep{2009MNRAS.398.1601F}.

\subsection{Cylindrical power spectra }
\label{Sec:Cylindricalpowerspectra}

In addition to obtaining estimates of the spherical power spectral coefficients, to further analyse the spatial structure of the signal, we calculate the cylindrical power spectrum corresponding to the maximum likelihood set of signal coefficients, $\hat{\mathbfit{a}}=\sSigma^{-1}\bar{\mathbfit{d}}$.

When constructing the maximum likelihood cylindrically averaged power spectrum of the signal, we average over cylindrical shells in $k$-space. We calculate the quantity $k_{\perp} = \sqrt{k_x^2 + k_y^2}$ as a function of $k_{\parallel}$ ($k_{\parallel}\equiv k_{z}$) and we define a set of bins in this quantity spaced uniformly with width double that of our $k_{\perp}$-space model pixel widths, $\sDelta k_{\perp}=2\pi\sDelta u/D_M(z)$. Here $\sDelta u$ is the inverse of the image domain field of view. Our model for the cylindrically averaged power spectrum $\bm{\varphi}_\mr{cyl}$ will then be a set of parameters $\varphi_\mr{cyl}(k_{\perp, i}, k_{\parallel, j})$, one for each $k_{\perp}$ bin $i$ and $k_{\parallel}$ bin $j$.

\subsection{Two-dimensional spatial power spectra }
\label{Sec:2DSpatialpowerspectra}

The posteriors for the two-dimensional spatial power spectrum of the sky components can be derived in an equivalent manner as those for the three-dimensional $k$-space power spectrum with appropriate redefinition of the variables for the two-dimensional case. The resulting posteriors for the analytically marginalized two-dimensional spatial power spectrum can then be expressed in an identical form to \autoref{Eq:Margin}. 

When estimating the two-dimensional spatial power spectrum we define our data vector, $\mathbfit{d}_\mr{s}$, as the two-dimensional DFT of our image-domain simulation at a single frequency. We have corresponding two-dimensional redefinitions, $\overline{\mathbfit{d}}_\mr{s}=(\mathbfss{F}^{-1}\mathbfss{F}^{\prime})^{\dagger}\mathbfss{N}^{-1}\mathbfit{d}_\mr{s}$ and $\sSigma_\mr{s}= (\mathbfss{F}^{-1}\mathbfss{F}^{\prime})^{\dagger}\mathbfss{N}^{-1}(\mathbfss{F}^{-1}\mathbfss{F}^{\prime}) + \sPsi^{-1}$. 

Additionally, we define the parameters for our spatial power spectrum in $({u^{\prime}, v^{\prime}})$-space and calculate the physical rather than dimensionless power spectrum. This enables broader use of our two-dimensional spatial power spectral results and simpler comparison with existing results in the literature\footnote{For example the angular power spectrum, which is related to the $uv$-space power spectrum by the coordinate transformation $l=\pi \abs{\mathbfit{u}}$, and the $(k_x, k_y)$-space equivalent, which is related by the coordinate transformations given in \autoref{Eq:CoordinateConversion}.}.

When estimating the two-dimensional spatial power spectrum we make the approximation of spatial isotropy of the sky components at the scales of interest and bin in circular annuli in $uv$-space. We define our model for the two-dimensional power spectrum, $\bm{\varphi}_\mr{s}$, over a set of annuli in $\abs{\mathbfit{u}} = \sqrt{u^{\prime 2} + v^{\prime 2}}$. We optimise the binning for the two-dimensional power spectral decomposition of the Galactic spatial power spectrum (see \autoref{Sec:GalacticForegroundEmission} for further details). We use three bins over the sub-harmonic grid with log-uniform spacing between $0.17~\lambda$ and $2.4~\lambda$ and seven bins with log-uniform spacing between $2.4~\lambda$ and $53~\lambda$ over the coarse grid.

\section{The cosmological signal}
\label{Sec:CosmologicalSignal}

We simulate the brightness temperature of the 21-cm signal we seek to estimate using the seminumerical simulation 21cmFAST \citep{2007ApJ...669..663M, 2011MNRAS.411..955M}. 21cmFAST generates three-dimensional realizations of evolved density, ionization, peculiar velocity and spin temperature fields, using them to compute 21-cm brightness temperature. While 21cmFAST uses approximate methods in its treatment of the physical processes, comparison with large scale hydrodynamic simulations shows good agreement on scales pertinent to upcoming observations, and power spectra from 21cmFAST agree well with those generated using full numerical simulations \citep{2011MNRAS.411..955M}.

In simulating our EoR signal cube using 21cmFAST, we assume a Lambda Cold Dark Matter ($\sLambda$CDM) universe with $\Omega_{\sLambda}=0.72$, $\Omega_{\mr{m}}=0.28$, $\Omega_{\mr{b}}=0.046$, $\mr{n}=0.96$, $\sigma_{8}=0.82$, $\mr{H}_{0}=0.73$ \citep{2011ApJS..192...18K}. We initialize 21cmFAST at $ z =300$ on a $2048^{3}$ grid with physical dimensions of one comoving $\mr{Mpc}^{3}$ per voxel and form the resulting brightness temperature cube on a $512^{3}$ lower resolution grid. We select the redshift of the centre of our signal cube to be at $z=10.26$ such that, for our input parameters, the neutral fraction is close to a half and the variance of the cosmological signal is close to maximum (see e.g. \citealt{2008ApJ...680..962L}).

These parameters correspond to a field of view with $\theta_{x}=\theta_{y}\approx 12\fdg0$, a pixel resolution, $\Delta \theta_{x}=\Delta \theta_{y}\approx 1.4~\mr{arcmin}$  and a 21-cm frequency depth $\sDelta{\nu}\approx107\ \mr{MHz}$. The observational central frequency of our cube is $\nu_{\mr{c}}=f_{21}/(1+ z)=126\ \mr{MHz}$ for $z=10.26$, where the conversion from cosmological to observational units is given by (e.g. \citealt{2004ApJ...615....7M}),
\begin{eqnarray}
\label{Eq:CoordinateConversion2}
\sDelta \theta_{x}&=&\dfrac{\sDelta r_{x}}{D_\mr{M}( z )}, \nonumber \\
\sDelta \theta_{y}&=&\dfrac{\sDelta r_{y}}{D_\mr{M}( z )}, \\
\sDelta{\nu}&\approx&\dfrac{H_{0}f_{21}E( z )}{c(1+ z )^{2}} \sDelta r_{ z }. \nonumber 
\end{eqnarray} 
Here $\sDelta r_{x}$, $\sDelta r_{y}$ and $\sDelta r_{z}$ are transverse comoving separations of the cube at the redshift of the observation and the remaining variables are defined as in \autoref{Eq:CoordinateConversion}. 

The Bayesian estimation of the power spectrum, described previously, estimates the signal of interest with the assumption that it is fixed within the simulated region. Fulfilling this condition places no constraints on the large scale angular extent of the EoR cube, since cosmological isotropy implies stationary angular dependence of the signal. However, the EoR signal is non-stationary in time, evolving as the universe transitions from a neutral to ionised state. The frequency bandwidth over which the evolution of the signal is expected to have a minimal impact on the power spectrum is of the order of 10 MHz \citep{2012MNRAS.424.1877D, 2014MNRAS.442.1491D}. An analysis seeking to estimate the power spectrum of the EoR at a single point in its evolution is therefore restricted to estimations across a frequency interval within this bound.

The simulation output from 21cmFAST is a coeval cube containing the brightness temperature distribution in a region of space at a single look-back time. It does not, therefore, suffer from this frequency restriction. However to reflect this constrained bandwidth over which a non-evolving EoR power spectrum can be estimated from observational data, we select a 38 channel subset of the cube with channel width 209 kHz, giving us a total bandwidth of 7.94 MHz. Since minimal evolution of the three-dimensional $k$-space power spectrum of the signal is expected over this bandwidth, this subset is a good approximation to the observable signal of interest.

\begin{figure}
	\centerline{\includegraphics[width=\columnwidth]{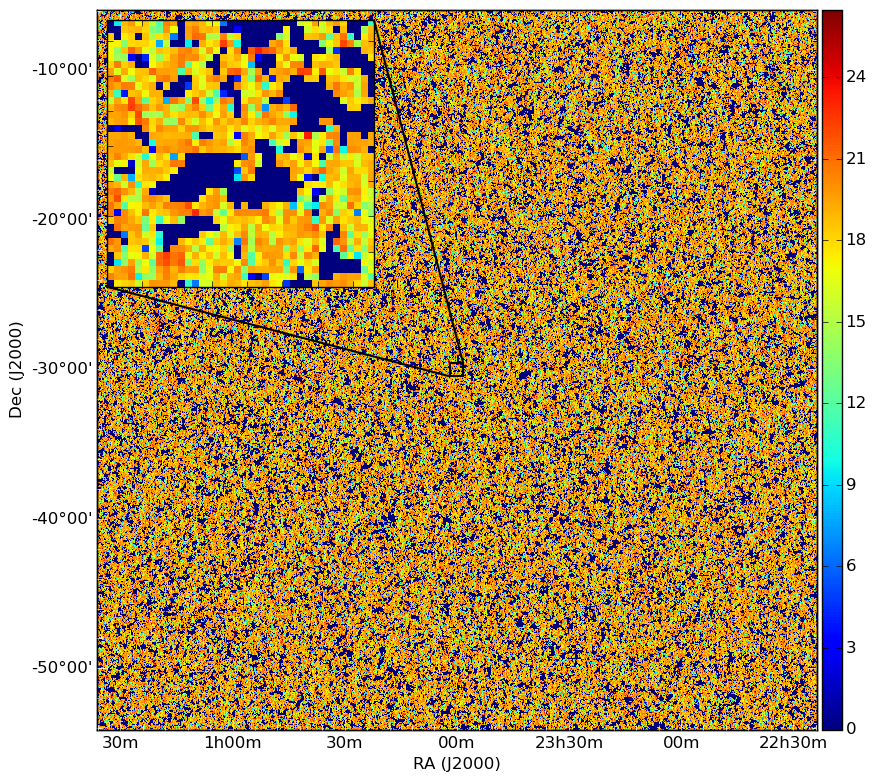}}
	\caption{The $48\fdg0 \times 48\fdg0$ field derived using slices from a $\sim12\fdg0$-cubed 21cmFAST simulation of the differential brightness between the 21-cm spin temperature and CMB temperature at redshift $z=10.26$. The simulation cube has a side of width 2048 $\mr{h^{-1}Mpc}$. The inset shows a spatial region at the centre of the field with $20\times$ magnification. The colourscale is in mK.}
	\label{Fig:Signal}
\end{figure}

We construct our final 21-cm simulation cube by tiling layers from the full coeval cube in the $\theta_{x}$ and $\theta_{y}$ directions to produce a tiled cube with dimensions $(\theta_{x}, \theta_{y}, \sDelta{\nu}) = (48\fdg0, 48\fdg0 , 7.94~\mr{MHz})$. We select this angular extent so that when, in \autoref{Sec:SimulatedHERAObservations}, we construct simulated observed data we obtain a smooth fall-off to zero from the centre to edge of the primary beam which we approximate as a Gaussian with FWHM of order $8\fdg0$. This replaces what would otherwise be a step function at $\sim1.4~\sigma$ from the beam centre in the untiled cube (see \autoref{Sec:SimulatedHERAObservations} for a more detailed description of the simulated observation). Channel 20 of the cube at redshift 10.26 and with central frequency 126 MHz is shown in \autoref{Fig:Signal}. 

\section{Galactic foreground emission}
\label{Sec:GalacticForegroundEmission}

Emission from the Galaxy in the frequency range relevant to detection of the redshifted 21-cm signal from the EoR is dominated by Galactic diffuse synchrotron emission. We start here by constructing our GDSE simulation then, in \autoref{Sec:FreeFreeModel}, follow it with our free--free emission simulation which we produce in a similar manner.

\begin{figure}
	\centerline{\includegraphics[width=\columnwidth]{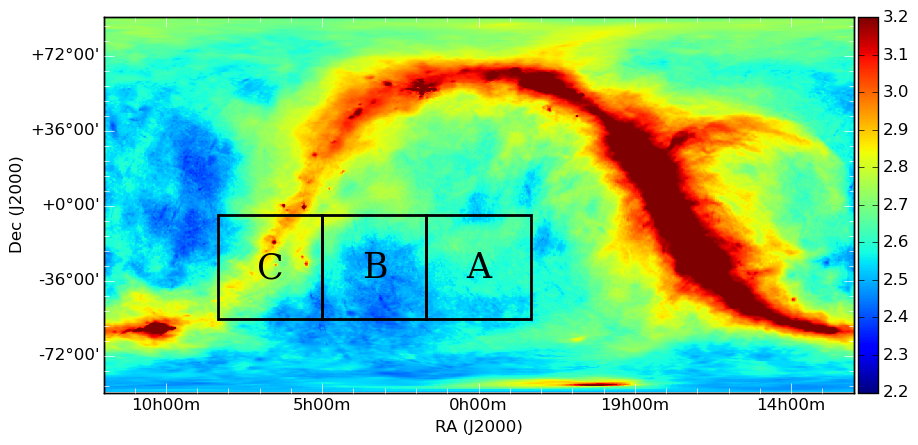}}
	\caption{Partially desourced and destriped Haslam all-sky brightness temperature map scaled from 408 MHz to 126 MHz with temperature spectral index $\beta=2.52$. Three $48\fdg0 \times 48\fdg0$ regions of the map A, B and C, used in the construction of our GDSE simulations and centred on (RA, Dec) = ($0\fdg0,\ -30\fdg0$), ($50\fdg0,\ -30\fdg0$) and ($100\fdg0,\ -30\fdg0$) respectively are shown bordered in black. The colourscale shows log-brightness-temperature with brightness temperature in kelvin. The scale range is chosen to highlight the map intensity structure in our target simulation regions.}
	\label{Fig:remazeilles_haslam408}
\end{figure}

We construct our simulations to match the $48\fdg0 \times 48\fdg0$ field of view and $\sim1.5$ arcminute resolution of our simulated 21-cm signal cube. We consider three regions (labelled A, B and C and bordered in black in \autoref{Fig:remazeilles_haslam408}) to investigate the impact of spatial structure and temperature dependence of the GDSE emission on the power spectrum. Regions A and B lie out of the plane of the Galaxy, are centred on $(\mr{RA}=0\fdg0,\ \mr{Dec}=-30\fdg0)$ and $(\mr{RA}=50\fdg0,\ \mr{Dec}=-30\fdg0)$ respectively and have mean brightness temperatures at $126~\mr{MHz}$ of $T_{\mr{A},126} = 392~\mr{K}$, and $T_{\mr{B},126} = 357~\mr{K}$. Region C lies in the plane of the Galaxy, is centred on $(\mr{RA}=100\fdg0,\ \mr{Dec}=-30\fdg0)$ and has mean brightness temperature $T_{\mr{C},126} = 528~\mr{K}$. We use region A as a reference when describing construction of our GDSE model but apply identical procedures to regions B and C. 

We use the \citet{2015MNRAS.451.4311R} (hereafter R15) bright-source-removed and destriped Haslam 408 MHz all-sky map (displayed in \autoref{Fig:remazeilles_haslam408} and hereafter referred to as RH408) as the starting point of a spatial template for our GDSE simulation (scaled to 126 MHz as described below). RH408 is a version of the original Haslam 408 MHz survey \citep{1981A&A...100..209H} that has been processed to remove the brightest point sources and large scale striations associated with instrumental noise in the survey. 

To construct our GDSE emission simulation at the $\sim1.5$ arcminute resolution of our simulated 21-cm signal cube we require an emission template at higher resolution that the original Haslam survey. In addition to a high fidelity map, at the resolution of the Haslam survey, R15 have also made available an extrapolated high resolution version of the all-sky Haslam map. However, in common with its standard resolution counterpart, it contains multiple power spectral components. The emission in both maps is comprised of three dominant components: i) extended GDSE, ii) emission from Galactic and extragalactic sources below the level to which removal has been attempted, as well as iii) source-subtraction residuals and noise. As outlined in \autoref{Sec:Foregroundsimulations}, to correctly account for spatial and spectral differences between emission components and determine their contribution to the total power spectrum, we seek to model emission components independently. Here, we are interested in inferring the most likely synchrotron-only component of the emission from the sum of these components in the available map. We therefore estimate the most likely synchrotron-only component of the power spectrum of the data first, via a power spectral decomposition. We then apply this estimate to enhance the resolution of the corresponding map in a manner free from contamination by the other power spectral components.

Using this procedure, we can later introduce EGS independently into our complete foreground simulation enabling us to model differences in the spectral structure of the two components. Further, since the GDSE spatial power spectrum varies across the sky (see e.g. \citealt{2008A&A...479..641L}) this has the advantage of allowing us to perform spatial extrapolation based on the spatial power spectral characteristics of the region under consideration. This will be of interest for our analysis in \autoref{Sec:SpatialPowerSpectra}, and in general in experiments designed to detect the EoR where, we argue, the ratio of the spatial power spectrum of the EoR signal and the foregrounds in the observed region can enable targeted estimation of the three-dimensional $k$-space power spectrum of the EoR signal on spatial scales that will minimise foreground bias.

\subsection{Power-spectral decomposition and resolution enhancement}
\label{Sec:HR}

\begin{figure}
\centerline{\includegraphics[width=\columnwidth]{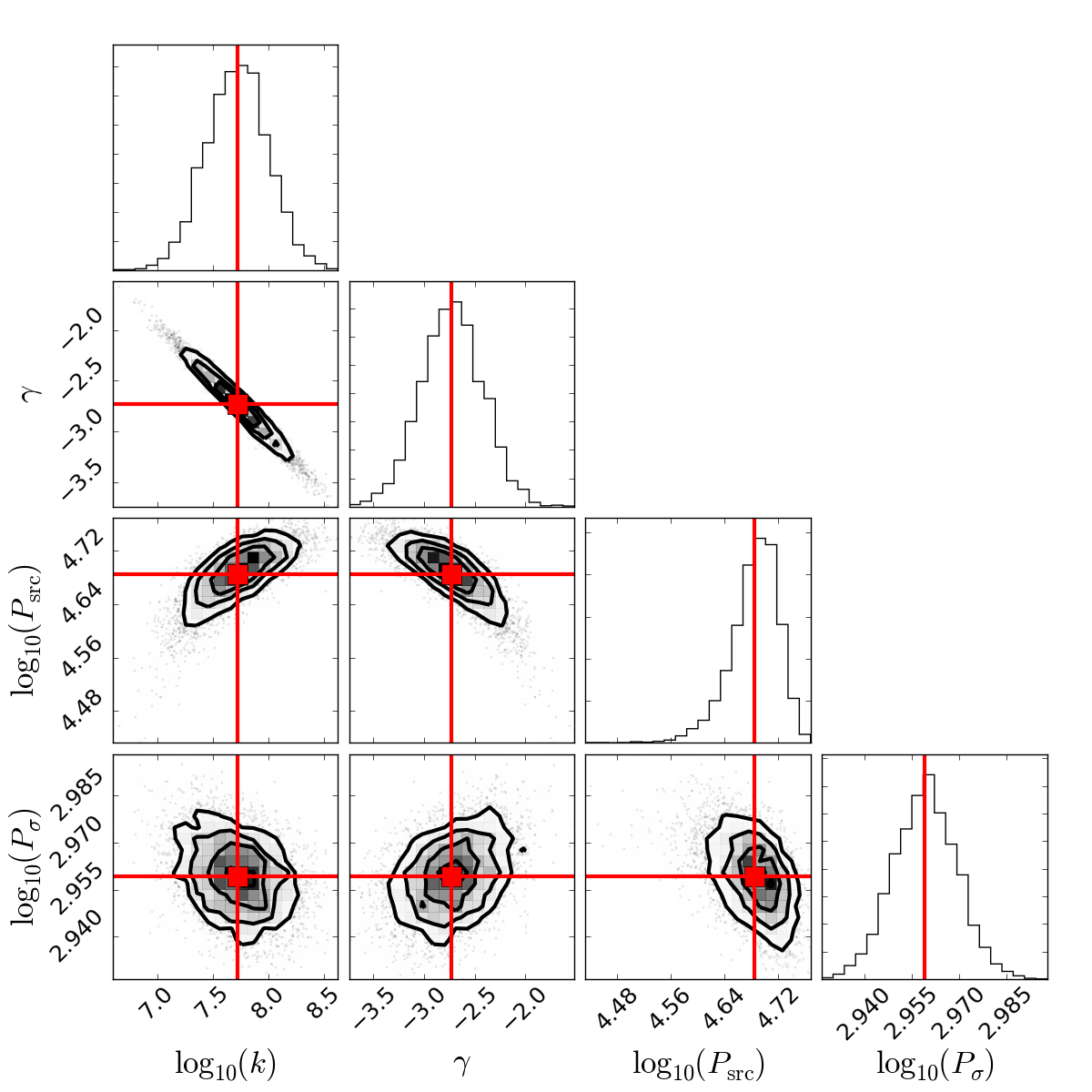}}
	\caption{One-dimensional and two-dimensional marginalized posteriors for the 4 parameters of our 3-component power spectral decomposition given by \autoref{Eq:SpatialPSComp}. The red horizontal and vertical lines indicate the posterior means: $\gamma = 2.7 \pm 0.3$, $\log_{10}(k) = 9.7 \pm 0.3$, $\log_{10}(P_{\mr{src}}) = 6.69 \pm 0.04$ and $\log_{10}(P_{\sigma}) = 4.96 \pm 0.01$. Contours in the two-dimensional plots show steps of 0.5-$\sigma$ between 0.5-$\sigma$ and 2-$\sigma$, with 1 and 2-$\sigma$ representing 68\% and 95\% confidence levels.}
	\label{Fig:HaslamDecompositionPosteriors}
\end{figure}

\begin{figure}
	\centerline{\includegraphics[width=\columnwidth]{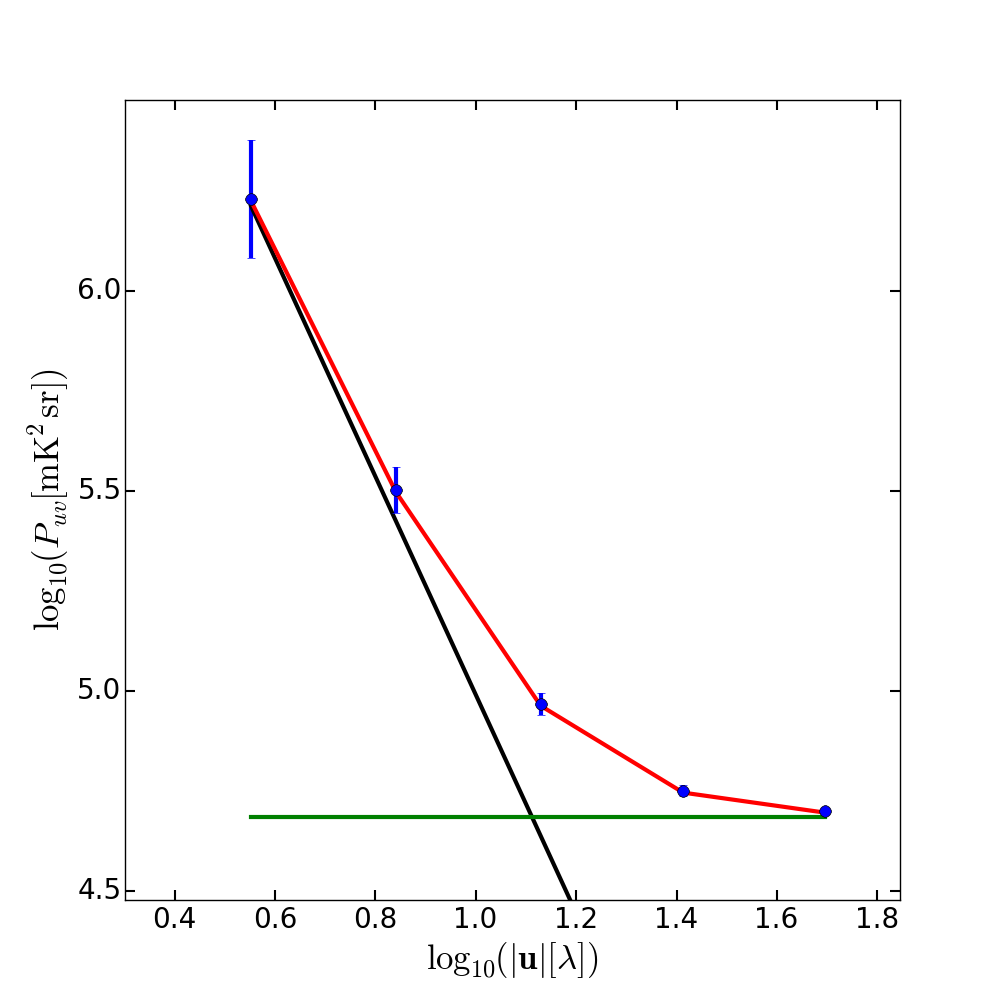}}
	\caption{Modelled power spectral decomposition of the RHS126 power spectrum into its component parts. The data points denote $(P_{\mr{map}} - P_{\sigma})/A(\mathbfit{u})^2$ with the plotted error bars being the standard deviation of the posteriors for $\mr{Pr}(\bm{\varphi}_\mr{s}|\mathbfit{d}_\mr{s})$. The black and green lines show the models derived from the mean of the posteriors for the synchrotron and source components of the power spectrum respectively, with the red line their sum.}
	\label{Fig:CompFit}
\end{figure}

We extract from RH408 the $48\fdg0 \times 48\fdg0$ subsections shown in \autoref{Fig:remazeilles_haslam408}. For each region we obtain a single channel $N_{\rm{pix}} \times N_{\rm{pix}}$ map where $N_{\rm{pix}}=150$ pixels and the width of each pixel is $1/3$ of a degree. We extrapolate the brightness temperature map to 126 MHz using a single temperature spectral index\footnote{With sign convention for the temperature spectral index: $T \propto \nu^{-\beta}$}, $\beta=2.5$, consistent with the mean spectral index between 100 MHz and 200 MHz as measured by \citet{2008AJ....136..641R}; hereafter we will refer to this extrapolated subset of the map as RHS126. The astrophysical emission in RHS126 results from two components: GDSE, which is visibly dominant on large spatial scales, and point source emission, which is of increasing importance on smaller spatial scales and is comprised of three sub-components, Galactic and extragalactic sources below the level to which sources have been removed from the map and source-subtraction residuals. 

The result of the instrumental response of single-dish observations, such as those that produced RH408, is to convolve the sky brightness distribution we seek to estimate with the instrumental response. The effective beam response across the map has been found to be well modelled as a Gaussian with a full width at half maximum (FWHM) of $56.0 \pm 1.0$ arcmin \citep{2015MNRAS.451.4311R}. In addition, the measurements will be contaminated with instrumental noise which, following \citet{2008A&A...479..641L} (hereafter LP08), we will assume is approximately white. We note that the noise is added to the sky signal after it has been convolved by the instrumental response (e.g. \citealt{1997ApJ...480L..87T}). By the convolution theorem, the Fourier transform to the convolution of the image by the instrumental response is equal to the product of the Fourier transform of the image with the Fourier transform of the instrumental response. For our power spectral model we approximate the power spectrum of the map $P_{\mr{map}}$, 
\begin{equation}\label{Eq:SpatialPSComp}
P_{\mr{map}} = (P_{\mr{GDSE}} + P_{\mr{src}})A(\mathbfit{u})^2 + P_{\sigma} \ ,
\end{equation}
where $P_{\mr{GDSE}}$, $P_{\mr{src}}$ and $P_{\sigma}$ are the two-dimensional spatial power spectra of the GDSE, sources and noise in the map respectively and $A(\mathbfit{u}) = \exp(-2\pi^2\abs{\mathbfit{u}}^2\sigma_{P}^2)$ is the Fourier transform of the effective image domain instrumental response, with $\mr{FWHM} = 2\sqrt{2\ln(2)}\sigma_{P}$.

We calculate $P_{\mr{map}}$ using the approach described in \autoref{Sec:2DSpatialpowerspectra}. We sample directly from the posteriors $\mr{Pr}(\bm{\varphi}_\mr{s}|\mathbfit{d}_\mr{s})$ using {\sc{MultiNest}} with evaluation of the likelihood accelerated by the use of {\sc{MAGMA}} (Matrix Algebra on GPU and Multicore Architectures) libraries\footnote{http://icl.cs.utk.edu/magma/
}. When evaluating $\mathbfss{F}^{\prime}$ and $\mathbfss{F}^{-1}$ we use a coarse visibility grid with spacing $\sDelta u_{\mr{c}} \sim 1.2~\lambda$ corresponding to the $48\fdg0$ field of view of the simulation. The sub-harmonic grid has spacing $\sDelta u_{\mr{sh}} = \sDelta u_{\mr{c}}/7$ modelling large scale spatial power on scales between 1 and 7 times the simulation field of view. We select this value by using tests of smaller sky regions for which we find a negligible impact on the coarse-grid parameter estimates following further reduction of the sub-harmonic pixel widths. 

We define $\bm{\varphi}_\mr{s}$ over a set of annuli in $\abs{\mathbfit{u}}$, with 3 bins over the sub-harmonic grid with log-uniform spacing between $0.17~\lambda$ and $2.4~\lambda$ and 7 bins with log-uniform spacing between $2.4~\lambda$ and $53~\lambda$ over the coarse grid. We calculate the maximum likelihood visibility plane $\hat{V}(\mathbfit{u}_{i},\nu_{i})$ of RHS126 as $\sSigma^{-1}\bar{\mathbfit{d}_\mr{s}}$, where in constructing $\sSigma^{-1}$ we populate $\mr{diag}(\sPsi^{-1})$ with the inverse power spectrum determined by the mean of the sampled posteriors. We then seek to deconstruct $P_{\mr{map}}$ into its components. Writing the mean of the posteriors for the power spectrum of RHS126 as a vector $\mathbfit{P}_{\bm{\varphi}_\mr{s}}$ we define a Gaussian log-likelihood function of the form,
\begin{equation}
\label{Eq:DecompLogLike}
\log(\mathcal{L}) \propto -\dfrac{1}{2}\left(\mathbfit{P}_{\bm{\varphi}_\mr{s}} - \mathbfit{m}_\mr{P}(\sTheta_\mr{P})\right)^{T}\mathbfss{N}_\mr{P}^{-1}\left(\mathbfit{P}_{\bm{\varphi}_\mr{s}} - \mathbfit{m}_\mr{P}(\sTheta_\mr{P})\right) \ .
\end{equation}
Here $\mathbfss{N}_\mr{P}^{-1}$ is the covariance matrix of $\bm{\varphi}_\mr{s}$ and $\mathbfit{m}_\mr{P}(\sTheta_\mr{P})$ is a model vector described by \autoref{Eq:SpatialPSComp}. Following LP08, we model the synchrotron power spectrum as a power law $P_{\mr{GDSE}}=k\abs{\mathbfit{u}}^{\gamma}$. For the point source component we assume the source statistics can be well approximated as Poissonian and therefore that $P_{\mr{src}}=\mr{constant}$. We therefore have four parameters comprising $\sTheta_\mr{P}$: $k$, $\gamma$, $P_{\mr{src}}$ and $P_{\sigma}$.

Theoretical power spectrum estimates assuming isotropic turbulence with a $k^{-11/3}$ Kolmogorov spectrum for the Galactic magnetic
field imply an expected spatial power spectral coefficient, $\gamma \sim -2.4$ (e.g. \citealt{2000ApJ...530..133T}). We take this estimate as the centre of our prior which we assume to be uniform in the range $\gamma \in [-3.9, -0.9]$. For the remaining parameters we apply minimal prior constraints, using log-uniform priors in the range -5 to 0. 

We perform the sampling using {\sc{MultiNest}}. The resulting two-dimensional and one-dimensional posteriors for the power spectral decomposition of region A are displayed in \autoref{Fig:HaslamDecompositionPosteriors}. The mean parameter estimates and uncertainties obtained for each of regions A, B and C using the decomposition given by \autoref{Eq:SpatialPSComp} are shown in \autoref{Tab:GalPSDparams}. The spatial power spectrum varies as a function of position on the sky, therefore a single region cannot be compared directly to the average power spectrum over the sky calculated by R15 (their Figure 14). We find that after accounting for extrapolation of the maps to 126 MHz, the average power spectrum is below but consistent with $P_{\mr{map}}$ within uncertainties for our region C and above that of our (relatively cold) regions A and B. The spatial power spectral coefficients vary between regions and are consistent within uncertainties with the best--fitting value of $\gamma=-2.7$ for the Haslam map between $l=30$ and $l=90$ quoted by R15. A plot of $(P_{\mr{map}} - P_{\sigma})/A(\mathbfit{u})^2$ is displayed in \autoref{Fig:CompFit} with an overlay of $P_{\mr{GDSE}}$, $P_{\mr{src}}$ and $(P_{\mr{GDSE}} + P_{\mr{src}})$ generated using the means of the posteriors for the parameters.

\begin{table}
\caption{Mean posterior parameter estimates and 1-$\sigma$ uncertainties for power spectral decompositions of region A, B and C. Variables $k$, $P_{\mr{src}}$ and $P_{\sigma}$ are in units $\mr{mK^2sr}$}.
\centerline{
\begin{tabular}{l l l l l}
\toprule
Region & $\log_{10}(k)$ & $\gamma$       & $\log_{10}(P_{\mr{src}})$ & $\log_{10}(P_{\sigma})$    \\
\midrule
A      & $7.7 \pm 0.3$       & $-2.7 \pm 0.3$      & $4.68 \pm 0.04$     & $3.00 \pm 0.01$            \\
B      & $7.8 \pm 0.3$       & $-2.5 \pm 0.3$      & $6.60 \pm 0.08$     & $2.87 \pm 0.01$            \\
C      & $9.2 \pm 0.2$       & $-2.9 \pm 0.2$      & $5.75 \pm 0.04$     & $2.44 \pm 0.52$            \\
\bottomrule
\end{tabular}
}
\label{Tab:GalPSDparams}
\end{table}

We calculate the vector of maximum likelihood gridded visibilities corresponding to the three power spectral components as $\hat{V}'(\mathbfit{u}_{i}) = \sSigma_\mr{s}'^{-1}\bar{\mathbfit{d}'_\mr{s}}$, where $\bar{\mathbfit{d}'_\mr{s}}$ is a triple concatenation of $\bar{\mathbfit{d}_\mr{s}}$ and $\sSigma_\mr{s}'^{-1}$ is a triple concatenation of $\sSigma^{-1}_{\mr{s,comp}}$, where the subscript `comp' refers to each of the three components (synchrotron, source and noise power). For each $\sSigma^{-1}_{\mr{s,comp}}$ our power spectra are modelled by the diagonal matrix $\sPsi^{-1}_{\mr{comp}}$, where diag($\sPsi^{-1}_{\mr{comp}}$) is populated by the mean posterior estimate for the power spectrum of the given model component. The joint maximum likelihood solution $\hat{V}'(\mathbfit{u}_{i})$ has three components, $\hat{V}'_{\mr{GDSE}}$, $\hat{V}'_{\mr{scr}}$ and $\hat{V}'_{\sigma}$ which are the joint maximum likelihood visibility planes for the synchrotron, source and noise power spectral components respectively.

To construct our enhanced resolution simulation, we extend the maximum likelihood GDSE power spectrum to higher inverse-spatial scales by populating the $uv$-plane with a zero-mean Hermitian Gaussian random field.
This procedure is commonly used to construct a fully random realisation of a sky signal following a prescribed power law (see e.g. \citealt{2008MNRAS.389.1319J, 2009A&A...507.1087S}). Here we adapt its use to populate the outer part of the $uv$-plane while retaining the component of the $uv$-plane well sampled by the data. In so doing we enhance the resolution of our previously derived maximum likelihood map corresponding to the most probable GDSE component of the power spectrum inferred from RHS126 while accurately representing large scale structure present in the data.
We construct this field with variance as a function of $uv$-distance such that it follows the power law defined by the mean of the posteriors for the synchrotron component of the fit (see black line in \autoref{Fig:CompFit}). We extend the $uv$-plane in this manner out to a maximum $uv$-distance corresponding to $1/1.5~\mr{arcmin}^{-1}$, the spatial resolution of our 21-cm simulation. 

\begin{figure*}
	\centerline{
	\includegraphics[width=0.49\textwidth]{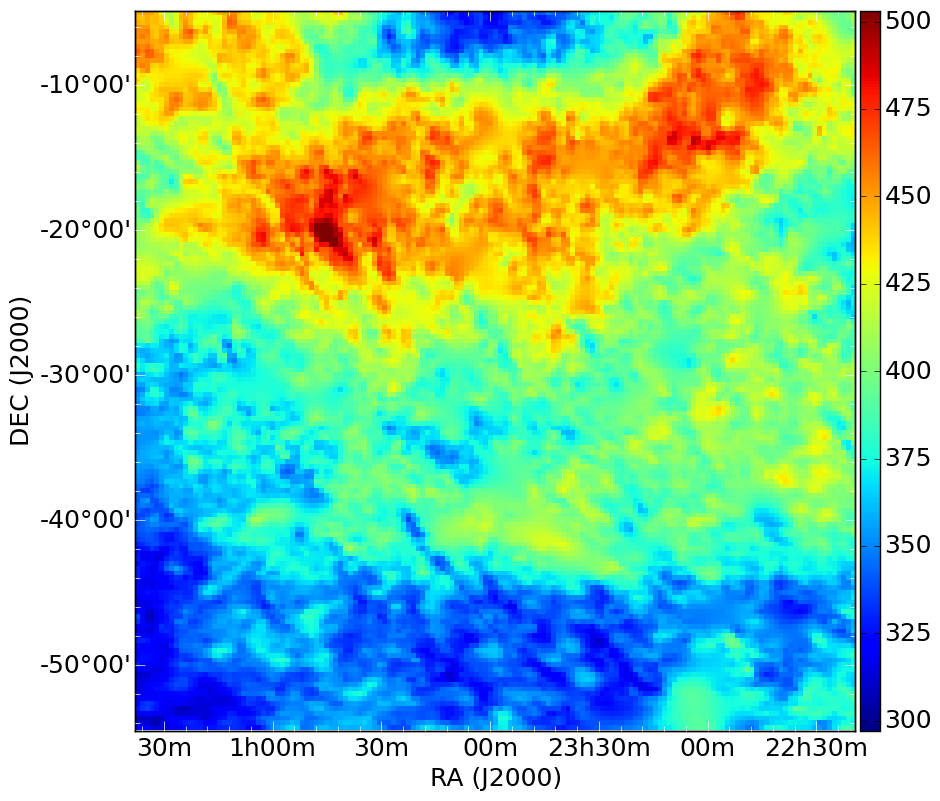}
	\includegraphics[width=0.49\textwidth]{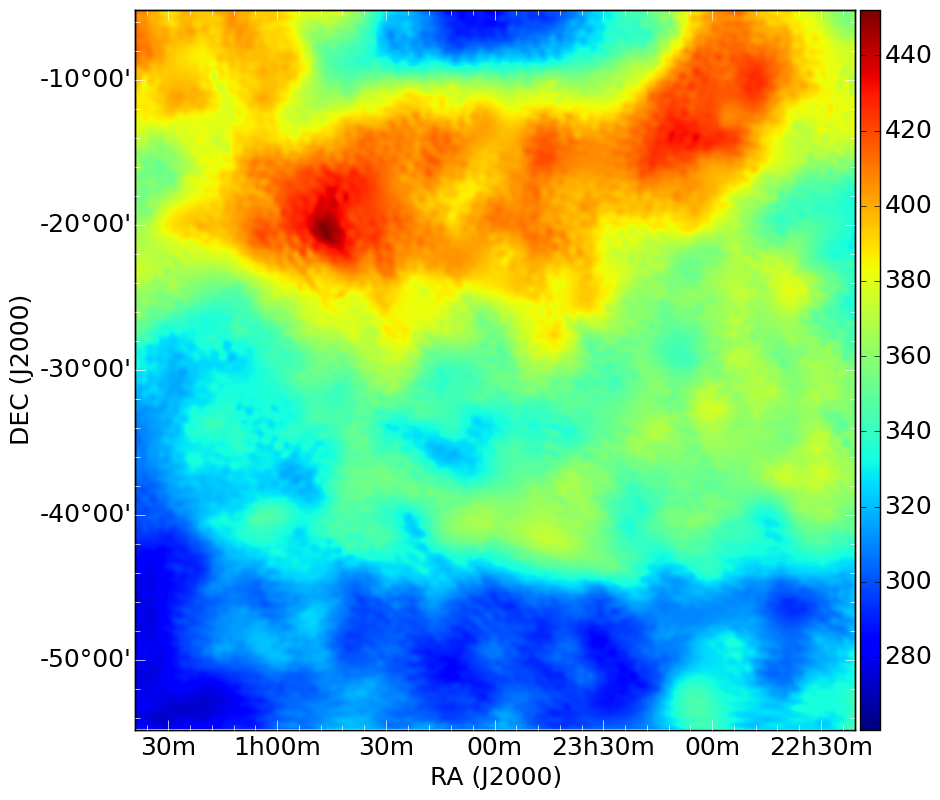}
	}
	\caption{[Left] RHS126: 48 degree subsection of the partially desourced and destriped Haslam 408 MHz map following extrapolation of flux-density scale to 126 MHz using a temperature spectral index $\beta=2.52$ \citep{2008AJ....136..641R}. [Right] Synchrotron emission model inferred from the map region shown left. The map is given by the Fourier transform of the maximum likelihood GDSE $uv$-plane $\hat{V}(\mathbfit{u}_{i})_{\mr{GDSE}}$. Assuming self-similar GDSE spatial structure, it contains power given by the extrapolation of the synchrotron component of our model for the power spectral decomposition of the map (shown in \autoref{Fig:CompFit}) between the resolution of the Haslam all-sky survey ($\sim1\fdg0$) and that of our 21-cm simulation ($\sim1.5$ arcmin). The colourscales have units of kelvin.
	}
	\label{Fig:HaslamDecomposition}
\end{figure*}

Owing to the power law nature of the synchrotron spatial power spectrum, the maximum likelihood synchrotron image estimated from the power spectrum of the synchrotron component of the emission in RHS126 and our high resolution simulation in the image domain are visually similar. That is, the extrapolation of the synchrotron $uv$-plane following the mean a posteriori synchrotron parameters acts as a minor perturbation on the large scale synchrotron structure visible in the image domain. As such, we display here the enhanced resolution model only, shown in \autoref{Fig:HaslamDecomposition}, right. Our high resolution model has dimensions $N_{\rm{pix,e}}$ by $N_{\rm{pix,e}}$ where $N_{\rm{pix,e}}=2048$. It covers an equal sky area and has identical spatial structure to our initial synchrotron model on large angular scales, but possesses additional self-similar structure on smaller scales down to the smallest probed by our 21-cm simulation.

\subsection{Frequency structure}
\label{Sec:FrequencyStructure}

To match the frequency extent of our cosmological model we consider GDSE across an $8~\mr{MHz}$ bandwidth, $126 \pm 4$ MHz. We assume that the emission is optically thin, approximating the spectrum along a line of sight as a single power law with temperature spectral indices $\beta$. To construct our spectral index distribution, we assume that it can be well approximated as Gaussian with mean temperature spectral index $\bar{\beta}_\mr{g}=2.5$ and width of the distribution, $\sigma_\mr{g}=0.1$, in agreement with observational constraints in the relevant frequency range (e.g. \citealt{2008AJ....136..641R}),
\begin{equation}
\label{Eq:SID1}
N(\bar{\beta}_\mr{g}, \sigma_\mr{g}^{2})=\dfrac{1}{\sigma_\mr{g}\sqrt{2\pi}}\exp\left[-\dfrac{1}{2}\dfrac{(\beta_\mr{g}-\bar{\beta}_\mr{g})^{2}}{\sigma_\mr{g}^{2}} \right] \ .
\end{equation}
The temperature and temperature spectral index, $\beta$, of synchrotron emission along a line of sight are both dependent on the energy distribution of relativistic electrons producing the emission (e.g. \citealt{2011hea..book.....L}). Observations of the structure of spatially dependent spectral index fluctuations at high resolution are scarce. The large scale dependence of the spectrum on Galactic latitude, however, is well established (e.g. \citealt{1988A&AS...74....7R}) and thus makes reasonable the assumption that there is some level of correlation between the total intensity along the line of sight and spectral index. However the expected level of correlation is a priori unknown. As such, we create multiple GDSE models with varying levels of temperature--spectral index correlation in order, in \autoref{Sec:Analysis}, to study its impact on the power spectrum.

We apply an algorithmic approach to constructing $N(\bar{\beta}_\mr{g},\sigma_\mr{g}^{2})$ such that the spectral index in a given simulation pixel is correlated with its temperature to a known degree. We achieve this by summing realisations from two intermediately constructed Gaussian distributions. The first is uncorrelated with temperature, with elements $\beta_{\mr{uc},i}$ where $i$ runs from 1 to $N_{\rm{pix,e}}^{2}$, with mean, $\bar{\beta}_{\mr{uc}}$, and width $\sigma_{\mr{uc}}$. The second with elements $\beta_{\mr{c},i}$ where $i$ runs from 1 to $N_{\rm{pix,e}}^{2}$, with mean, $\bar{\beta}_{\mr{c}}$, and width $ \sigma_{\mr{c}}$ has elements sorted in order of pixel temperature in the map.

\begin{figure*}
	\centerline{
	\includegraphics[width=0.33\textwidth]{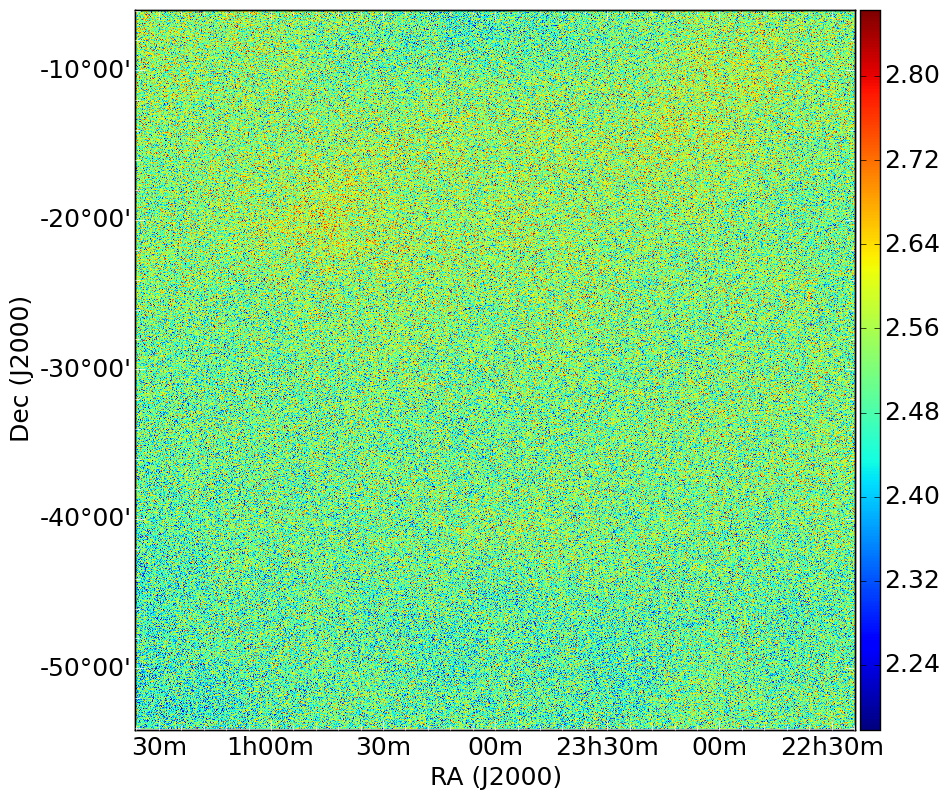}
	\includegraphics[width=0.33\textwidth]{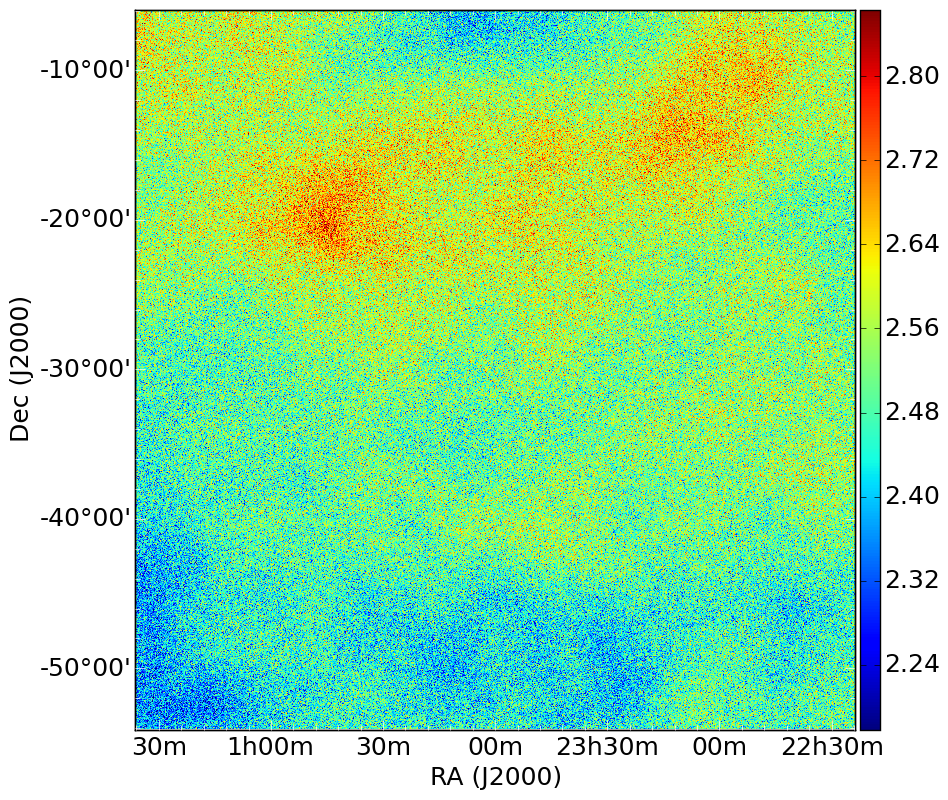}
	\includegraphics[width=0.33\textwidth]{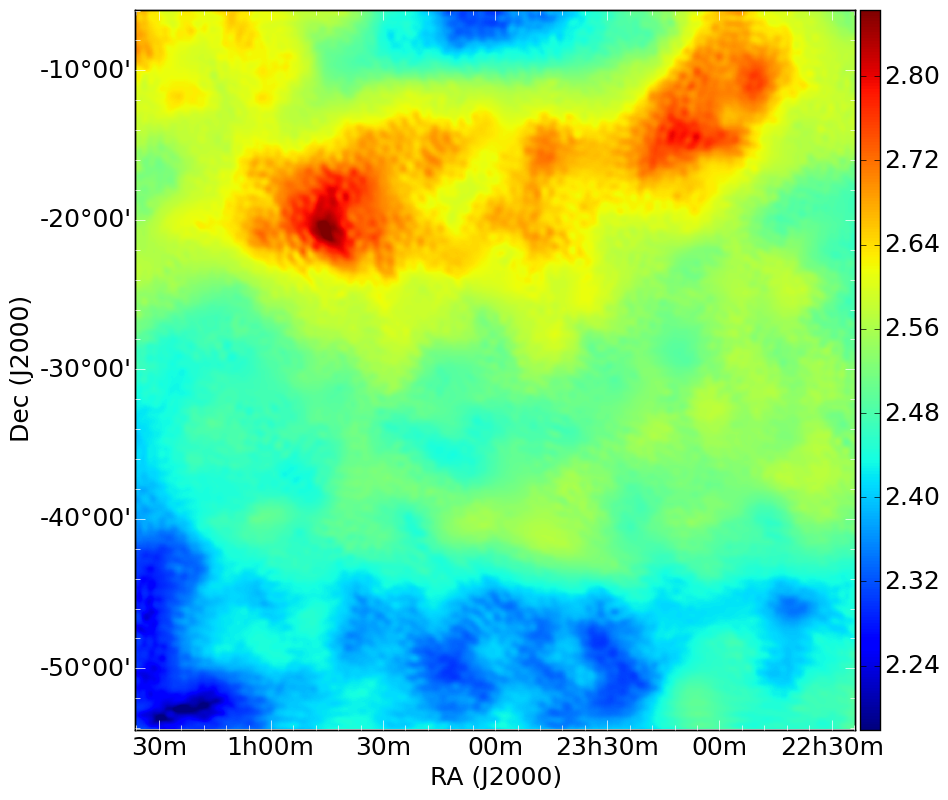}}
	\centerline{
	\hspace{0.5cm}\includegraphics[width=0.33\textwidth]{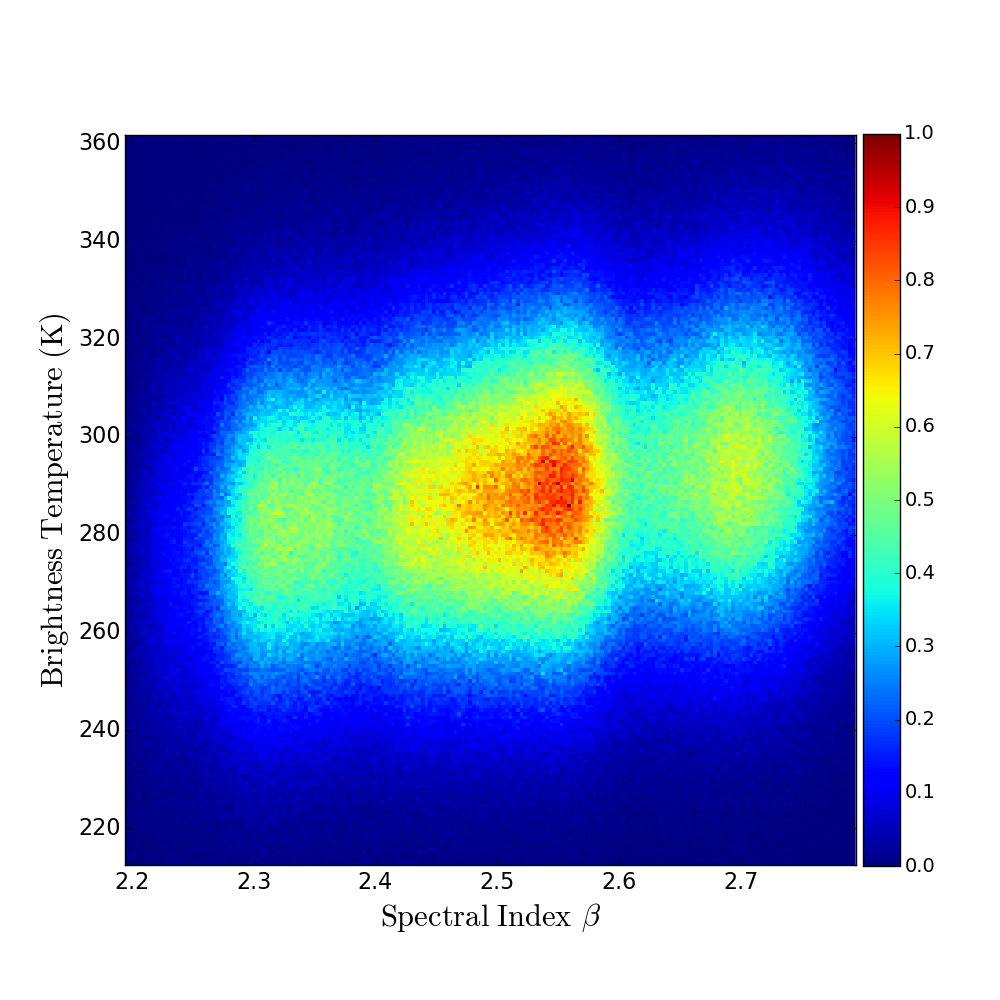}
	\includegraphics[width=0.33\textwidth]{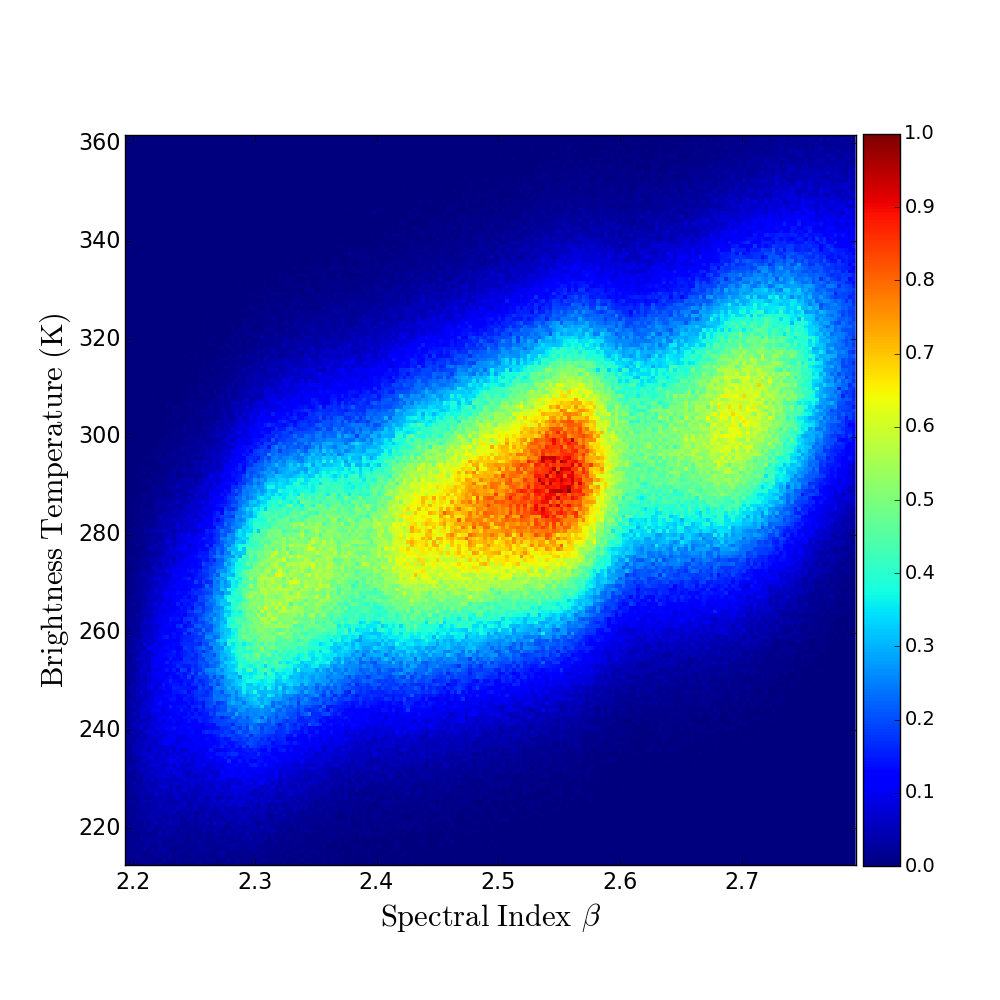}
	\includegraphics[width=0.33\textwidth]{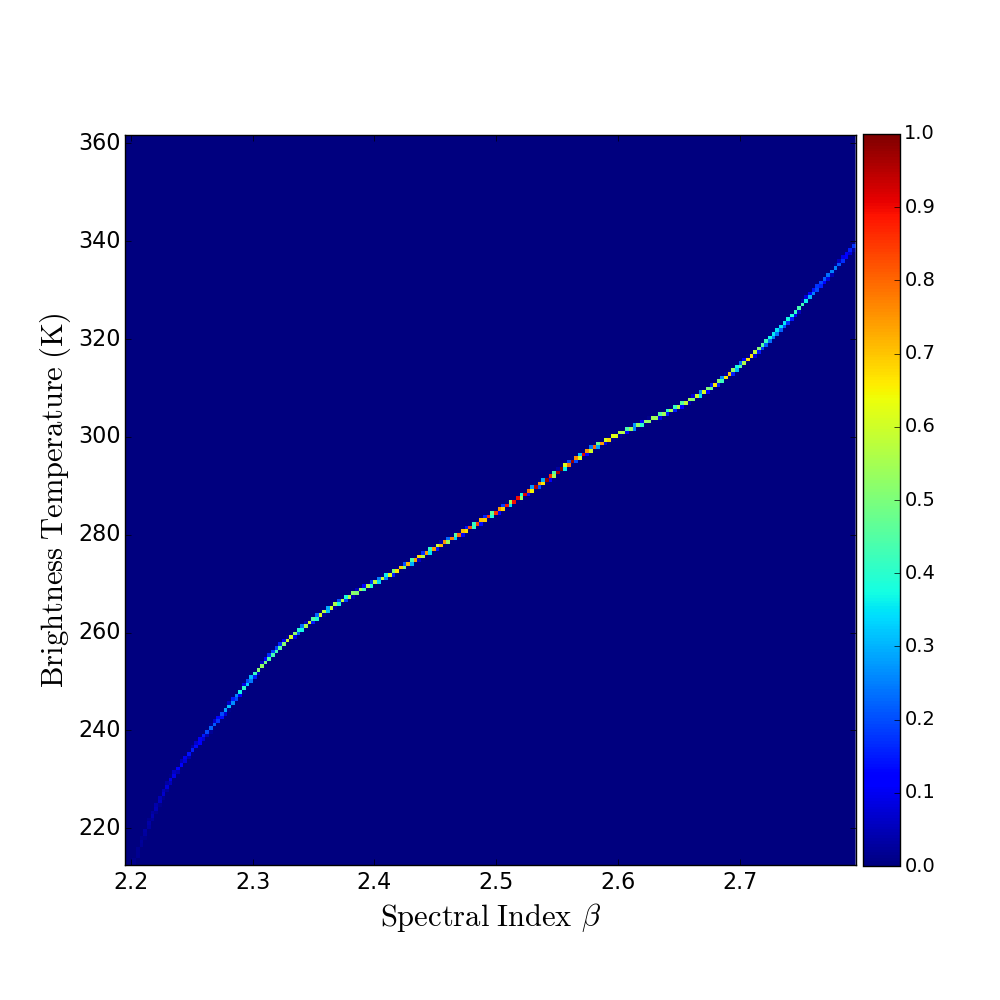}}
	\caption{[Top] Temperature spectral index maps for $20\%$ (left), $60\%$ (middle) and $100\%$ (right) brightness temperature -- spectral index correlation. Spectral indices are calculated between two channels of our enhanced resolution GDSE region A brightness temperature map. The colourscale denotes temperature spectral index $\beta$, where brightness temperature, $T_\mr{B}\propto \nu^{-\beta}$. [Bottom] two-dimensional joint probability distributions between the brightness temperatures of our GDSE simulation and the spectral index maps above. The distributions are normalised to a peak of 1.}
	\label{Fig:SImaps}
\end{figure*}

In order for \autoref{Eq:SID1} to describe the sum of our two intermediate distributions, we require:
\begin{align}\label{Eq:Constraints}
\beta_{\mr{g}}&=\beta_{\mr{c}} + \beta_{\mr{uc}}\\
\sigma_{\mr{g}}^2&=\sigma_{\mr{c}}^2 + \sigma_{\mr{uc}}^2.
\end{align}
We assign $\beta_{\mr{uc}}=0$ and therefore require $\beta_{\mr{c}}=2.5$. This yields a simple relation for the correlation coefficient between the temperatures in the map and $N(\bar{\beta}_\mr{g}, \sigma_\mr{g}^{2})$:
\begin{equation}
\label{Eq:CorrelationCoef}
\rho_{T,\beta_\mr{g}} = \dfrac{\sigma_{\mr{c}}}{\sigma_{\mr{uc}}+\sigma_{\mr{c}}} \ .
\end{equation}
We solve simultaneously \autoref{Eq:Constraints} and \autoref{Eq:CorrelationCoef} and test correlation levels between $\beta_\mr{g}$ and ${T}$ of $20$, $60$, and $100\%$. Temperature spectral index maps for these three correlations, along with joint probability distributions between each of the maps and our brightness temperature simulation, are shown in \autoref{Fig:SImaps}.

Our final GDSE simulation cube is constructed with frequency structure given by,
\begin{equation}
\label{Eq:SynchFrequencyExtrapolation}
I(\beta_\mr{g,j}, \nu_{k}) = I_{126} \left(\dfrac{\nu_{k}}{126~\mr{MHz}}\right)^{-\beta_\mr{g,j}} \ ,
\end{equation}
where $I(\beta_\mr{g,j}, \nu_{k})$ is the brightness temperature of voxel ($j$,$k$), $j$ runs from 1 to $N_{\rm{pix},e}^{2}$ and $k$ runs from 1 to $N_{\mr{chan}}=38$. The channel-width is chosen to match our 21-cm simulation ($\approx 200\ \mr{kHz}$).

\subsection{Free--free emission}
\label{Sec:FreeFreeModel}

Thermal bremsstrahlung radiation resulting from scattering of free electrons in diffuse \HII regions within the Galaxy is expected to account for approximately $1\%$ of the sky temperature at $150~\mr{MHz}$ \citep{1999A&A...345..380S}. \HII regions are optically thin in the frequency range of interest for reionization experiments and have a well determined power law spectrum with temperature spectral index $\beta=2.15$.

Significant spatial correlation between dust and warm ionized gas has been observed \citep{1996ApJ...460....1K, 1997ApJ...482L..17D}. Maps of dust emission therefore constitute a useful tracer for free--free emission. At high Galactic latitudes, $\mr{H}\alpha$ and free--free emission from diffuse gas clouds are both proportional to the emission measure. Galactic $\mr{H}\alpha$ surveys therefore provide a second independent tracer. Since free--free emission is the least dominant of the foreground components in total intensity and spectrally most simple, either tracer is sufficient for our purposes (this assumption is confirmed in \autoref{Sec:Analysis}). Therefore we elect to use the simulated full-sky maps of Galactic diffuse interstellar dust emission \citep{1999ApJ...524..867F} simplifying calculation of the spatial power spectrum (as the map is free of instrumental effects and noise, the power spectrum can be estimated directly rather than requiring further decomposition into components).

In particular, we select the section of the full-sky simulated map of emission from the diffuse interstellar dust in the Galaxy \citep{1999ApJ...524..867F} corresponding to region A of our GDSE simulation as a tracer for the spatial structure for our free--free emission simulation. We assume at the central frequency of our band, free--free emission contributes 1 percent of the total Galactic emission temperature (see e.g. \citealt{1999A&A...345..380S}). We scale the map of diffuse interstellar emission to a mean brightness temperature of 3.9 K accordingly. This is a conservative estimate but remains comparable to others such as, for example, to the 2.2 K contamination at 120 MHz resulting from free--free emission used in the foreground simulations of \citet{2008MNRAS.389.1319J}.

\begin{figure}
	\centerline{\includegraphics[width=\columnwidth]{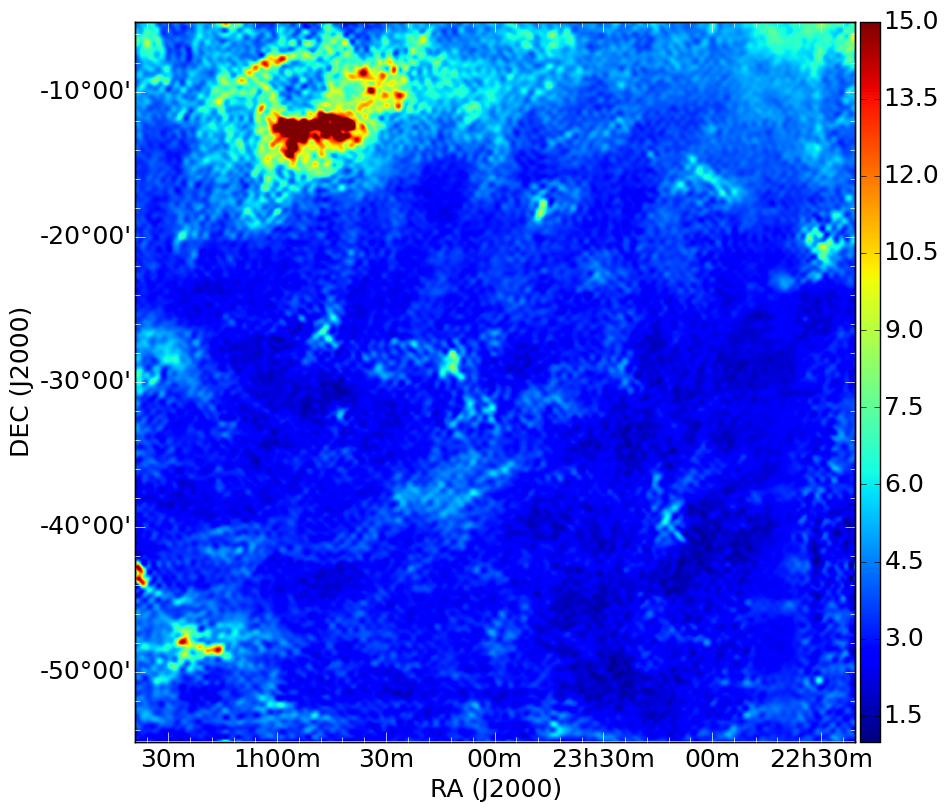}}
	\caption{Enhanced resolution brightness temperature map of a 48 degree subsection of our free--free emission simulation. The flux-density scale of the map has been scaled to contribute 1 percent of Galactic emission at $126~\mr{MHz}$ and has power law spectral structure with temperature spectral index $\beta=2.1$. The $\sim1.5$ arcmin resolution of the map is an increase by a factor of 14 relative to the original. This is achieved assuming self-similar spatial structure on small physical scales via extrapolation of the $uv$-domain power spectral fit. The colourscale has units of kelvin.}
	\label{HRimageFreeFree}
\end{figure}

In order to construct our model at equal resolution to our foreground cube, we perform an equivalent power law extrapolation of the spatial power spectrum of the diffuse interstellar dust map to that described for the GDSE in \autoref{Sec:HR}. The resulting estimate for the free--free spatial power spectral index yields $\gamma = 2.59 \pm 0.04$. An image of our free--free emission simulation at 126 MHz is shown in \autoref{HRimageFreeFree}. For the spectral structure of our free--free simulation, we assume a spatially uncorrelated and constant temperature spectral index, $\beta = 2.15$.

\section{Extragalactic foreground emission}
\label{Sec:ExtragalacticForegroundEmission}

\subsection{Introduction}
\label{Sec:EGSintro}

In this section we model the point source population of EGS.
The EGS population can be expected to become increasingly dominant component of the $k$-space power spectrum on smaller spatial scales. As with Galactic emission we construct our EGS simulation cube at the resolution of our EoR signal cube.

As the basis of our EGS simulation we use the Square Kilometre Array (SKA) Simulated Skies ($\mr{S^{3}}$) Simulation of Extragalactic Sources ($\mr{S^{3}}$-SEX) \citep{2008MNRAS.388.1335W}. $\mr{S^{3}}$-SEX is a simulation of extragalactic radio continuum sources in an $20\fdg0 \times 20\fdg0$ sky area out to a redshift of $z = 20$. The simulation models contributions from radio-quiet active galactic nuclei (AGN), radio-loud AGN of the Fanaroff--Riley type I (FR I) and type II (FR II) structural classes, and star-forming galaxies. The latter population consists of quiescent and starbursting galaxies. The simulation method is `semi-empirical', meaning sources were drawn from observed (or extrapolated) luminosity functions and grafted onto an underlying dark matter density field with biases which reflect their measured large scale clustering. Sources are modelled down to a flux-density limit $S_\mr{lim}=10$ nJy. For further details see \citet{2008MNRAS.388.1335W}. 

For each source, radio fluxes can be retrieved at multiple frequencies with the closest frequencies to that of the desired $122$ -- $130~\mr{MHz}$ frequency range of our other foreground simulations being $151~\mr{MHz}$ and $610~\mr{MHz}$. Initially we extrapolate the $\mr{S^{3}}$-SEX flux-densities from 151 MHz to 126 MHz on a per source basis using power law of the form $S_{126}=S_{151} (126/151)^{-\alpha}$, with $\alpha = \log(S_{151}/S_{610})/\log(151/610)$. We take a $12\fdg0 \times 12\fdg0$ subset of the $126~\mr{MHz}$ catalogue and, following rotation though a random angle drawn from a uniform distribution between $0$ and $2\pi$, we tile copies 4 times in both RA and Dec in order to produce a $48\fdg0 \times 48\fdg0$ field. This allows us to maintain the characteristic extragalactic clustering scales modelled by the $\mr{S^{3}}$ simulation while matching the angular extent of our 21-cm emission, GDSE and free--free model cubes.

At this point, it is of interest to consider, for a choice of instrument model, the number of sources in our simulation that will be unresoved in observations. That is, those sources with flux densities below the (spatial resolution dependent) classical confusion noise limit. These are the sources that will be least constrained by the data and contribute most significantly to extragalactic contamination of the EoR power spectrum. The RMS confusion $\sigma_\mr{c}$ resulting from the flux-density of point sources fainter than some limiting signal-to-noise ratio $q = S_0 / \sigma_{\rm c}$ can be calculated analytically for a power law approximation to the differential source count distribution, $\mr{d}N/\mr{d}S = k S^{-\gamma}$, as (\citealt{1974ApJ...188..279C, 2012ApJ...758...23C}), 
\begin{equation}\label{Eq:CondonConfusion}
\sigma = \left(\dfrac{q^{3-\gamma}}{3 - \gamma}\right)^{1/(\gamma-1)} (k \Omega_\mr{e})^{1/(\gamma - 1)} \ ,
\end{equation}
where,  $\Omega_\mr{e} = \Omega_\mr{b}/(\gamma  -1)$ and $\Omega_\mr{b}$ is the solid angle of the synthesised beam. To estimate $\sigma_\mr{c}$  we take the \citet{2002ApJ...564..576D} power law fit of the high flux-density ($S>0.88~\mr{Jy}$) differential source count at 151 MHz, ($\mr{d}N/\mr{d}S(151~\mr{MHz}) = 4\times10^{3} (S/1~\mr{Jy})^{-\gamma}~\mr{Jy^{-1}sr^{-1}}$, with $\gamma=2.51$) and extrapolate to $126~\mr{MHz}$ assuming a mean source spectral index, $\alpha=0.82$ (see \autoref{Sec:OpticallyThin}). At $126~\mr{MHz}$ this gives, $\mr{d}N/\mr{d}S(126~\mr{MHz}) = 4.6\times10^{3} (S/1~\mr{Jy})^{-\gamma}~\mr{Jy^{-1}sr^{-1}}$. Taking $q = 5$ for reliable source detection, a maximum baseline length $b_\mr{max} = 87~\mr{m}$ for HERA in 37 antenna configuration (see \autoref{Sec:SimulatedHERAObservations} for details) and a corresponding beam solid angle $\Omega_\mr{b} \simeq (\lambda/b_\mr{max})^{2}$, gives a confusion noise estimate for our instrument model, $\sigma_\mr{c} = 4~\mr{Jy~beam^{-1}}$. This corresponds to a minimum flux-density for reliably detectable and individually countable point sources, $S>20~\mr{Jy}$.

In what follows, we include in our EGS simulation only confusion limited sources below this flux density limit. This is equivalent to assuming that sources with flux-densities greater than this limit can be precisely characterised by the instrument and, in the limit of negligible uncertainties on their amplitudes and positions, can be removed from the data in the visibilities. The total number of sources as a function of source types used in our full simulation are listed in \autoref{Tab:SCubedSources}. We note that, ideally, one would include prior information on resolved extragalactic point source population in the power spectral likelihood. The amplitude of the signal would then be estimated given this prior. This allows the available information on sources to be accounted for in the estimation of the power spectrum in a statistically robust manner. In future work we will investigate the level to which including realistic uncertainties on resolved source parameters as well as including priors on unresolved sources (for example, from external source catalogues constructed using high resolution instruments) impacts the power spectral constraints obtainable.

\subsection{Synchrotron self-absorption}
\label{Sec:SSA}

Synchrotron emission results from the acceleration of high-energy electrons around magnetic field lines in the source medium. At low frequencies, as the unabsorbed brightness temperature of synchrotron radiation approaches the effective temperature of the emitting electrons, the cross section for photon scattering becomes large and the emitting region becomes optically thick to synchrotron radiation (see e.g. \citealt{1970RvMP...42..237B}). The result is synchrotron self-absorption and is characterised by a turnover in the synchrotron spectrum at frequency $\nu_{\mr{t}}$, below which it asymptotes to $S_{\nu} \propto \nu^{2.5}$.
 
At sufficiently low frequencies, all synchrotron spectra will undergo self-absorption, however, for sources with a turnover frequency significantly below the frequency range spanned in our simulations, the effects of self-absorption will be negligible and their simulated spectra are well approximated as optically thin. For this transition frequency we use $\nu_\mr{tr} = 122 - W/2~\mr{MHz}$, where $122~\mr{MHz}$ is the lower limit on our observing band and $W/2$ is half of the synchrotron self-absorption turnover width. For local ($z=0$) sources with rest-frame turnover frequencies $\nu_\mr{t}<\nu_\mr{tr}$, the impact of self absorption on the observed spectrum will be small; it will be less still for sources at higher redshifts. Assuming a typical turnover width of a synchrotron self-absorbed source, $W \sim 100~\mr{MHz}$ (e.g. \citealt{2012MNRAS.424.2562Z}), we use $\nu_\mr{tr} = 70~\mr{MHz}$. We therefore split the sources into two categories. Sources with a rest frame synchrotron self-absorption turnover frequencies, $\nu_{\mr{t}} < 70(1+z)~\mr{MHz}$ are considered to be optically thin across the $122$ -- $130~\mr{MHz}$ frequency range of our foreground simulations. Remaining sources with an observed turnover frequency in excess of $70~\mr{MHz}$ are modelled as optically thick.

\begin{table}
\caption{Types and numbers of galaxies included in our full EGS simulation.}
\centerline{
\begin{tabular}{l l}
\toprule
Source Type              & Number of Galaxies ($10^{6}$)  \\
\midrule
Radio-Quiet AGN          & 207.9     \\
FR I                     & 137.1     \\
FR II                    & 0.0135    \\
Normal Galaxies          & 1196.9    \\
Starburst Galaxies       & 41.8      \\
\bottomrule
\end{tabular}
}
\label{Tab:SCubedSources}
\end{table}

\subsection{Optically thin}
\label{Sec:OpticallyThin}

The spectra of those sources in the optically thin regime are calculated as power laws with spectral indices $\alpha$ drawn from the experimentally derived spectral index distribution of \citet{2014MNRAS.440..327L}, their Figure 7. The distribution has been calculated from a power law fit to the flux-density of sources matched between the NRAO VLA Sky Survey (NVSS) catalogue at 1.4 GHz \citep{1998AJ....115.1693C} and Very Large Array low-frequency Sky Survey redux (VLSSr) catalogue at 73.8 MHz and has a mean $\lan\alpha\ran=0.82$ and standard deviation $\sigma_{\alpha}=0.19$. There have been conflicting results in the literature regarding whether there is a flattening of the average spectral index at low flux-densities in this frequency range (e.g. \citealt{2008AJ....136.1889O, 2009MNRAS.397..281I}). The more recent results of \citet{2012MNRAS.421.1644R} suggest there is no statistically significant evidence for a trend towards flatter spectral indices with decreasing flux density. Here, following these findings, we assume no significant evolution of the spectral index distribution with flux-density and draw spectral indices from the distribution of \citet{2014MNRAS.440..327L} for all sources in our EGS simulation.

\subsection{Optically thick}
\label{Sec:OpticallyThick}

Data on the fraction of sources expected to have a rest frame synchrotron self-absorption frequency turnover in the 100--1000 MHz frequency range is collated in \citet{1998PASP..110..493O}. They find that the fractional abundance of compact steep spectrum and gigahertz peaked sources (GPS) displaying absorption features in bright source populations are between $8.5\%$ and $31\%$ across a range of flux-density limited samples and selection frequencies.

To produce a physically motivated distribution of synchrotron self-absorption turnover frequencies for the radio-loud AGN population, we adopt the procedure of \citet{2008MNRAS.388.1335W}. This links turnover frequency with redshift dependent physical source size, $D_{\mr{true}}$, using the parametrisation of \citet{1998PASP..110..493O}: $\log(\nu_{\mr{t}}[\mr{GHz}]) = -0.21 - 0.65\log(D_{\mr{true}}[\mr{kpc}])$. $D_{\mr{true}}$ is drawn from $[0, D_{0}(1+z)^{-1.4}]$ with $D_{0} = 1~\mr{Mpc}$.
Combined with our optically thick / thin categorisation given in \autoref{Sec:EGSintro}, this results in $\sim 18\%$ of the radio-loud AGN in our EGS simulation being classified and modelled as optically thick.

In the rest frame of the source, the synchrotron self absorbed emission spectrum, $S_\mr{ab}$, resulting from electrons in a randomly orientated magnetic field, $B$, with a power-law energy distribution (of the form $N(E)\mr{d}E=\kappa E^{-p}\mr{d}E$, where $N(E)\mr{d}E$ is the number density of electrons in the energy interval $E$ to $E+\mr{d}E$ and $\kappa$ is the electron density distribution) can be parametrised as (see e.g. \citealt{2011hea..book.....L}),
\begin{equation}\label{Eq:SSASpectrum2}
S_\mr{ab}(\nu, p, l, B, \kappa)= \dfrac{S_{\nu}}{4\pi\chi_{\nu}}[1-\exp(-\chi_{\nu}l)] \ .
\end{equation}
Here $S_{\nu}$ is the optically thin power law spectrum, $l$ is the path length through the emitting region and $\chi_{\nu} = \chi_{0} \nu^{-(p+4)/2}$ is the synchrotron absorption coefficient with $\chi_{0} \propto \kappa B^{(p+2)/2}$.

One option for constructing $S_\mr{ab}(\nu, p, l, B, \kappa)$ is to ascribe distributions from which each of $p, l, B$ and $\kappa$ can be drawn. The electron energy density power law index, $p$, is related to the spectral index $\alpha$ by $p=2\alpha +1$. We draw $\alpha$ from the experimentally derived spectral index distribution of \citet{2014MNRAS.440..327L} as described in \autoref{Sec:OpticallyThin}. We equate $l$ with the physical size of the source, $D_{\mr{true}}$, described previously. Rather than ascribing distributions to $\kappa$, and $B$, we make use of the fact that $\chi_{0}$ mediates their impact on the source spectrum, and that $\chi_{0}$, in turn, is fixed if the source self-absorption turnover frequency (in addition to $p$ and $l$) is specified.
The source turnover frequency, $\nu_\mr{t}$, is equal to the frequency maximum of the synchrotron spectrum,
\begin{equation}\label{Eq:SSADerivative}
\left. \dfrac{\partial S_\mr{ab}(\nu, p, l, B, \kappa)}{\partial \nu} \right|_{\nu=\nu_{\mr{t}}}=0 \ .
\end{equation}
We therefore draw values for $\nu_\mr{t}$, $p$ and $l$ from their respective distributions, solve \autoref{Eq:SSADerivative} for $\chi_{0}$ numerically and construct the corresponding spectrum, $S_\mr{ab}(\nu, p, l, \chi_{0})$, on a per-source basis.

\subsubsection{Redshift dependence}
\label{Sec:SSARedshift}

\begin{figure*}
	\centerline{\includegraphics[width=0.5\textwidth]{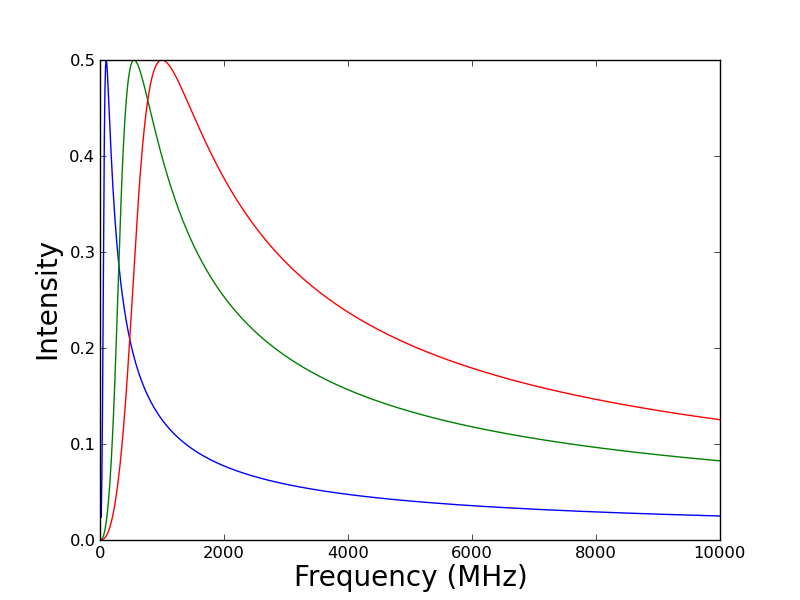}
	\includegraphics[width=0.5\textwidth]{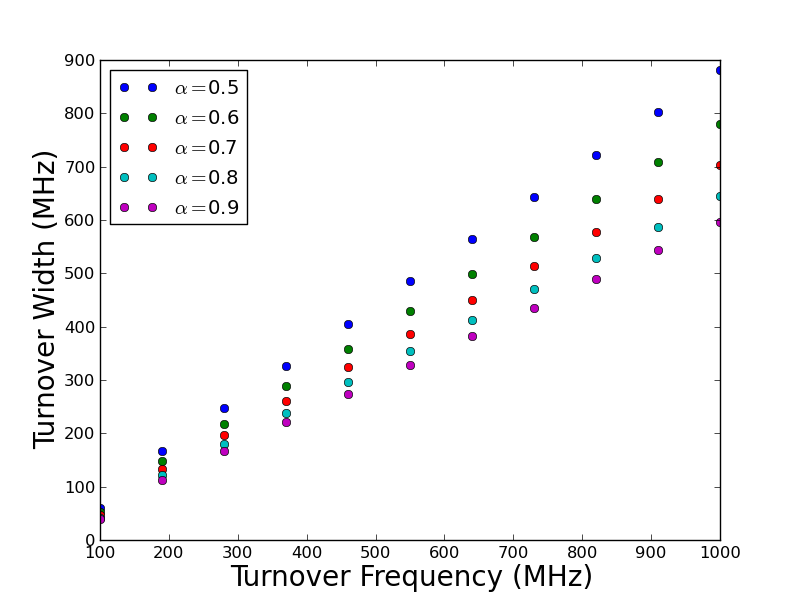}}
	\caption{[Left] Spectra of synchrotron sources exhibiting self-absorption at multiple rest frame turnover frequencies $\mr{d}I_{\nu,i}/\mr{d}\nu |_{\nu=\nu_{\mr{t, i}}}=0$. 
	[Right] Synchrotron self-absorbed spectral width (in MHz) -- defined as the width of the spectrum between points at intensities $1/\mr{e}$ times the maximum, $S_{\nu}=S(\nu_{t})$ -- as a function of turnover frequency for a range of source spectral indices. Turnover width increases linearly with turnover frequency. Flux-limited source samples will be biased towards observed turnover frequencies in close proximity to the observation frequency range due to the rapid ($\alpha=2.5$) fall-off of flux-density with frequency below the turnover. Therefore the sources detected in the simulated sample that have high rest frame turnover frequencies are high redshift ones where the observed turnover frequency has been redshifted towards the observation frequency.}
	\label{Fig:SSASpectrumVsTurnoverFrequency}
\end{figure*}

The impact of self-absorbed spectra on our ability to isolate the EoR power spectrum derives from the fact that absorption makes foreground spectra less smooth and thus more similar to the rapidly fluctuating 21-cm emission spectrum. The sharper the turnover, the more problematic absorption effects become. This was noted by \citet{2012MNRAS.424.2562Z} who studied the impact of synchrotron self-absorption, approximated as a broken power law, on foreground removal. 

We define the rest frame characteristic width of the synchrotron self-absorbed spectra, $W_{\nu}$, as the frequency interval asymmetrically about $\nu_{t}$ between the two spectral points at which the flux-density of the source has fallen to $1/\mr{e}$ of its peak, $S(\nu_{t})$. The width $W_{\nu}$ is a function of the spectral index $\alpha$ and turnover frequency $\nu_{t}$ and observationally can be seen to be of the order of $100~\mr{MHz}$ (e.g. \citealt{1995A&AS..113..409S}). The observed characteristic width, $W_{\nu,\mr{o}}$, is redshift narrowed by a factor of $(1+z)$ where $z$ is the redshift of the source. Therefore, all other things being equal, it would appear that high redshift sources have the potential to be the most problematic synchrotron self-absorbed foregrounds when spectral characterisation is being relied upon to distinguishing the EoR signal from foregrounds. A faint source at redshift 9, for example, will have $W_{\nu,\mr{o}}$ reduced by a factor of 10 relative to $W_{\nu}$. This would bring an $\mathcal{O}(100~\mr{MHz})$ characteristic width down to $\mathcal{O}(10~\mr{MHz})$ which is the same order of magnitude as the coherence length of the EoR signal (see e.g. \citealt{2008AJ....136..641R}). However, there exists a second effect resulting from biasing due to source selection which acts in opposition to spectral narrowing, as described next.

\subsubsection{Spectral selection biasing}
\label{Sec:SSASelectionBiasing}

The characteristic width of the synchrotron self-absorbed spectra, $W_{\nu}$, can be shown to be linearly dependent on turnover frequency $\nu_{\mr{t}}$ as illustrated in \autoref{Fig:SSASpectrumVsTurnoverFrequency}. For a source at redshift $z$ to exhibit synchrotron self-absorption in our observing frequency range, it is required to have a rest frame turnover frequency $\nu_{\mr{t, min}} \ge (1+z)\nu_\mr{tr}$. Therefore, high redshift sources are constrained to have high turnover frequencies and correspondingly large rest frame characteristic synchrotron self-absorbed widths. Since at redshift $z$ the mean broadening resulting from this selection bias is greater than or equal to $(1+z)$ and the corresponding redshift narrowing is inversely proportional to the same factor, the two effects cancel leaving the characteristic width of the observed synchrotron self-absorbed spectra from a flux-limited sample approximately constant.

We use an $8~\mr{MHz}$ bandwidth for our EGS simulation, matching that of our EoR simulation. On these short spectral scales, the synchrotron self-absorbed emission sources remain comparatively smooth. That is, deviations from a constant emission gradient in log-frequency space, along a line of sight, are of the same order of magnitude to sources in the optically thin regime. The major impact, therefore, of synchrotron self-absorption over the $8~\mr{MHz}$ bandwidth is to produce spectra with effective spectral indices drawn from a wider distribution encompassing both positive and negative (GPS sources) spectral indices.

\section{Simulated observations}
\label{Sec:SimulatedHERAObservations}

For our instrumental model, we consider two planned configuration of the HERA experiment with 37 and 331 antennas. HERA is a drift scanning interferometer currently under construction in the low RFI environment of the SKA-South Africa site in the Karoo desert (see e.g. \citealt{2015ApJ...800..128B} for additional details). Here we consider estimation of the power spectrum using data from the simulated observation, with HERA, of the field defined by our Galactic region A (see \autoref{Fig:remazeilles_haslam408}).

\begin{figure*}
	\centerline{
	\includegraphics[width=0.49\textwidth]{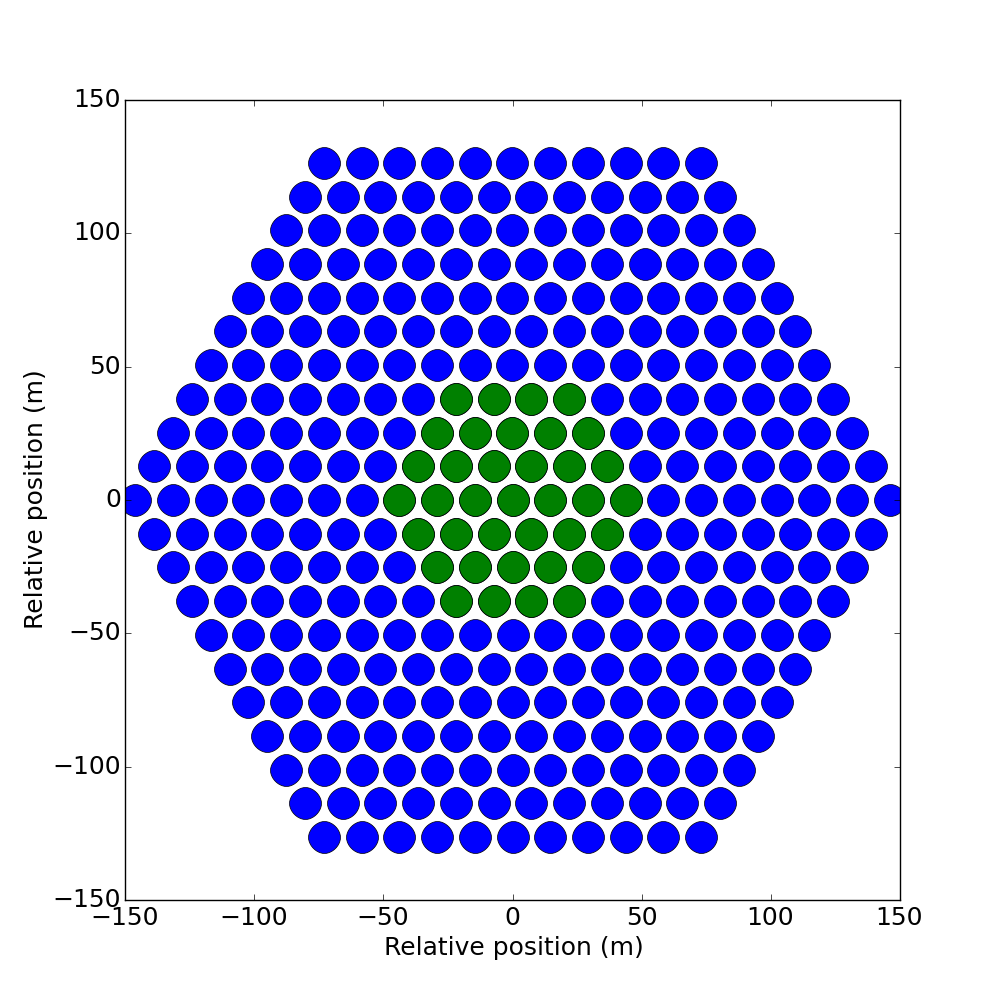}
	\includegraphics[width=0.49\textwidth]{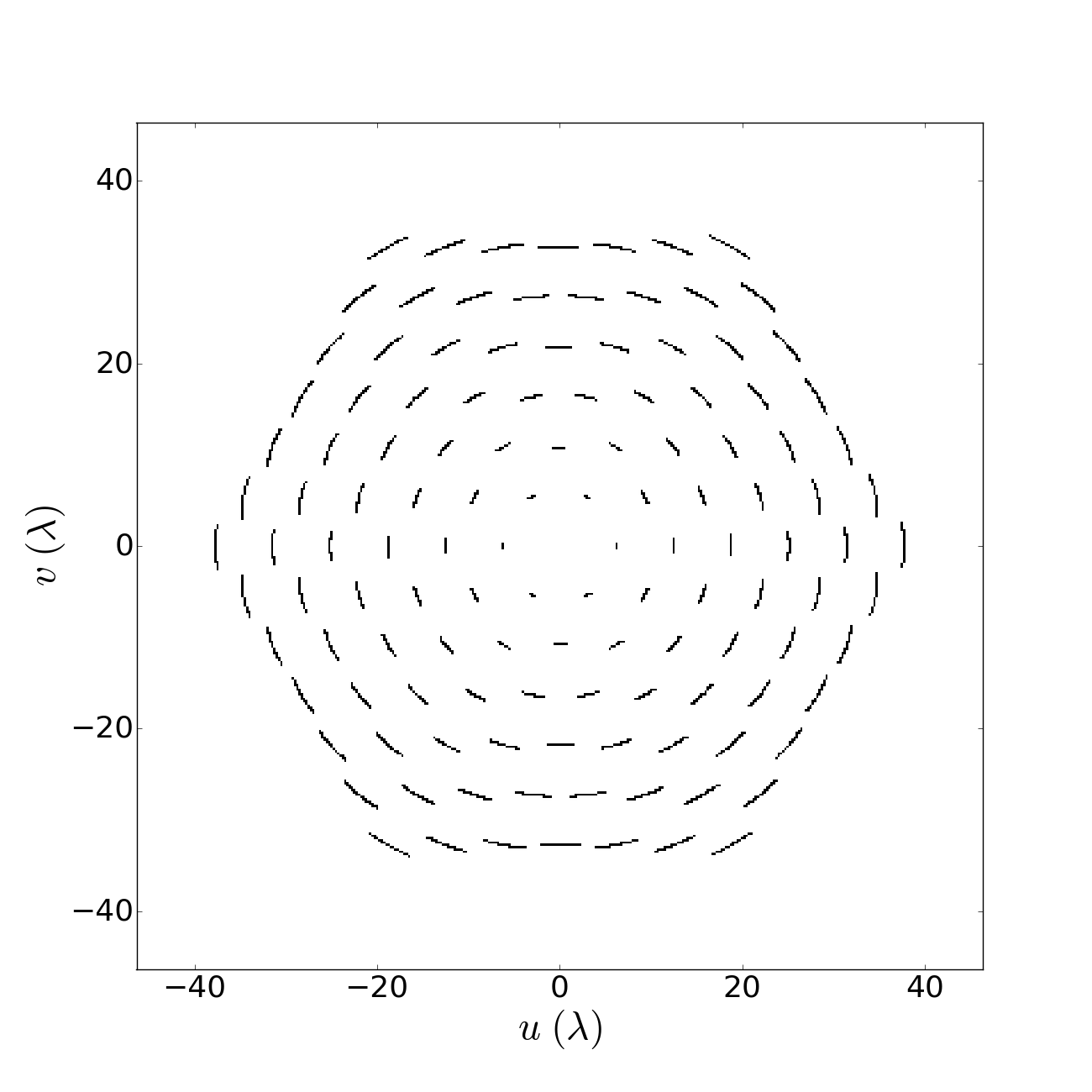}}
	\caption{[Left] HERA 37 (green) and 331 (green $+$ blue) antenna configurations. Antenna positions plotted relative to $30\fdg721$S $21\fdg411$E at the SKA-South Africa site, Karoo desert. [Right] HERA 37 sampled $uv$-coordinates at 126 MHz for 30 minutes of observation time with an integration time of 30 seconds. Within the spatial resolution range sampled by HERA 37 ($2.5~\lambda < u < 37~\lambda$), HERA 331 samples the same $uv$-coordinates but with increased frequency as a result of the high redundancy of the hexagonal close packed configuration of the antennas.}
	\label{Fig:AntennaPlusuvCoverage}
\end{figure*}

HERA is being constructed from $14~\mr{m}$ diameter dishes in a hexagonal configuration (see \autoref{Fig:AntennaPlusuvCoverage}, left) which can be shown \citep{2012ApJ...753...81P} to maximise the total number of redundant baselines. We use the array configuration for HERA in 37 and 331 antenna configurations as an inputs to the Common Astronomy Software Applications (CASA\footnote{http://casa.nrao.edu}) simobserve tool to obtain the set of sampled $(u,v)$ visibility coordinates (see \autoref{Fig:AntennaPlusuvCoverage}, right) corresponding to a 30 minute transit observation of our simulated sky centred on (RA, Dec) = ($0\fdg0,\ -30\fdg0$) and comprised of 60, 30 second integrations for 38, $\sim 200~\mr{kHz}$ channels spanning the range 122.17--129.90 MHz.

Our focus, here, is calculating the impact of the intrinsic structure of the foregrounds and the EoR on their power spectra. We therefore avoid a number of instrumental complications to calculating the power spectrum by applying some simplifying assumptions. We do, however, model fundamental attributes of an interferometer including frequency dependent $uv$-coverage and primary beam. We approximate the HERA primary beam $\mathbf{P}$ as a Gaussian with FWHM $8\fdg0 \nu_{i}/122.17~\mr{MHz}$ where $122.17~\mr{MHz}$ is the central frequency of our lowest frequency channel and we assume that the data under consideration has been perfectly calibrated. Computation time of the likelihood when calculating the power spectrum scales as number of $uv$-cells cubed. Therefore, for the larger HERA 331-antenna configuration we restrict our analysis to baselines within the $2.5~\lambda < u < 37~\lambda$ resolution range of HERA in 37-antenna configuration. This enables us to take advantage of the increased sensitivity of the larger instrument in this resolution range without additional computational cost. We will address the effect on the power spectrum of deviations from these assumptions in a future paper.

We perform a non-uniform DFT of the zero-noise, primary beam multiplied model image\footnote{
By the convolution theorem, this is identical to performing a grid-to-grid DFT from the image to $uv$-domain followed by degridding, with a convolution kernel equal to the aperture function of the instrument (the Fourier transform of the primary beam), to the sampled visibilities.}
to the $uv$-coordinates sampled by HERA for each of the $i=38$ channels during the 30 minute simulated observation, obtaining for each channel the sampled visibilities (see \autoref{Sec:PowerSpectralModelandAssumptions} for details). We add uncorrelated white noise to the real and imaginary component of each of the sampled visibilities independently. The noise level on a visibility resulting from a pair of identical antennas individually experiencing equal system noise can be shown to be (e.g. \citealt{1999ASPC..180.....T}),
\begin{equation}\label{Eq:VisabilityNoise}
\sigma_{V}=\dfrac{1}{\eta_{s}}\dfrac{SEFD}{\sqrt{2\Delta\nu\tau}} \ ,
\end{equation}
where $\eta_{s}$ is the system efficiency, $\Delta\nu$ is the frequency width of the observation, $\tau$ is the integration time and $SEFD$ is the System Equivalent Flux Density given by,
\begin{equation}\label{Eq:SEFD}
SEFD=\dfrac{2k_{B}T_{\mr{sys}}}{\eta_{a}A} \ ,
\end{equation}
where $T_{\mr{sys}}$ is the system noise temperature, $k_{B}=1.3806\times10^{-23}~\mr{JK^{-1}}$ is Boltzmann's constant, $\eta_{a}$ is the antenna efficiency and $A$ is the area of a 14 m diameter HERA dish.

\begin{table}
\caption{Instrumental and observational parameters.}
\centerline{
\begin{tabular}{l l l}
\toprule
Parameter                & Description                   & Value  \\
\midrule
$\eta_{s}$               & System efficiency             & 1    \\
$\Delta\nu$              & Channel width                 & $200~\mr{kHz}$     \\
$\tau$                   & Integration time              & $30~\mr{s}$  \\
$\eta_{a}$               & Antenna efficiency            & 1    \\
$A$                      & Antenna area                  & $150~\mr{m^2}$\\
\bottomrule
\end{tabular}
}
\label{Tab:NoiseParameters}
\end{table}

Using the parameter values given in \autoref{Tab:NoiseParameters} and assuming a $2000~\mr{hr}$ (4000 transits of the centre of the simulation cube each of 30 minute duration) observation and an approximately constant system noise temperature, $T_{\mr{sys}}=550~\mr{K}$, across our $8~\mr{MHz}$ bandwidth, \autoref{Eq:VisabilityNoise} yields $0.045~\mr{Jy}$ for the RMS of the uncorrelated Gaussian random noise. We add this RMS independently to the real and imaginary component of each of the sampled visibilities.

\section{Analysis}
\label{Sec:Analysis}

We organise the results of our analysis into four subsections. Firstly, in \autoref{Sec:SpatialPowerSpectra}, we calculate the intrinsic spatial power spectrum across the full spatial resolution range probed by our simulations. We investigate the ratio of EoR to foreground power as a function of position and determine where it is maximised.
In \autoref{Sec:Results3D} we calculate the spherically averaged intrinsic three-dimensional $k$-space power spectrum at a fixed spatial resolution range corresponding to that probed by HERA in 37-antenna configuration (see \autoref{Fig:AntennaPlusuvCoverage}). We investigate the dependence of the power spectrum on the level of correlation between brightness temperature and spectral index in our GDSE simulation. For fully correlated brightness temperature -- spectral index distributions we calculate the power spectrum of GDSE in three Galactic regions (regions A, B and C in \autoref{Fig:remazeilles_haslam408}), EGS, free--free emission and our EoR simulation. We calculate the fraction of total recovered power accounted for by EoR emission both in the spherically averaged and corresponding maximum likelihood cylindrically averaged power spectra.
In \autoref{Sec:ObservedPowerSpectra} we estimate the three-dimensional $k$-space power spectrum of our sky model from visibilities corresponding to its simulated observation with HERA in 37-antenna configuration and HERA in 331-antenna configuration using baselines restricted to those in the resolution range of HERA 37. For our sky model we assume region A is typical of the relatively cold Galactic regions favoured for observations from which to estimate of the EoR power spectrum, and take the model in this region as input for our simulated observations.
In \autoref{Sec:Comparison} we compare our results with alternative power spectral estimation methodologies.
 
Throughout this section, when calculating intrinsic power spectra, we add a minimum level of noise compatible with numerical stability of the likelihood at double precision. The posteriors are thus signal dominated and so the maximum a posteriori and maximum likelihood parameter estimates are consistent. When plotting the intrinsic power spectra we use the maximum a posteriori parameter estimates. We use the same binning for the $k$-space intrinsic and observed power spectra (see \autoref{Sec:ObservedPowerSpectra}) and for the intrinsic power spectra we linearly interpolate between bin centres to produce the displayed figures.

\subsection{Spatial power spectra }
\label{Sec:SpatialPowerSpectra}

In this subsection we estimate the two-dimensional spatial power spectrum, $P_{uv}(\nu)$, of each of our simulated emission components. First, we consider the relation between $P_{uv}$ and the three-dimensional $k$-space power spectrum, $P(k)$. If $P(k)$ is assumed to be spatially separable, we can write, $P(k_{\perp}, k_{\parallel}) = P_{k_{\perp}}(k_{\perp})P_{k_{\parallel}}(k_{\parallel})$ and it follows that $P_{k_{\perp}} \propto P_{uv}$. The spatial separability of $P(k)$ requires the functional form of the spatial power spectrum to be independent of frequency\footnote{The spatial separability of the foreground three-dimensional $k$-space power spectra can also be assessed directly through calculation of their cylindrical power spectra (see \autoref{Sec:MCPS})}. That is, we require, $P_{uv}(\nu_{1}, \abs{\mathbfit{u}}) = C P_{uv}(\nu_{2}, \abs{\mathbfit{u}})$ where $C(\nu_{1}, \nu_{2})$ is a constant with respect to $\abs{\mathbfit{u}}$. Across the $8~\mr{MHz}$ bandwidth of our foreground simulations this is a reasonable approximation with fractional fluctuations (as a function of $\abs{\mathbfit{u}}$) of $\Delta C / C \lesssim 0.1$. As such, foreground specific characterisation of $P_{uv}$ is a method through which bias in estimates of the three-dimensional $k$-space power spectrum of the EoR signal can be reduced by targeting estimates in regions of $k_{\perp}$-space where contaminating foreground power is minimised.

\begin{figure}
	\centerline{\includegraphics[width=\columnwidth]{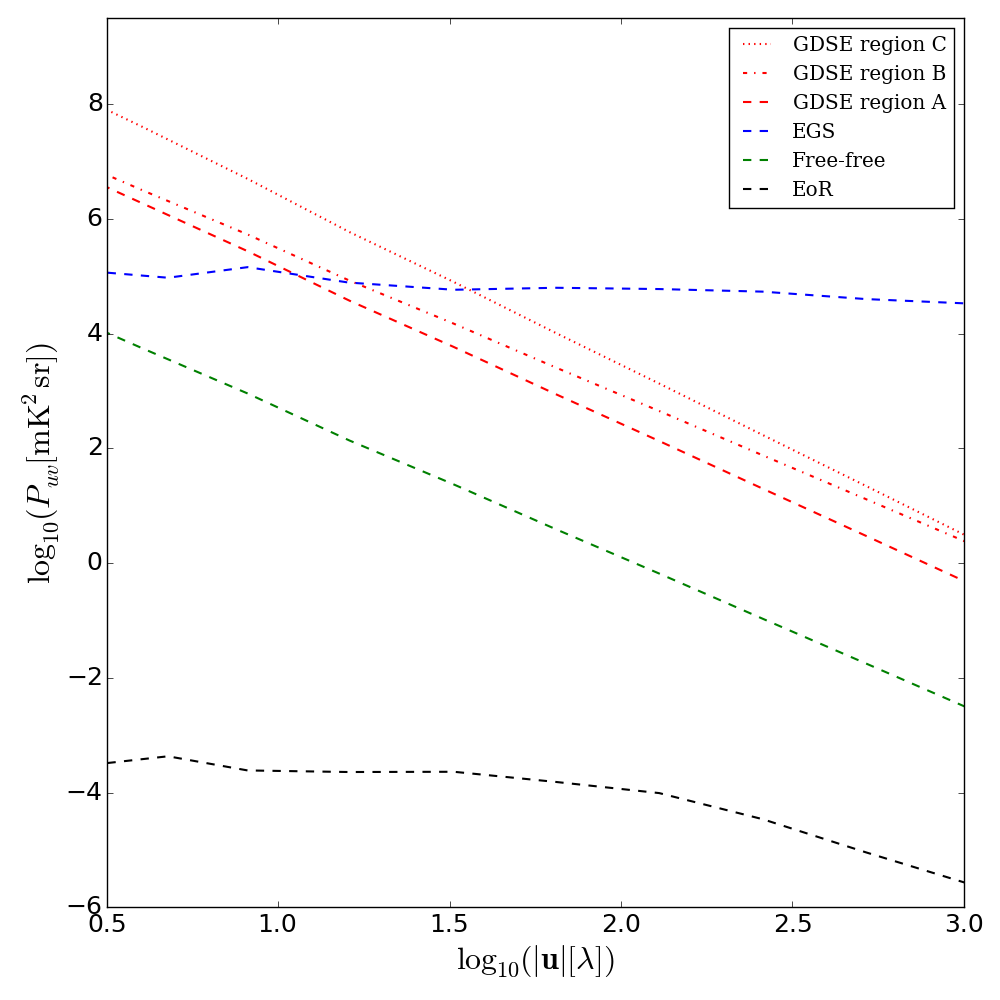}}
	\caption{Spatial power spectra at 126 MHz of the redshifted 21-cm signal (black dashed line), Galactic diffuse synchrotron emission in regions A, B and C (red dashed, dash--dotted and dotted lines respectively), diffuse free--free emission (green dashed line) and extragalactic sources (blue dashed line).}
	\label{Fig:SkyCompSpatialPS}
\end{figure}

In \autoref{Fig:SkyCompSpatialPS} we show the spatial power spectra, $P_{uv}$, between 3 and 1000 wavelengths\footnote{This corresponds to a $k_{\perp}$ range: $1.9 \times 10^{-3}~h\mr{Mpc^{-1}} \le k_{\perp} \le 6.5 \times 10^{-1}~h\mr{Mpc^{-1}}$ for the 21-cm signal at $z=10.26$.} 
at 126 MHz of the Galactic diffuse synchrotron emission in regions A, B and C (red dashed, dash--dotted and dotted lines respectively; regions defined in \autoref{Fig:remazeilles_haslam408}), extragalactic sources (blue dashed line), diffuse free--free emission (green dashed line) and the redshifted 21-cm signal (black dashed line). Of the foreground power spectral components, power in free--free emission from the Galaxy is sub-dominant on all spatial scales. GDSE varies as a function of position in the Galaxy. Regions A and B lie out of the plane of the Galaxy and are relatively cold in total intensity. Region C overlays the Galactic plane and is found to have the greatest power of the three regions on all sampled scales. For $\abs{\mathbfit{u}} \lesssim 30~\lambda$ ($\sDelta\theta \gtrsim 2\fdg0$), GDSE dominates the spatial power spectrum of the sky. At higher angular resolutions, point sources become the dominant source of foreground power. The relative importance of the different foreground components in \autoref{Fig:SkyCompSpatialPS} is consistent with previous work (see e.g. \citealt{1999A&A...345..380S, 2005ApJ...625..575S}).

There is a strong dependence of the functional form of the spatial power spectrum on emission component. Galactic emission is well approximated by spatial power laws and EGS are approximately flat, subject to perturbations due to source clustering. For our analysis of the spherically averaged three-dimensional $k$-space power spectra in \autoref{Sec:Results3D} we use a spherically symmetric prior (as outlined in \autoref{Sec:PowerSpectralModelandAssumptions}) which is the correct model for an EoR dominated $k$-space power spectrum. However, if significant foreground contamination is assumed, then incorporating the spatial power spectral results outlined in this section via more informative spatial priors when estimating the three-dimensional $k$-space power spectrum of the total signal provides a promising route towards isolating the component of the total power spectrum attributable to EoR emission. This can be achieved in an analogous manner to that used for our power spectral decomposition of the GDSE in \autoref{Sec:GalacticForegroundEmission} and is investigated separately in an upcoming paper (Sims et al. in prep.).

\begin{figure}
	\centerline{\includegraphics[width=\columnwidth]{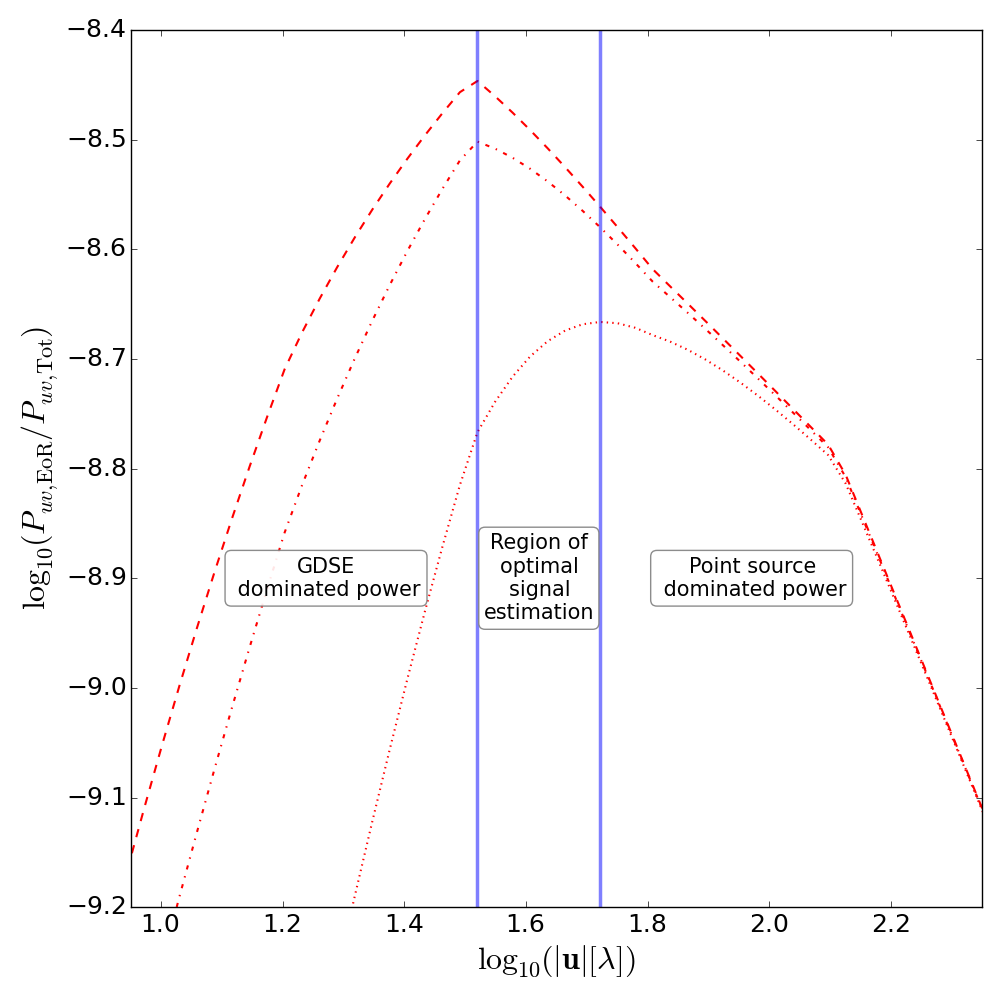}}
	\caption{Ratio of the spatial power spectrum of the EoR to the sum of spatial power in the foregrounds at 126 MHz in regions A, B and C (red dashed, dash--dotted and dotted lines respectively). We identify a model dependent region of optimal signal estimation in the range $35~\lambda \le \abs{\mathbfit{u}} \le 55~\lambda$, with lower and upper bounds equal to the spatial EoR-to-foreground power ratio maximum in region A and C respectively.}
	\label{Fig:EoRFGFractionalSpatialPS}
\end{figure}

In \autoref{Fig:EoRFGFractionalSpatialPS} we plot the ratio of the spatial power spectrum of our EoR simulation to the sum of spatial power in the foregrounds in regions A, B and C (red dashed, dash--dotted and dotted lines respectively). This suggests an alternative or complementary instrumental approach to minimising foreground bias when estimating the three-dimensional $k$-space power spectrum of the EoR. With knowledge of the spatial power spectrum of each emission component, the antenna layout in interferometric observations can be tuned for baselines that concentrate $uv$-sampling on scales where the spatial EoR-to-foreground power ratio (SEFPR) is maximised. In so doing, we can instrumentally minimise foreground bias in estimates of the three-dimensional $k$-space power spectrum of the EoR. Following this principle, for the foreground and EoR models used here, we identify a model dependent region of optimal signal estimation in the approximate wavelength range $35~\lambda \le \abs{\mathbfit{u}} \le 55~\lambda$ at $126~\mr{MHz}$, with lower and upper bounds equal to the SEFPR maximum in region A and C respectively. Also marked in \autoref{Fig:EoRFGFractionalSpatialPS} are the regions of GDSE and point source dominated foreground power previously mentioned.

The limits derived are a function of the input EoR and foreground models presented and demonstrate the calculation of the region of optimal signal estimation for our analysis with an EGS model corresponding to point sources unresolved by HERA in 37 antenna configuration. Ideally, prior information on both resolved sources and unresolved sources (for example from source catalogues compiled with higher resolution instruments)  would be included in the likelihood in order to optimally constrain this component of the foregrounds in a statistically robust manner. Since the number of sources that can be resolved is antenna configuration dependent, the region of optimal signal estimation will be instrument specific. Additionally, the foreground emission and region of optimal signal estimation in any particular target field will be a function of the relative intensities of point source emission and GDSE. As such, in observations, the optimal baselines on which to perform our Bayesian estimation of the EoR power spectrum will be field specific and should be estimated using the spatial power spectrum of the foregrounds measured from the data.

A second important factor in calculating the region of optimal signal estimation is the astrophysical parameter dependence of the shape and amplitude of the EoR power spectrum. For a foreground spatial power spectrum in a field measured from observational data, it is valuable to consider the generality with which a region of optimal signal estimation can be defined. The specific shape of the SEFPR, and with it the location of the region of optimal signal estimation, is a function of the form of the EoR power spectrum. The SEFPR is frequency dependent due to the evolution of the 21-cm signal over cosmic time. It is a generic prediction of 21-cm simulations that the amplitude of the 21-cm power spectrum peaks at $x_{H} \sim 0.5$ (e.g. \citealt{2008ApJ...680..962L}). This makes it a promising candidate for the neutral fraction at which a first detection of the EoR power spectrum will be made. For this reason, it is also the choice of neutral fraction for the EoR signal used in this paper. Although the power spectrum as a function of neutral fraction differs substantially amongst models, it is only the shape of the power spectrum, not its amplitude\footnote{While the amplitude of the EoR power spectrum does not alter the location of the maximum of the SEFPR, and thus the baselines on which the ratio of the EoR to foreground power is maximised, it does alter the absolute level of foreground bias in the spherically averaged three-dimensional $k$-space power spectrum, estimated on a fixed set of baseline.}, that defines the maximum of the SEFPR. For the EoR power spectrum at a fixed neutral fraction it is therefore valuable to consider the effect of varying astrophysical parameters on its shape. It is apparent from Figures \ref{Fig:SkyCompSpatialPS} and \ref{Fig:EoRFGFractionalSpatialPS} that, in order for the shape of the EoR power spectrum to alter the limits of the region of the optimal signal estimation, one of two features must be present: \begin{enumerate*} \item either, in the GDSE dominated region of the spatial power spectrum, the increase in EoR spatial power with increasing spatial scale must be more rapid than that of the spatial power spectrum of the GDSE, or \item the decline in EoR power with decreasing spatial scale must reverse in the EGS dominated region of the spatial power spectrum\end{enumerate*}. We consider the $\zeta$, $R_\mr{mfp}$, $T^\mr{Feed}_\mr{vir}$ parametrisation of the EoR signal made use of in \citet{2015MNRAS.449.4246G} (hereafter G15), with $\zeta$ the galactic ionizing efficiency, $R_{\rm mfp}$ the mean free path of ionizing photons in the IGM and $T^{\rm Feed}_{\rm vir}$ the minimum virial temperature of star-forming haloes. We match our parameter space to the observationally motivated ranges used by G15: $\zeta\in[5,100]$, $R_{\rm mfp}\in[5, 20]$~cMpc and $T^{\rm Feed}_{\rm vir}\in[10^{4},2\times10^{5}]$~K (see \citealt{2015MNRAS.449.4246G} for details) and consider models with a neutral fraction, $x_{H}$, in the range $0.45$ to $0.55$. In this parameter range, no EoR models exhibit an increase in spatial power with increasing spatial scale more rapid than that of the spatial power spectrum of our GDSE model. While some EoR models exhibit shallow intermediate maxima, the impact of this is minimal, resulting in shifts of the peak SEFPR by less than $15\%$. 

Finally, we note that the spherical $k$-space symmetry of the EoR signal implies that, unlike the foregrounds, the power spectrum of the EoR is not spatially separable. Rather, it is a function of $\abs{\mathbfit{k}} = (k_{\perp}^{2}+k_{\parallel}^{2})^{1/2}$. As a result, a difference between the three-dimensional power spectrum of the EoR in physical units, $P(k)$, and $P_{uv}$ could, in principle, alter the optimal spatial resolution range at which to estimate the EoR signal. The spatial power spectrum of the EoR represents a cross-section through $P(k)$ at $k_{\parallel}=0$. It follows from the expected spherical symmetry of the EoR signal in $k$-space, that $P(k)$ for non-zero $k_{\parallel}$ can be estimated from  $P_{uv}$ and that the functional form of $P(k)$ is described by $P_{uv}$, bar a change in coordinates. For this coordinate shift to alter the SEFPR for a specific target field, equivalent conditions must be present to those that can lead to a variation in the power spectrum with astrophysical parameters. That is, \begin{enumerate*} \item either, in the GDSE dominated region of the spatial power spectrum, the increase in $P(k)$ for the EoR with increasing spatial scale must be more rapid than that of the spatial power spectrum of the GDSE, or \item the decline in $P(k)$ for the EoR with decreasing spatial scale must reverse in the EGS dominated region of the spatial power spectrum\end{enumerate*}. For the range of EoR models described above, we find that neither of these features occur in estimates of $P(k)$ for the range of $k_{\parallel}$ coordinates sampled by our fiducial EoR simulation.

Together, this suggests that for a foreground spatial power spectrum estimated from the data for an observed area of sky, a relatively robust calculation of the optimal range of baseline lengths on which to target estimation of the three-dimensional spatial power spectrum of the EoR using our Bayesian analysis can be achieved.

\subsection{$k$-space power spectra }
\label{Sec:Results3D}

In this section we consider the dimensionless spherically and cylindrically averaged three-dimensional $k$-space intrinsic power spectra\footnote{It is important to point out that because the frequency axis of the foregrounds does not map to radial distance, as it does for the redshifted 21-cm signal, it is only the EoR power spectrum in this section that represents the physical power in the signal averaged over three spatial dimensions. The measurement space $(\mathbfit{u}_{i},\nu_{i})$ of the foregrounds and EoR are the same, and for convenience, we will discuss the analysis of the foregrounds in the same terms as the EoR signal, however, the reader should keep in mind that the foreground power spectra presented represent the level of contamination of the EoR power spectrum in the analysis rather than their physical $k$-space power spectra.} obtained using the Bayesian analysis framework outlined in \autoref{Sec:PowerSpectralModelandAssumptions}. Here we analyse these $k$-space power spectra across the spatial resolution range $2.5~\lambda < u < 37~\lambda$ ($1.6\times10^{-3}~h\mr{Mpc^{-1}} < k_{\perp} < 3.3\times10^{-2}~h\mr{Mpc^{-1}}$), matching the spatial resolution range of HERA in 37-antenna configuration (shown in \autoref{Fig:AntennaPlusuvCoverage}) and broadly corresponding to the region of maximum spatial sensitivity of current generation 21-cm experiments.

\subsubsection{Brightness temperature -- spectral index correlation}
\label{Sec:ResultsSIC}

\begin{figure}
	\centerline{
	\includegraphics[width=0.50\textwidth]{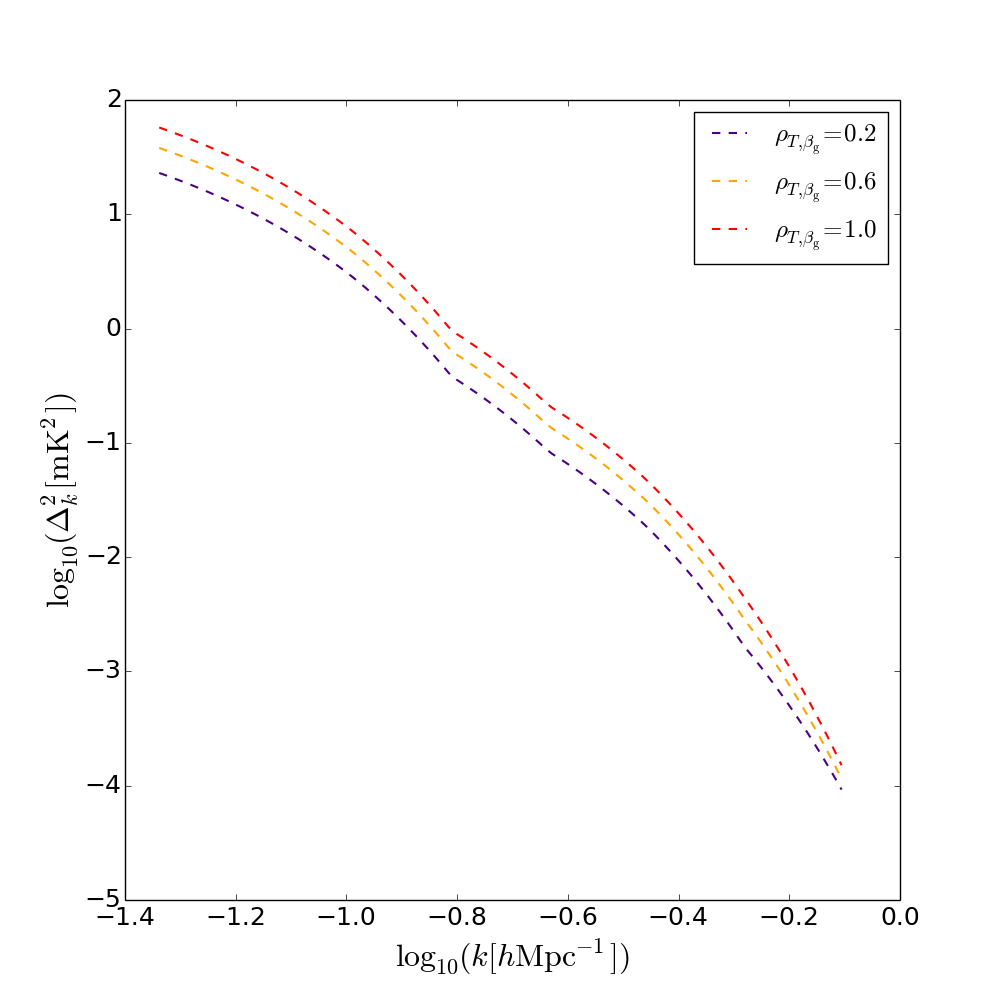}
	}
	\caption{Maximum a posteriori spherically averaged three-dimensional $k$-space dimensionless intrinsic power spectra of GDSE in region A for three levels of brightness temperature -- spectral index correlation. Power decreases with decreasing brightness temperature -- spectral index correlation. This is expected due to decreasing spatial coherence over the frequency band of the simulation with decreasing $\rho_{T,\beta_\mr{g}}$. This transfers spatial power to smaller spatial scales and, at the lower edge of the range, out of the spatial resolution range sampled by the spherical power spectrum.}
	\label{Fig:GalacticPSvsSIC}
\end{figure}

\autoref{Fig:GalacticPSvsSIC} shows the maximum a posteriori spherically averaged three-dimensional $k$-space dimensionless intrinsic power spectra obtained for GDSE with $20\%$ (blue dashed line), $60\%$ (yellow dashed line) and $100\%$ (red dashed line) levels of brightness temperature -- spectral index correlation. Power is observed to decrease with decreasing correlation. 

In the limit that the correlation between temperature and spectral index tends to zero, coherent structure in the emission (initially following the spatial power law illustrated in \autoref{Fig:SkyCompSpatialPS}) will tend towards a flat noise-like power spectrum for spectral extrapolation over a large frequency range. The impact of this is to shift power towards smaller spatial scales thereby shifting power out of the fixed spatial scale range probed in our analysis of the spherical power spectrum. This will occur similarly for non-zero correlation with the rate of decoherence inversely proportional to the level of correlation.

As detailed in \autoref{Sec:GalacticForegroundEmission}, our GDSE simulation is matched to our maximum a posteriori synchrotron spatial structure model derived from our Bayesian power spectral decomposition of the frequency scaled \citet{2015MNRAS.451.4311R} Haslam all-sky map at $126~\mr{MHz}$. The fractional frequency range corresponding to the $126 \pm 4~\mr{MHz}$ frequency band of our simulations is small (a $\sim 3.2\%$ shift between the centre and the edge of the band). Nevertheless, \autoref{Fig:GalacticPSvsSIC} demonstrates that decorrelation of spatial structure has a measurable impact even over this range. Therefore, we conservatively use $\rho_{T,\beta_\mr{g}} = 1$ for our GDSE model in the following analyses.

\subsubsection{Multi-component power spectra }
\label{Sec:MCPS}

\begin{figure*}
	\centerline{
	\includegraphics[width=0.50\textwidth]{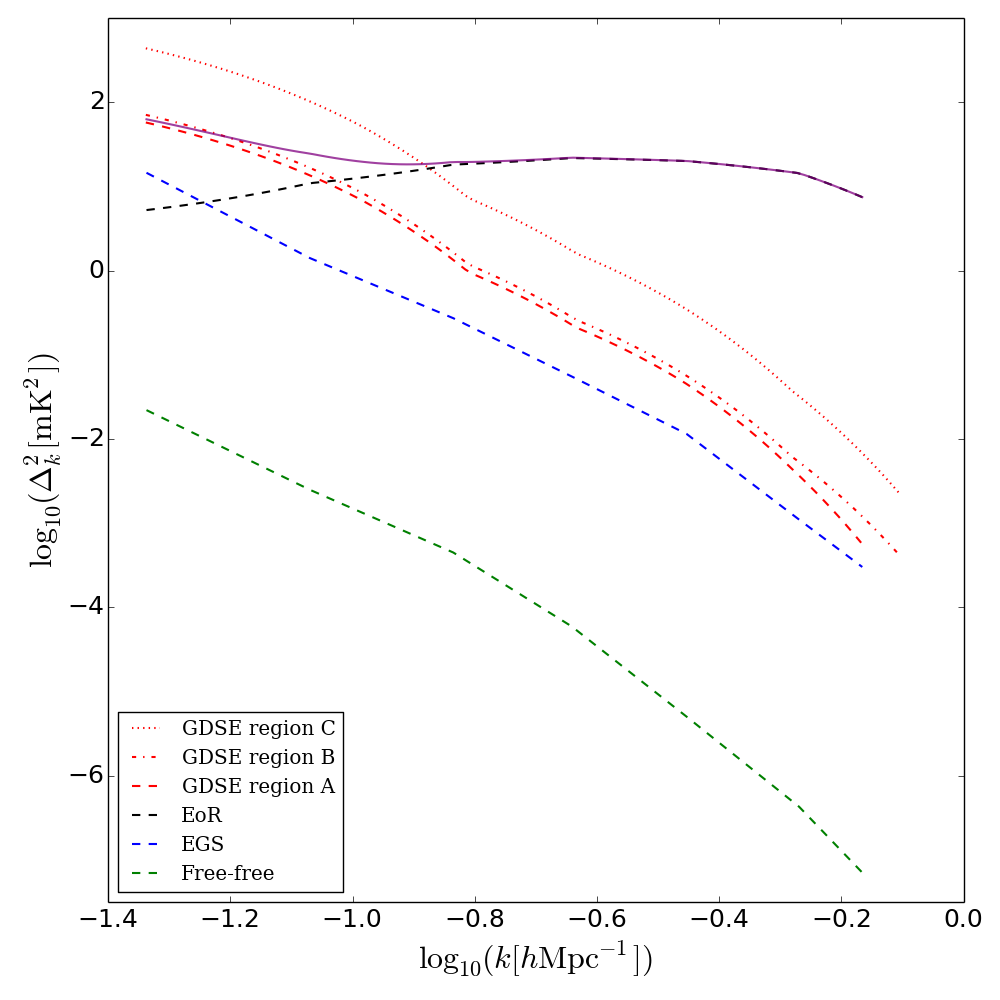}
	\includegraphics[width=0.50\textwidth]{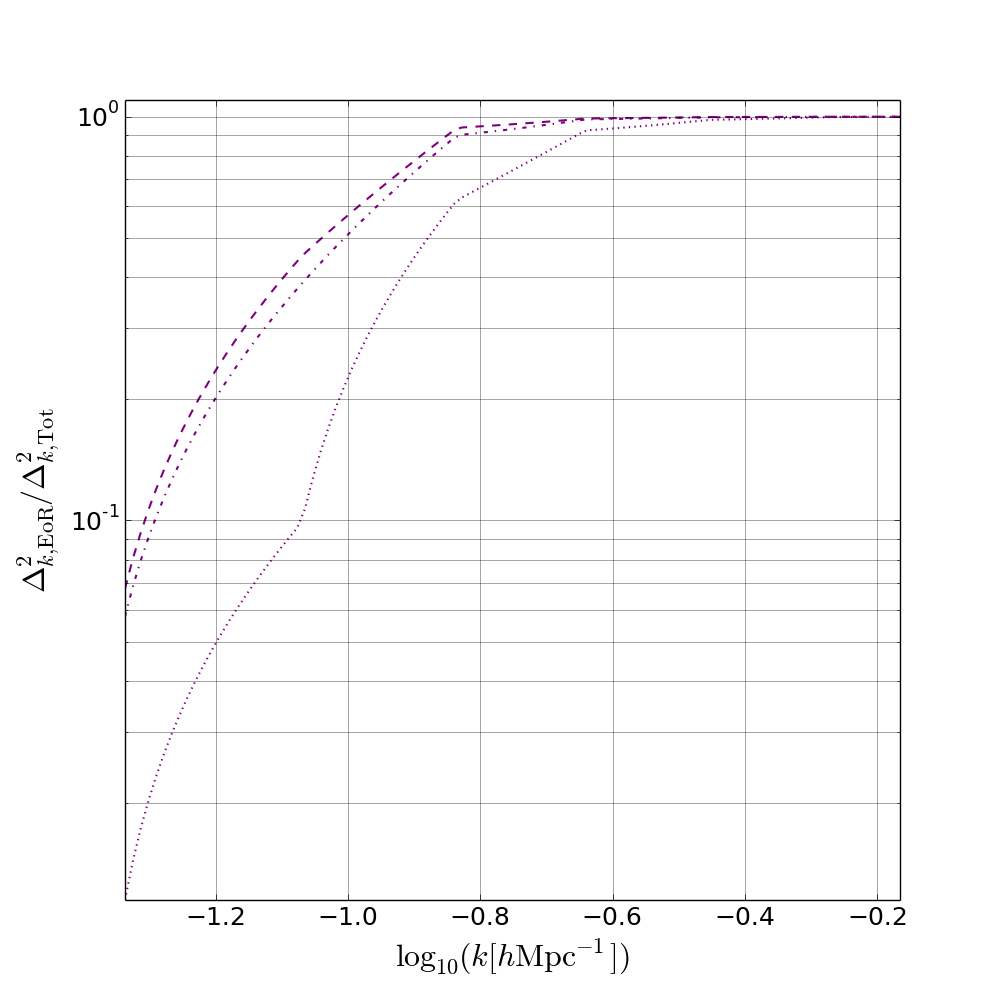}}
	\caption{[Left] Maximum a posteriori spherically averaged three-dimensional $k$-space dimensionless intrinsic power spectra obtained for the redshifted 21-cm signal (black dashed line), Galactic diffuse synchrotron emission in regions A, B and C (red dashed, dash--dotted and dotted lines respectively), extragalactic foregrounds (blue dashed line) and diffuse free--free emission (green dashed line). Total power summed over all simulation components assuming GDSE emission in region A is shown in purple. [Right] Fraction of total power in the spherical power spectrum accounted for by the EoR signal in region A (dashed line), region B (dash--dotted line) and region C (dotted line).
	}
	\label{Fig:SphericalIntrinsicPS}
\end{figure*}

\begin{figure*}
	\centerline{
	\includegraphics[width=0.50\textwidth]{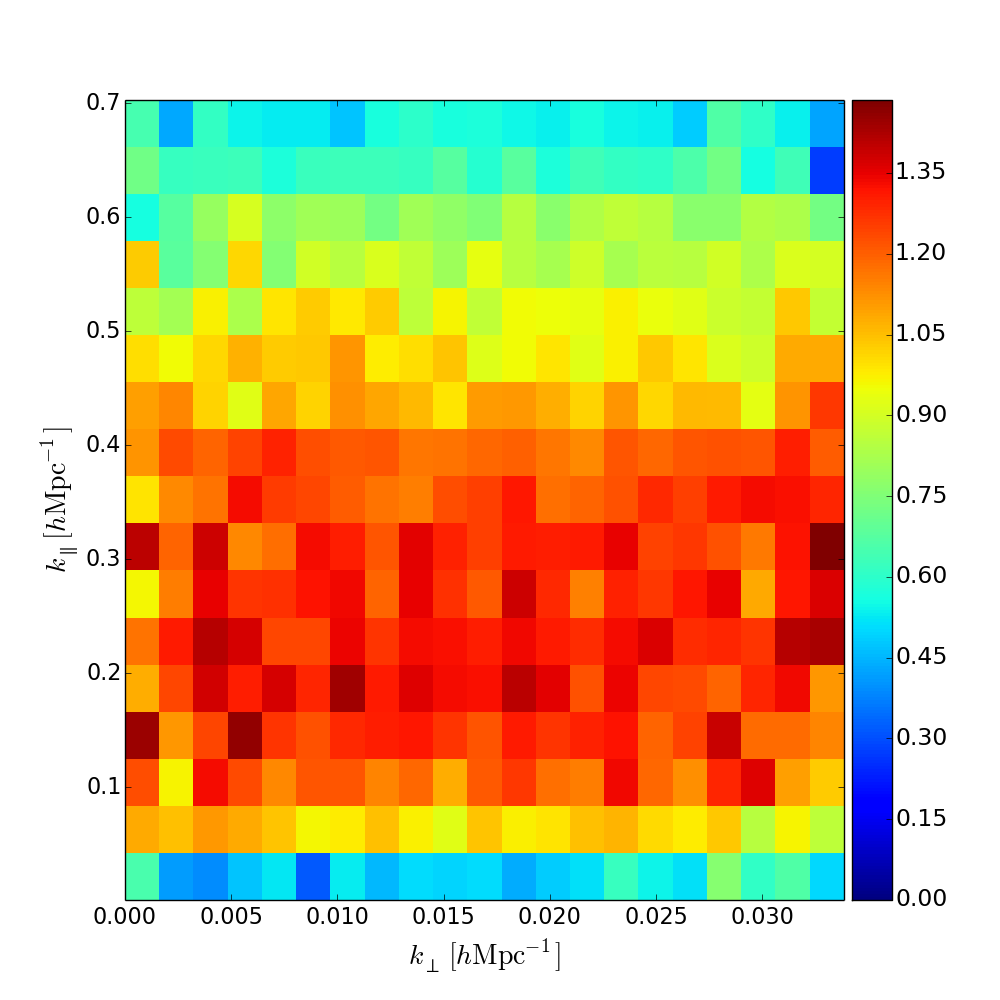}
	\includegraphics[width=0.50\textwidth]{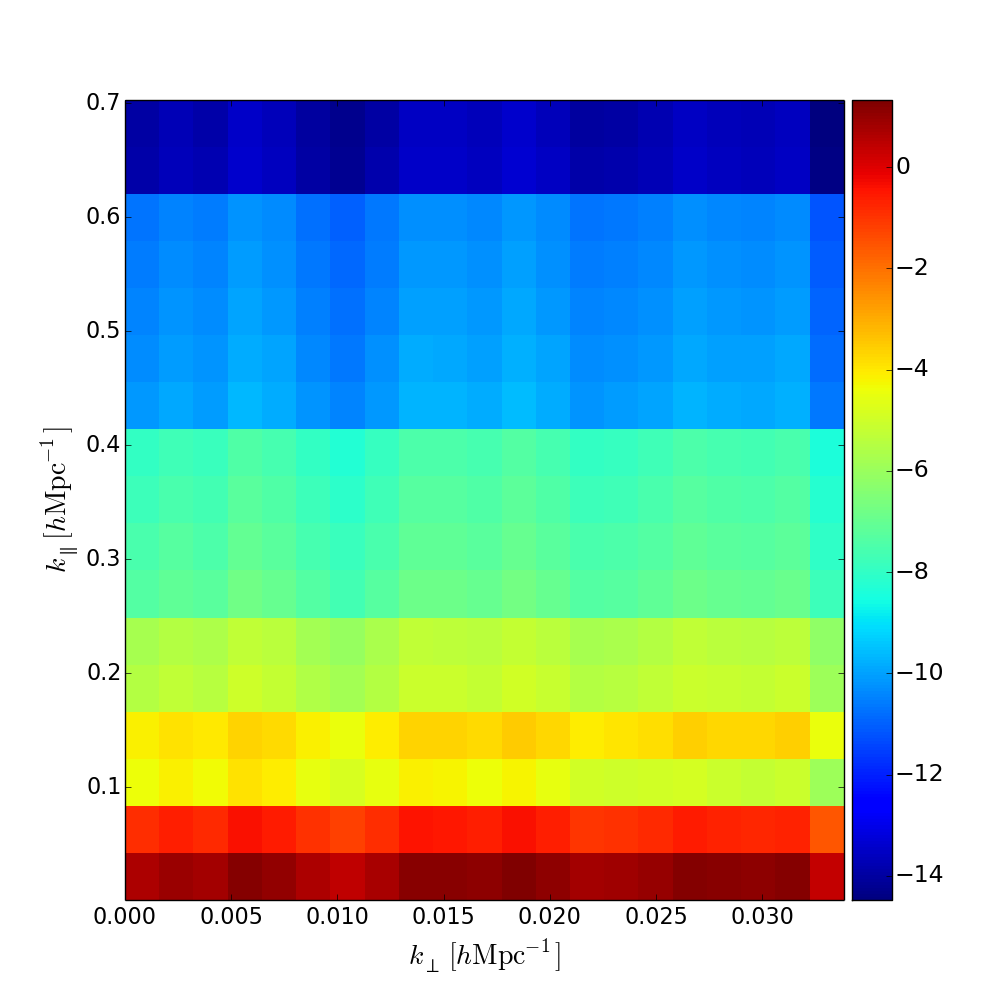}}
	\vspace{-0.25cm}
	\centerline{
	\includegraphics[width=0.50\textwidth]{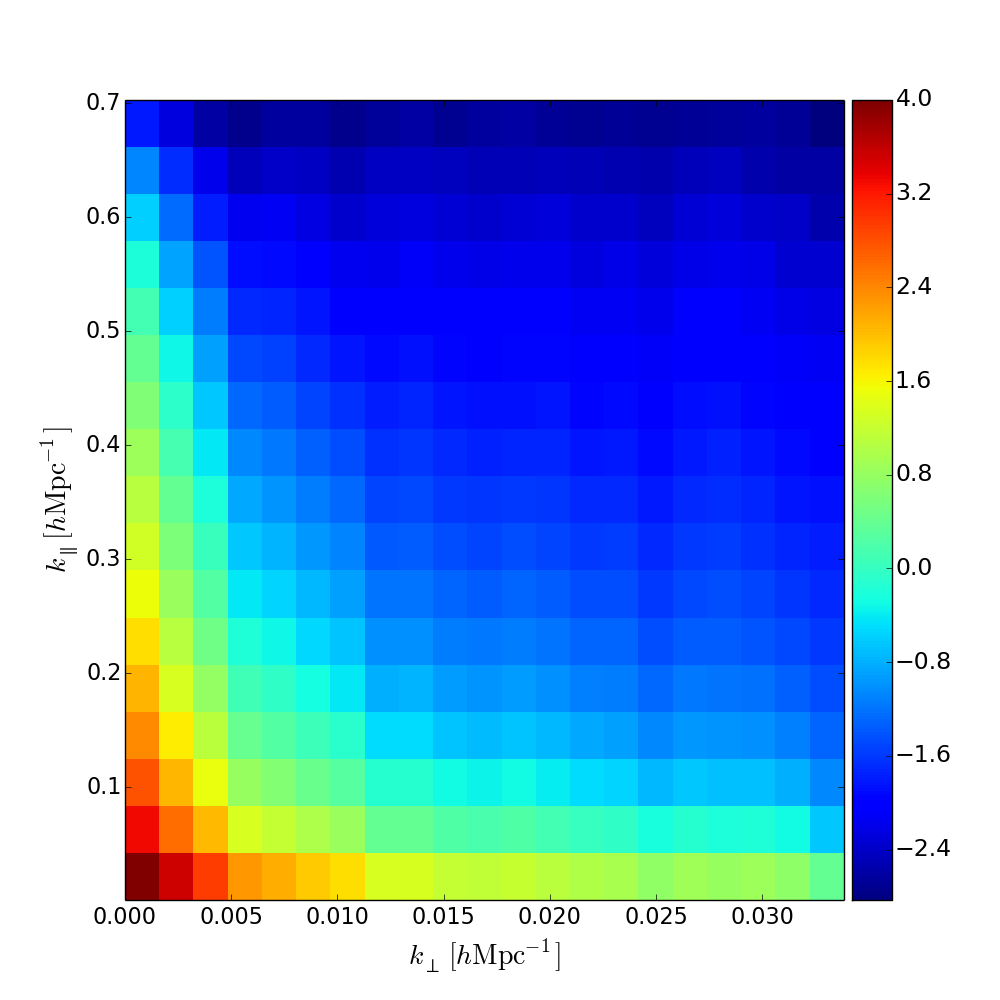}
	\includegraphics[width=0.50\textwidth]{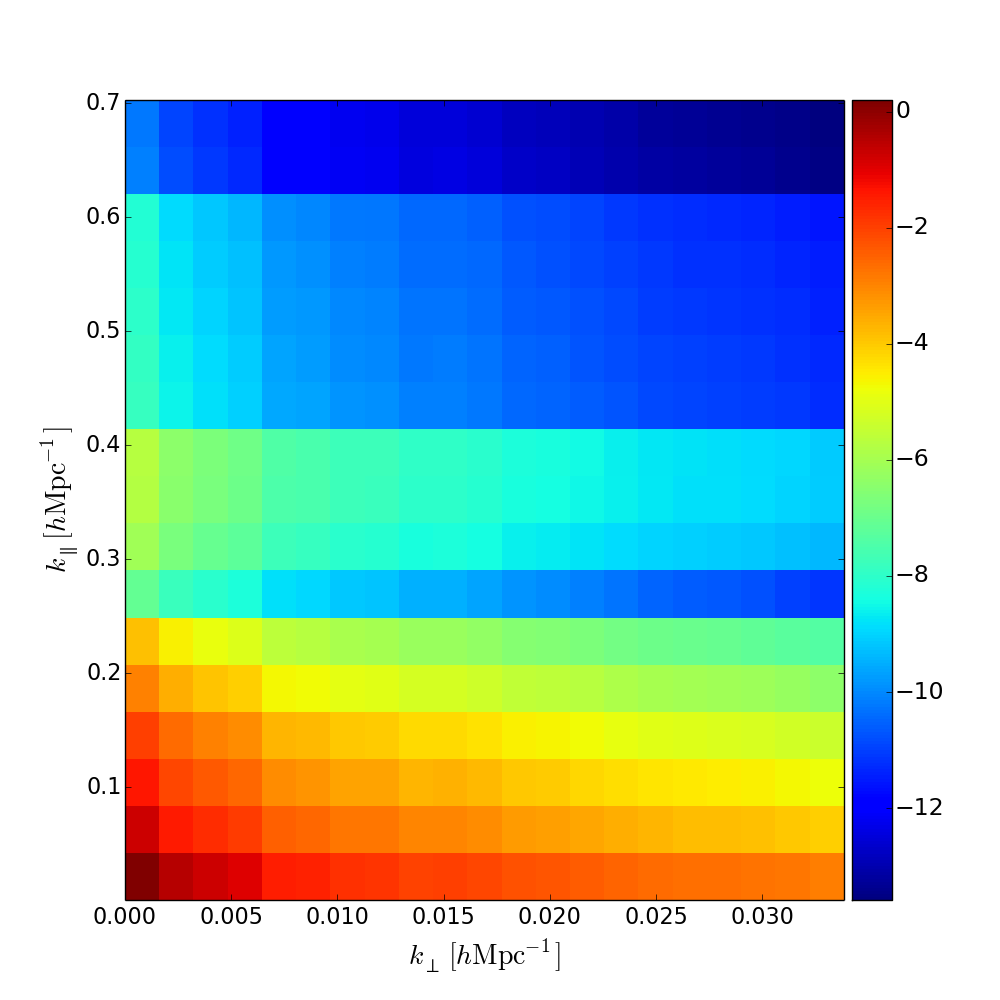}}
	\caption{Maximum likelihood dimensionless intrinsic cylindrical power spectra of the 21-cm signal [top left], extragalactic sources [top right], Galactic diffuse synchrotron emission (in region A; see \autoref{Fig:remazeilles_haslam408}) [bottom left] and diffuse Galactic free--free emission [bottom right], corresponding to the maximum a posteriori spherically averaged three-dimensional $k$-space dimensionless intrinsic power spectra of their respective emission simulations. Deviation from the spherical symmetry of the power spectral prior is evident in the GDSE and diffuse free--free Galactic foregrounds in the form of power spectral gradients as a function of $k_{\perp}$. This is expected given the steep spatial power spectra of these foreground components (c.f. \autoref{Fig:SkyCompSpatialPS}). The colourscale shows log-brightness-temperature-squared with brightness temperature in mK.}
	\label{Fig:CylindricalIntrinsicPS}
\end{figure*}

In \autoref{Fig:SphericalIntrinsicPS}, left, we show the maximum a posteriori spherically averaged three-dimensional $k$-space dimensionless intrinsic power spectra obtained for the EoR (black dashed line), Galactic diffuse synchrotron emission in regions A, B and C (red dashed, dash--dotted and dotted lines respectively), extragalactic foregrounds (blue dashed line) and diffuse free--free emission (green dashed line). The total sky power summed over contributions from the EoR and foregrounds and assuming GDSE emission in region A is shown in purple. Each of the simulated foreground components possesses non-zero power at all sampled $k$. 

In the $2.5~\lambda \le \abs{\mathbfit{u}} \le 37~\lambda$ spatial resolution range under consideration, power in the spherical power spectrum of the foregrounds is dominated by GDSE. This component can also be seen to dominate the total estimated power at low $k$. \autoref{Fig:SphericalIntrinsicPS}, right, emphasises this point and displays the fraction of total power in the spherical power spectrum accounted for by the EoR signal in each region. For $\log_{10}(k[h\mr{Mpc^{-1}}]) > -0.80$, $ -0.85$ and $ -0.55$ for regions A, B and C, respectively, the measured power spectrum is EoR dominated, with greater than $95\%$ of the estimated power attributable to EoR emission. However, in the complements to these intervals, at low $k$, the recovered power spectrum becomes rapidly dominated by foreground emission and, specifically, by large spatial scale power in the Galactic diffuse synchrotron emission.

In \autoref{Fig:CylindricalIntrinsicPS} we show the maximum likelihood dimensionless intrinsic cylindrical power spectra of the sky components in region A\footnote{As noted in \autoref{Sec:SpatialPowerSpectra} we select region A as being typical of the relatively cold Galactic regions favoured for EoR power spectral estimation from observational data.}. These provide a more nuanced view of the contamination of the EoR power spectrum by the foregrounds as a function of position in $k$-space. As can be anticipated from the spherical power spectrum of the foreground components, foreground power is dominated by GDSE across ($k_{\perp}$, $k_{\parallel}$)-space in the spatial resolution limit probed here. In each of the foreground components, the rapid drop-off in power as a function of $k_{\parallel}$ reflects the relative smoothness of their spectra. In the GDSE and diffuse free--free cylindrical power spectra, the steep power law structure of their spatial power spectra is apparent. This is despite the spherically symmetric prior we impose on the power spectrum. The relatively flat spatial power spectrum of the EGS is likewise reflected in its maximum likelihood cylindrical power spectrum.

As described in \autoref{Sec:SpatialPowerSpectra} in the context of the two-dimensional spatial power spectrum, this strong $k_{\perp}$ dependence of the free--free and GDSE foregrounds provides a method to distinguish them from the two-dimensional spatial power spectrum of the EoR. Incorporating this information into the prior will allow for enhanced component separation and can be particularly relevant in the low two-dimensional spatial resolution region of parameter space probed here where GDSE is the most significant source of foreground power on the scales of interest. 

\subsubsection{Intrinsic EoR window}
\label{Sec:IntEoRWin}

\begin{figure}
	\centerline{
	\includegraphics[width=0.50\textwidth]{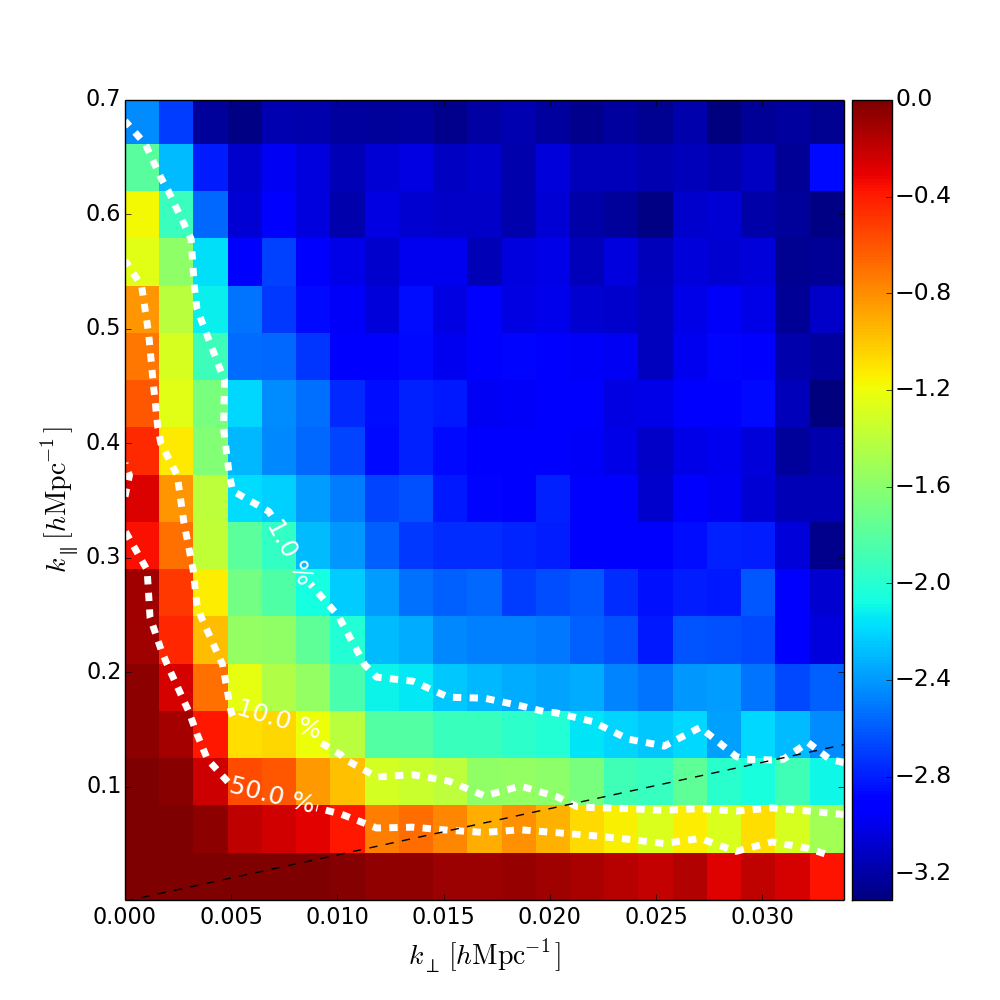}
	}
	\caption{Fractional foreground contamination of the total recovered power spectrum as a function of $k_{\perp}$ and $k_{\parallel}$. Dashed contours represent lines of constant foreground contamination at the levels of $1\%$, $10\%$, and $50\%$ as marked. In the $1.6\times10^{-3}~h\mr{Mpc^{-1}} < k_{\perp} < 3.3\times10^{-2}~h\mr{Mpc^{-1}}$ spatial resolution range analysed, foreground power is dominated by GDSE which has a power law drop off as a function of both $k_{\perp}$ and $k_{\parallel}$. This results in a foreground dominated region at low $k$ and an EoR window relatively free of intrinsic foreground contamination at larger $k$. The colourscale displays log-base-10 of the foreground-to-total-power ratio ($\log_{10}(\sDelta^{2}_{\mr{FG}} / \sDelta^{2}_{\mr{Tot}})$). The dashed black line displays the instrumental horizon at 126 MHz (see \autoref{Sec:ForegroundRemovalTechniques}) for reference when comparing with alternate power spectral estimation methodologies (see \autoref{Sec:Comparison}).}
	\label{Fig:FGContCylindricalIntrinsicPS}
\end{figure}

In \autoref{Fig:FGContCylindricalIntrinsicPS} we show the fractional contamination of the total recovered power by foreground emission as a function of $k_{\perp}$ and $k_{\parallel}$ derived as the ratio of the EoR to foreground intrinsic cylindrical power spectra in \autoref{Fig:CylindricalIntrinsicPS}. The power law drop-off of the dominant foreground in this regime, GDSE, as a function of both $k_{\perp}$ and $k_{\parallel}$ produces a region of intrinsic foreground power spectral contamination in ($k_{\perp}$, $k_{\parallel}$)-space at low $k$. Contours of constant foreground contamination are marked at the $50\%$, $10\%$ and $1\%$ levels. Outside of this region, at larger $k$, we find an, increasingly foreground-free, intrinsic EoR window in which the 21-cm signal dominates the intrinsic power spectrum. 

While the precise level of fractional contamination as a function of position in ($k_{\perp}$, $k_{\parallel}$)-space shown in \autoref{Fig:FGContCylindricalIntrinsicPS} is EoR\footnote{A more general version of the plot of fractional power spectral contamination shown in \autoref{Fig:FGContCylindricalIntrinsicPS} could be constructed by marginalising over the intrinsic cylindrical power spectra of a range of possible EoR models. This would be computationally intensive. It would also add little value to the proof-of-concept calculation shown here, since the resulting plot would, nevertheless, be foreground model specific. However, for an intrinsic EoR window estimated using the foreground power spectrum calculated from observations in a specific field with an accurately calibrated instrument, marginalising over the intrinsic cylindrical power spectra of a range of possible EoR models is the preferred approach.} and foreground model specific, the structure of the contaminated region results primarily from the contrast between the, observationally well known, power law nature of the GDSE spatial power spectrum and the more slowly varying EoR power spectrum. As such, an increasingly foreground-free, intrinsic EoR window in which the 21-cm signal dominates the intrinsic power spectrum at larger $k_{\perp}$ and $k_{\parallel}$ can be expected as a generic feature.

When applying this approach to observations, the region of intrinsic foreground contamination and its complement the intrinsic EoR window should be calculated from the data. This can be achieved by targeting observations at frequencies above those at which reionization completes (expected to occur at $z \lesssim 6$, $\nu \gtrsim 200~\mr{MHz}$; e.g. \citealt{2015ApJ...802L..19R, 2015MNRAS.447..499M}). By applying the power spectral analysis presented in this paper to observational data at redshifts below which reionization completes, assuming an accurately calibrated instrument, the intrinsic power spectrum of the foregrounds can be calculated independently of the redshifted 21-cm signal. Barring changes to the physical structure of the emission over the frequency range of interest, the $k$-space structure of the intrinsic foreground power spectral contamination can be expected to be relatively well determined\footnote{Evolution of the intrinsic foreground power spectrum can be assessed more rigorously through calculation of the intrinsic power spectrum in a number of frequency bands at $z < 6$. This can be utilised to improve extrapolation to lower frequencies via model fitting between bands.}. The variation of the amplitude of the contamination with frequency can be estimated using the mean spectral indices of the foreground components. With appropriate scaling, this could enable the intrinsic foreground power spectrum obtained at higher frequencies to provide a foreground prior when estimating the EoR signal at lower frequencies. Further, the extent of the complementary low-foreground-power region of $k$-space can be estimated from the data. Extrapolation of this region to the frequencies of interest for estimating the EoR power spectrum will allow the estimation of a field and instrument specific intrinsic EoR window analogous to that shown in \autoref{Fig:FGContCylindricalIntrinsicPS}.

\subsection{Observed power spectra}
\label{Sec:ObservedPowerSpectra}

In this subsection we consider power spectral estimation from interferometric observations. We assume that the instrument model is known perfectly and we forward model it precisely in the power spectral likelihood in the manner described in \autoref{Sec:PowerSpectralModelandAssumptions}. We further assume that the observations are free of calibration errors. Within this approximation, while the specific sampling of the $uv$-plane by the instrument (in combination with the details of the binning of the power spectrum) influences its sensitivity to different power spectral coefficients, given sufficient averaging down of the noise, the power spectral estimates obtained from the interferometric data using our analysis will be consistent with the instrument-free intrinsic power spectrum of the signal.
We now demonstrate this to be the case.

\begin{figure}
	\centerline{
	\includegraphics[width=0.50\textwidth]{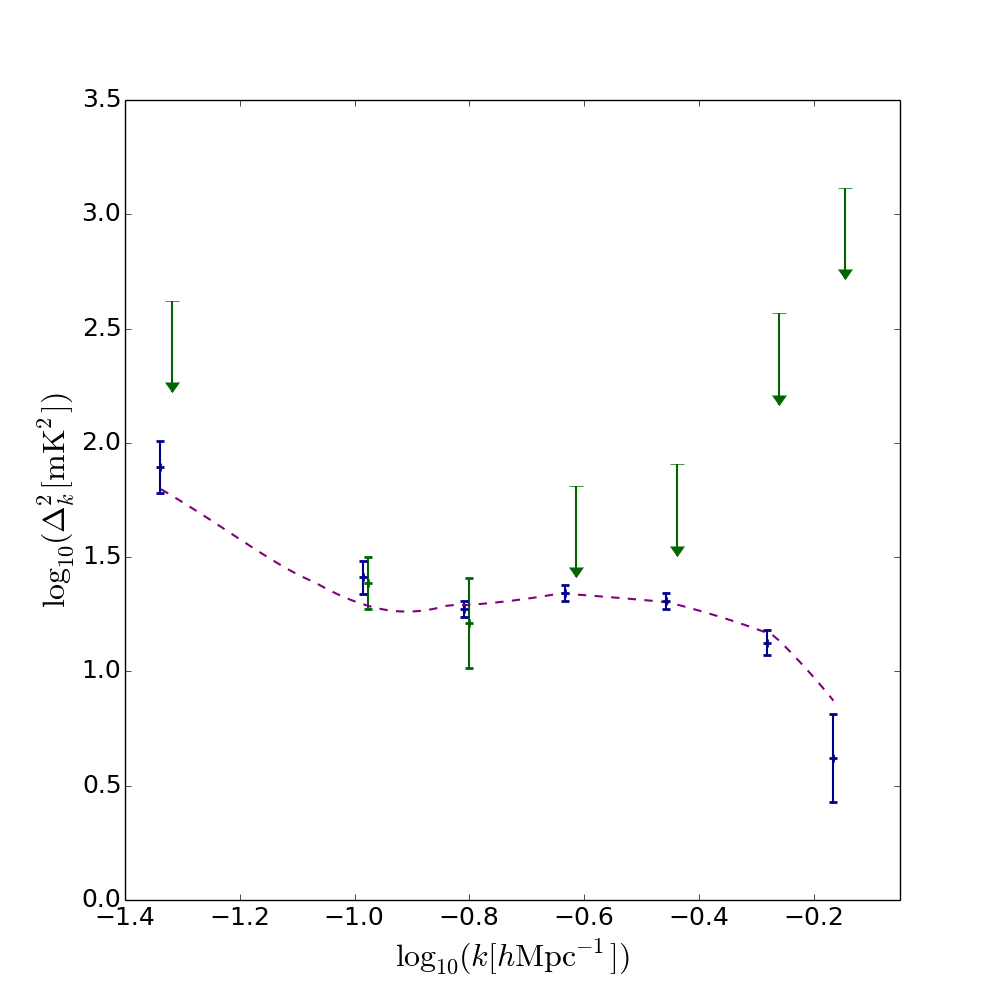}
	}
	\caption{Input (purple dashed line) and recovered values for the spherical power spectrum of the EoR plus foregrounds. Green and blue markers display estimates from simulated data generated using an instrumental configuration corresponding to HERA in 37-antenna configuration and HERA in 331-antenna configuration with baselines restricted to the spatial resolution range of HERA 37 (see \autoref{Sec:SimulatedHERAObservations}, for details) respectively. Arrows represent 2-$\sigma$ upper limits obtained using a uniform prior on the amplitudes of the coefficients. Points with 1-$\sigma$ error bars show power spectral detections obtained with a log-prior on the amplitudes of the coefficients. After 2000 hours of observation, two $k$-modes centred on $k=0.10~\mr{hMpc^{-1}}$ and $k=0.16~\mr{hMpc^{-1}}$ are detectable with HERA in 37-antenna configuration. With the same observation period all $k$-modes in the dataset are detected using HERA in 331-antenna configuration using baselines restricted to the spatial resolution range of HERA 37. In each case the power spectral estimates derived from simulated observation are fully consistent with intrinsic power spectrum of the input signal.}
	\label{Fig:SphericalInstrumentalPS}
\end{figure}

In \autoref{Fig:SphericalInstrumentalPS} we show the input versus recovered power spectrum from the analysis of simulated observations of the sum of our EoR and foreground models. Instrumental and observing parameters are as described in \autoref{Sec:SimulatedHERAObservations}. Power-spectral estimates shown in green are derived from simulated interferometric data generated using an instrumental model corresponding to HERA in 37-antenna configuration. Data points in blue use an instrumental model corresponding to HERA in 331-antenna configuration with baselines restricted to the spatial resolution range of HERA 37. Both use 2000 hours of observation time. We plot the inferred power with 1-$\sigma$ error bars where there is strong evidence of a detection using log-uniform priors on the amplitudes of the coefficients (see e.g. \citealt{2008arXiv0804.3173R, 2009arXiv0909.1008R}), otherwise we indicate 2-$\sigma$ upper limits derived using uniform priors.

Importantly, in each case, the detected power spectral estimates from the simulated observation are fully consistent with the intrinsic power spectrum of the input signal (dashed purple line in \autoref{Fig:SphericalInstrumentalPS} and solid purple line in \autoref{Fig:SphericalIntrinsicPS}). In particular, with HERA in 37-antenna configuration, two $k$-modes centred on $k=0.10~\mr{hMpc^{-1}}$ and $k=0.16~\mr{hMpc^{-1}}$ are detected. The level of power in the remaining coefficients is consistent with the noise on the visibilities and we place two sigma upper limits on the intrinsic power in the signal (green arrows). For the same observation period, all $k$-modes are detected in the dataset generated using instrumental parameters modelled on HERA in 331-antenna configuration.

\subsection{Comparison with alternative power spectral estimation methodologies}
\label{Sec:Comparison}

Recently \citet{2015ApJ...809...61A} have reported power spectral limits on 21-cm emission from the EoR at redshift $z=8.4$ making use of the foreground avoidance approach to power spectral estimation (see \autoref{Sec:ForegroundRemoval}). The limits were derived using selected 30 m baselines of the 64-antenna deployment of the PAPER array. In addition, a number of power spectral detections several orders of magnitude above the power predicted by models for emission from the EoR are also established. On large spatial scales ($k \sim 0.1$) these detections are attributed to foreground emission. In \autoref{Fig:SphericalIntrinsicPS} we calculate the intrinsic power spectra of the foregrounds. At $k=0.1~h\mr{Mpc^{-1}}$ power resulting from GDSE in our region C (see \autoref{Fig:remazeilles_haslam408}) is $\sDelta^{2}_{k} \sim 10^{2}$ when calculated using $uv$-sampling corresponding to HERA in 37-antenna configuration. Lesser, but still significant, contamination is also found in regions A and B. Our analysis therefore suggests that, at least in part, this emission is attributable to foregrounds. Two differences between the data used by \citet{2015ApJ...809...61A} and the analysis considered here will exacerbate the relative level of foreground contamination they report. As illustrated in \autoref{Sec:SpatialPowerSpectra} intrinsic foreground contamination can be expected to be more significant on shorter baselines and, as such, foreground power in the spherically averaged $k$-space power spectrum will be more significant when sampled on these baselines\footnote{The foreground and EoR models presented in this paper suggest that estimation of the power spectrum on longer baselines, within the region of optimal signal estimation for a target field  of the observation (see \autoref{Fig:EoRFGFractionalSpatialPS} for an example with the models used in the this paper), indicate a potential to mitigate this contamination. However to take advantage of this requires an accurate model of the instrument to be incorporated within the power spectral estimation framework. Without this, the resulting power in low $k$-modes will be masked by the instrumental foreground wedge (see \autoref{Fig:FGContCylindricalIntrinsicPS}).}. Additionally, calibration imperfections inherent in observations are liable to increase spectral structure in the foregrounds, enhancing their power spectra. 

Our analysis shows that, in the limit of a perfect instrumental model and calibration, the preferred approach is to include a model for the instrument within the power spectral analysis. For an accurate but not perfectly precise instrument model or calibration, as will be the case in observations, the constraints on the power spectrum will be less stringent than in this idealised case. However, the instrumental response acts purely to increase the foreground contamination in regions of $k$-space with minimal intrinsic contamination. Therefore inclusion in the analysis of an instrument model with uncertainties will improve upon the results obtainable without one as long as the model is unbiased.

In its simplest form, the instrumental foreground avoidance technique relies on an uncontaminated EoR window existing above a wedge of contaminating foreground emission. Contaminating emission in the wedge results from convolution of the spectrally flat component of the foregrounds (in $(\mathbfit{u}, \nu)$-space), with the instrumental response\footnote{The instrumental response, here, refers to two functions of the instrument. Firstly, the instrumental sampling. That is, the fact that the coordinate $\mathbfit{u}$ sampled by an interferometric baseline are frequency dependent with smaller spatial scales probed by a baseline with increasing $v$. As a result, an intrinsically spectrally flat source (in $(\mathbfit{u}, \nu)$-space) with non-flat spatial structure will have non-flat spectral structure along a baseline. Secondly, the aperture function of the telescope is similarly frequency dependent, thus altering the correlation between visibilities in the uv-domain as a function of frequency.}. We have found that the intrinsically non-spectrally flat component of the foregrounds in $(\mathbfit{u}, \nu)$-space (the component not fit by the quadratic basis vectors in our analysis) can contaminate the instrumental EoR window. This will reduce the region of $k$-space accessible for unbiased estimation of the EoR power spectrum. Convolution of this intrinsic foreground contamination with the instrumental response, as will occur in power spectral analyses in which the instrument is not modelled, will further expand this contaminated region. The impact of this on power spectral contamination could be investigated using the approach discussed in \autoref{Sec:IntEoRWin} adapted for the instrumental foreground avoidance technique. Experiments using this approach can analyse the power spectrum at frequencies above those at which reionization is expected to complete ($\nu \gtrsim 200~\mr{MHz}$). Assuming precise instrumental calibration, this will yield an estimate of the intrinsic foreground power spectrum convolved with the instrumental response. In the same manner as described in \autoref{Sec:IntEoRWin}, the region of $k$-space which is minimally contaminated can be identified. EoR power spectral estimation can then be targeted in the $k$-space volume occupied by the extrapolation of this region to the frequency range of interest.

Our analysis also indicates that while the EoR power spectrum is intrinsically contaminated by the foregrounds at low $k_{\perp}$ (which will complicate estimation of the EoR power spectrum in general), estimating the low-$k$ EoR power spectrum with reduced intrinsic foreground contamination can be achieved by targeting estimates at higher $k_{\perp}$ using longer baselines (those within the region of optimal signal estimation, calculated for a specific field and neutral fraction). More generally, considering the power spectral contamination by foregrounds as a function of $k_{\perp}$ and $k_{\parallel}$, this can be achieved through estimating the EoR power spectrum in the intrinsic EoR window determined from observations in the manner described in \autoref{Sec:IntEoRWin}. In the instrumental foreground avoidance power spectral estimation framework, on longer baselines, the region of interest will be contaminated by the instrumental foreground wedge (the region below the dashed black horizon line in \autoref{Fig:FGContCylindricalIntrinsicPS}). In this case, estimation of the EoR power spectrum at low-$k$ ($k \lesssim 0.1~h\mr{Mpc^{-1}}$) will be possible only if foreground power can be removed from this region.

In the $1.6\times10^{-3}~h\mr{Mpc^{-1}} < k_{\perp} < 3.3\times10^{-2}~h\mr{Mpc^{-1}}$ spatial scale range we have found that each of the simulated foreground components possesses non-zero power at all sampled $k$. For $\log_{10}(k[h\mr{Mpc^{-1}}]) < -0.80$, $ -0.85$ and $ -0.55$, for regions A, B and C respectively, the recovered power spectrum becomes rapidly dominated by foreground emission and, specifically, by large spatial scale power in Galactic diffuse synchrotron emission. Since the components of both the EoR and foreground emission present in the intrinsic power spectra at a given scale are entirely covariant, an independent foreground subtraction step in a power spectral estimation framework (such as methods described in \autoref{Sec:ForegroundRemoval}) will at best suppress power in both the EoR and the foregrounds on scales where foreground power is significant. More probably, the fine-tuning of the number of fitted basis vectors required by foreground subtraction methods in combination with subtraction being performed as an independent step to power spectral estimation will result in one of two scenarios. Either, modes where there is transition to EoR power dominance will be suppressed, resulting in unnecessary information loss, or under-subtraction will occur, leaving foreground emission as assumed EoR power and thus biasing the power spectrum and any derived cosmological and astrophysical parameter estimates. Consequently, in either case, it will at best hinder and more likely bias estimation of EoR power spectral uncertainties. As such, joint estimation of the EoR and foreground signals will be important for unbiased estimation of the EoR power spectrum.

\section{Conclusions and further work}
\label{Sec:Conclusions}

We have argued that independent estimation of the level of foreground and EoR power as a function of position in $k$-space provides a means of assessing EoR power spectrum contamination. We have developed foreground simulations comprising spatially correlated extragalactic and diffuse Galactic emission components to investigate this claim. In constructing the foreground simulations we include: accurate modelling of large scale Galactic structure (relative to the simulation region), Galactic diffuse synchrotron emission (GDSE) simulations for a range of levels of brightness temperature -- spectral index correlation, and a physically motivated spectral absorption model for compact extragalactic sources.

We have used the Bayesian power spectral estimation framework detailed in Lentati et al. 2016 (in prep.) to perform unbiased estimation of the `intrinsic' (instrument-free) two-dimensional spatial power spectrum, $P_{uv}$, the three-dimensional spherically averaged power spectrum, $P(k)$, and the maximum likelihood cylindrically averaged power spectrum, $P(k_{\perp}, k_{\parallel})$, of each of the foregrounds and of the EoR. Unbiased estimation is achieved by performing a joint estimate for the power spectrum on scales fulfilling the Nyquist criterion, given the data sampling rate, with models for both large and small scale power. Our large spatial and spectral scale power models are given by a sub-harmonic grid sampled on a set of 10 log-uniformly spaced spatial scales between the size of the image and 10 times the size of the image and a set of quadratics respectively. Power from emission components at higher frequencies than the Nyquist sampling rate of the data manifests itself as an additional source of noise that we model in the data covariance matrix. We additionally assume that the redshifted 21-cm signal is spatially isotropic and, over the redshift interval under consideration, homogeneous (assuming the power spectrum of the 21-cm signal is approximately constant) and uncorrelated between spatial scales. As such, we impose a spherical prior on the three-dimensional $k$-space power spectrum.

Comparison of $P_{uv}$ calculated for our EoR and foreground simulations has lead us to identify a model dependent region of optimal signal estimation within which the ratio of the spatial power in the EoR signal to foregrounds is maximised. This assumes no uncertainty on the flux-density distribution resulting from sources with flux-densities in excess of five times the confusion noise limit for HERA in 37-antenna configuration. The region of optimal signal estimation extends over the spatial resolution range $35~\lambda \le \abs{\mathbfit{u}} \le 55~\lambda$ at $126~\mr{MHz}$, with lower and upper bounds corresponding to the optimal spatial resolutions for estimation of the EoR signal in a cold region out of the plane of the Galaxy and a region of intense GDSE emission in the Galactic plane respectively.

We discuss the application of this procedure to observational data via calculation of the spatial power spectrum of the foregrounds in a specific target field. We find that for a fixed EoR neutral fraction, it is possible to calculate well defined limits on the region of optimal signal estimation, dominated by the structure of the spatial power spectrum. We consider the $\zeta$, $R_\mr{mfp}$, $T^\mr{Feed}_\mr{vir}$ parametrisation of the EoR signal (see e.g. \citealt{2015MNRAS.449.4246G}), with $\zeta$ the galactic ionizing efficiency, $R_{\rm mfp}$ the mean free path of ionizing photons in the IGM and $T^{\rm Feed}_{\rm vir}$ the minimum virial temperature of star-forming haloes, and find that the limits on the region of optimal signal estimation can be defined to within uncertainties of $15\%$, for our foreground simulations and EoR power spectra across an observationally motivated astrophysical parameter range and with neutral fractions, $x_{H}$, in the range $0.45$ to $0.55$.

For a spatially separable three-dimensional foreground power spectrum (which is a good approximation across the $\sim 8~\mr{MHz}$ bandwidth considered here), this provides a method for estimating the optimal baseline lengths (those on which intrinsic contamination of the EoR power spectrum by foreground emission is minimal), on which to perform our three-dimensional $k$-space Bayesian power spectral estimation of the EoR signal, subject to a choice of EoR neutral fraction, target field and instrument model. When performing this calculation, and in general when estimating the power spectrum of the EoR from observational data, ideally, one would include prior information on the resolved extragalactic point source population in the power spectral likelihood. The amplitude of the signal would then be estimated given this prior. This allows the available information on sources to be accounted for when estimating the power spectrum in a statistically robust manner. In future work we will investigate the level to which including realistic uncertainties on resolved source parameters, as well as including priors on unresolved sources (for example, from external source catalogues constructed using high resolution instruments), impacts the power spectral constraints that can be obtained.

Extending our analysis to the three-dimensional power spectrum, we have calculated the intrinsic spherically averaged $k$-space power spectrum of our EoR simulation and of each foreground simulations in the spatial resolution range $1.6\times10^{-3}~h\mr{Mpc^{-1}} < k_{\perp} < 3.3\times10^{-2}~h\mr{Mpc^{-1}}$, broadly matching the region of maximum spatial sensitivity of current generation 21-cm experiments. For our GDSE simulation we find that power on the scales of interest for EoR power spectral estimation varies as a function of position in the Galaxy and with the level of brightness temperature -- spectral index correlation of the emission. In all cases, we find GDSE is the dominant foreground component and that contaminating power in our GDSE simulations exceeds the power in the EoR emission by up to several orders of magnitude on large spatial scales (for $k \lesssim 0.15~h\mr{Mpc^{-1}}$ in cold regions, and $k \lesssim 0.3~h\mr{Mpc^{-1}}$ for emission in the Galactic plane).

The fact that the intrinsic power spectra of the foregrounds are significant across a range of spatial scales of interest for estimating the EoR power spectrum (specifically, at low-$k$) demonstrates that without a priori knowledge of the complexity and covariance between the foregrounds and the EoR signal, foreground subtraction prior to power spectral estimation is liable to bias the power spectral estimates, as well as resulting in incorrect uncertainties. This highlights the importance of joint estimation of a model for the foregrounds and the EoR power spectrum if unbiased estimates of the EoR power spectrum are to be obtained.

We have further investigated the distribution of contamination of the EoR power spectrum by foregrounds as a function of position in $k$-space by calculating the maximum likelihood intrinsic cylindrically averaged power spectra corresponding to the intrinsic spherically averaged power spectra of a reference region. For this region we select a relatively cold patch of sky, lying out of the plane of the Galaxy (centred on $\mr{RA}=0\fdg0,\ \mr{Dec}=-30\fdg0$), as being typical of preferred regions for estimation of the EoR signal from observational data. We find that foreground power is dominated by GDSE across ($k_{\perp}$, $k_{\parallel}$)-space in the spatial resolution range probed. In each of the foreground components we observe a rapid drop-off in power as a function of $k_{\parallel}$ which reflects the relative smoothness of their spectra. In the free--free and GDSE cylindrical power spectra, steep power law structure is apparent spatially. This structure is well described by the spatial power spectra of those foregrounds. We additionally calculate the fractional contamination of the total recovered power by foreground emission as a function of $k_{\perp}$ and $k_{\parallel}$. We find that the power law drop-off of the dominant foreground in this regime, GDSE, as a function of both $k_{\perp}$ and $k_{\parallel}$ produces significant contamination of the total power in ($k_{\perp}$, $k_{\parallel}$)-space evident at low $k$. Outside of this region, we find a relatively foreground-free intrinsic EoR window in which the 21-cm signal dominates the intrinsic power spectrum.

We propose a method for estimating the intrinsic foreground contamination as a function of position in $k$-space and its complement, the intrinsic EoR window, from observational data in a target field. By estimating the intrinsic power spectrum in observations at sufficiently high frequencies for the reionization signal to be negligible ($\nu \gtrsim 200~\mr{MHz}$), the intrinsic foreground power spectrum can be independently estimated. In the simplest case, the mean spectral index of the foreground components can be used to extrapolate from the measured high frequency intrinsic foreground contamination to the frequencies of interest for estimating the redshifted 21-cm power spectrum. We note that, by supplementing the spherically symmetric prior on the power spectrum with physically motivated priors on the power spectra of the foregrounds, better separation of the foreground and EoR $k$-space power spectral estimates will be possible. With appropriate extrapolation, the measured foreground power spectral contamination obtained at high frequencies can provide such a foreground prior at the frequencies of interest. Extrapolation of the high frequency intrinsic foreground power spectrum to the frequencies of interest in combination with numerical modelling of the EoR power spectrum can also enable the estimation of an intrinsic EoR window within which the ratio of EoR to foreground power in $k$-space is maximal. A balance between sensitivity and predicted power spectral bias can then allow a preferred region of $k$-space within which to estimate the EoR power spectrum to be determined.

Finally, we estimate the power spectrum from simulated interferometric observations incorporating frequency dependent $uv$-coverage and primary beam. 
We consider two test cases: firstly, in the low signal-to-noise regime, using $uv$-sampling relevant to HERA in 37-antenna configuration and secondly, in the high signal-to-noise regime, using the $uv$-sampling of HERA in 331-antenna configuration for a restricted set of baselines with lengths in the HERA 37 range (which enables us to take advantage of the increased sensitivity of the larger instrument in this resolution range without additional computational cost). In both cases power spectral detections are fully consistent with power predicted by the intrinsic power spectra of the emission components calculated with a filled $uv$-plane and in the absence of instrumental effects.

\section*{Acknowledgements}

This work was performed using the Darwin Supercomputer of the University of Cambridge High Performance Computing Service (http://www.hpc.cam.ac.uk/), provided by Dell Inc. using Strategic Research Infrastructure Funding from the Higher Education Funding Council for England and funding from the Science and Technology Facilities Council.
We thank the anonymous referee for comments that improved the clarity of the text. PHS thanks Irina Stefan, Jonathan Pober and Gianni Bernardi for valuable discussions and Dave Green for helpful comments on a draft of this manuscript.



\label{lastpage}


\begin{thebibliography}{99}

\bibitem[\protect\citeauthoryear{Ali et al.}{2015}]{2015ApJ...809...61A} Ali Z.~S., et al., 2015, ApJ, 809, 61

\bibitem[\protect\citeauthoryear{Alonso et al.}{2015}]{2015MNRAS.447..400A} Alonso D., Bull P., Ferreira P.~G., Santos M.~G., 2015, MNRAS, 447, 400 

\bibitem[\protect\citeauthoryear{Asad et al.}{2015}]{2015MNRAS.451.3709A} Asad K.~M.~B., et al., 2015, MNRAS, 451, 3709 

\bibitem[\protect\citeauthoryear{Beardsley et al.}{2015}]{2015ApJ...800..128B} Beardsley A.~P., Morales M.~F., Lidz A., Malloy M., Sutter P.~M., 2015, ApJ, 800, 128 

\bibitem[\protect\citeauthoryear{Bernardi, McQuinn \& Greenhill}{2015}]{2015ApJ...799...90B} Bernardi G., McQuinn M., Greenhill L.~J., 2015, ApJ, 799, 90 

\bibitem[\protect\citeauthoryear{Blumenthal \& Gould}{1970}]{1970RvMP...42..237B} Blumenthal G.~R., Gould R.~J., 1970, Rev. Mod. Phys., 42, 237 

\bibitem[\protect\citeauthoryear{Bonaldi \& Brown}{2015}]{2015MNRAS.447.1973B} Bonaldi A., Brown M.~L., 2015, MNRAS, 447, 1973 

\bibitem[\protect\citeauthoryear{Bowman, Morales \& Hewitt}{2009}]{2009ApJ...695..183B} Bowman J.~D., Morales M.~F., Hewitt J.~N., 2009, ApJ, 695, 183 

\bibitem[\protect\citeauthoryear{Chapman et al.}{2012}]{2012MNRAS.423.2518C} Chapman E. et al., 2012, MNRAS, 423, 2518 

\bibitem[\protect\citeauthoryear{Chapman, Zaroubi \& Abdalla}{2014}]{2014arXiv1408.4695C} Chapman E., Zaroubi S., Abdalla F., 2014, preprint (arXiv:1408.4695) 

\bibitem[\protect\citeauthoryear{Chapman et al.}{2015}]{2015aska.confE...5C} Chapman E., et al., 2015, aska.conf, 5

\bibitem[\protect\citeauthoryear{Condon}{1974}]{1974ApJ...188..279C} Condon J.~J., 1974, ApJ, 188, 279 

\bibitem[\protect\citeauthoryear{Condon et al.}{1998}]{1998AJ....115.1693C} Condon J.~J., Cotton W.~D., Greisen E.~W., Yin Q.~F., Perley R.~A., Taylor G.~B., Broderick J.~J., 1998, AJ, 115, 1693 

\bibitem[\protect\citeauthoryear{Condon et al.}{2012}]{2012ApJ...758...23C} Condon J.~J., et al., 2012, ApJ, 758, 23 

\bibitem[\protect\citeauthoryear{Datta, Bowman \& Carilli}{2010}]{2010ApJ...724..526D} Datta A., Bowman J.~D., Carilli C.~L., 2010, ApJ, 724, 526 

\bibitem[\protect\citeauthoryear{Datta et al.}{2012a}]{2012MNRAS.424..762D} Datta K.~K., Friedrich M.~M., Mellema G., Iliev I.~T., Shapiro P.~R., 2012a, MNRAS, 424, 762 

\bibitem[\protect\citeauthoryear{Datta et al.}{2012b}]{2012MNRAS.424.1877D} Datta K.~K., Mellema G., Mao Y., Iliev I.~T., Shapiro P.~R., Ahn K., 2012b, MNRAS, 424, 1877 

\bibitem[\protect\citeauthoryear{Datta et al.}{2014}]{2014MNRAS.442.1491D} Datta K.~K., Jensen H., Majumdar S., Mellema G., Iliev I.~T., Mao Y., Shapiro P.~R., Ahn K., 2014, MNRAS, 442, 1491 

\bibitem[\protect\citeauthoryear{de Oliveira-Costa et al.}{1997}]{1997ApJ...482L..17D} de Oliveira-Costa A., Kogut A., Devlin M.~J., Netterfield C.~B., Page L.~A., Wollack E.~J., 1997, ApJ, 482, L17 

\bibitem[\protect\citeauthoryear{DeBoer et al.}{2016}]{2016arXiv160607473D} DeBoer D.~R., et al., 2016, arXiv, arXiv:1606.07473 

\bibitem[\protect\citeauthoryear{Di Matteo et al.}{2002}]{2002ApJ...564..576D} Di Matteo T., Perna R., Abel T., Rees M.~J., 2002, ApJ, 564, 576 

\bibitem[\protect\citeauthoryear{Di Matteo, Ciardi \& Miniati}{2004}]{2004MNRAS.355.1053D} Di Matteo T., Ciardi B., Miniati F., 2004, MNRAS, 355, 1053 

\bibitem[\protect\citeauthoryear{Dillon, Liu\& Tegmark}{2013}]{2013PhRvD..87d3005D} Dillon J.~S., Liu A., Tegmark M., 2013, Phys. Rev. D, 87, 043005 

\bibitem[\protect\citeauthoryear{Feroz, Hobson\& Bridges}{2009}]{2009MNRAS.398.1601F} Feroz F., Hobson M.~P., Bridges M., 2009, MNRAS, 398, 1601 

\bibitem[\protect\citeauthoryear{Field}{1958}]{1958PIRE...46..240F} Field G.~B., 1958, Proc. IRE, 46, 240 

\bibitem[\protect\citeauthoryear{Field}{1959}]{1959ApJ...129..525F} Field G.~B., 1959, ApJ, 129, 525 

\bibitem[\protect\citeauthoryear{Finkbeiner, Davis\& Schlegel}{1999}]{1999ApJ...524..867F} Finkbeiner D.~P., Davis M., Schlegel D.~J., 1999, ApJ, 524, 867 

\bibitem[\protect\citeauthoryear{Furlanetto \& Mesinger}{2009}]{2009MNRAS.394.1667F} Furlanetto S.~R., Mesinger A., 2009, MNRAS, 394, 1667 

\bibitem[\protect\citeauthoryear{Geil, Gaensler, \& Wyithe}{2011}]{2011MNRAS.418..516G} Geil P.~M., Gaensler B.~M., Wyithe J.~S.~B., 2011, MNRAS, 418, 516 

\bibitem[\protect\citeauthoryear{Ghosh et al.}{2015}]{2015MNRAS.452.1587G} Ghosh A., Koopmans L.~V.~E., Chapman E., Jeli{\'c} V., 2015, MNRAS, 452, 1587 

\bibitem[\protect\citeauthoryear{Greig \& Mesinger}{2015}]{2015MNRAS.449.4246G} Greig B., Mesinger A., 2015, MNRAS, 449, 4246 

\bibitem[\protect\citeauthoryear{Harker et al.}{2009}]{2009MNRAS.397.1138H} Harker G. et al., 2009, MNRAS, 397, 1138 

\bibitem[\protect\citeauthoryear{Haslam et al.}{1981}]{1981A&A...100..209H} Haslam C.~G.~T., Klein U., Salter C.~J., Stoffel H., Wilson W.~E., Cleary M.~N., Cooke D.~J., Thomasson P., 1981, A\&A, 100, 209 

\bibitem[\protect\citeauthoryear{Haslam et al.}{1982}]{1982A&AS...47....1H} Haslam C.~G.~T., Salter C.~J., Stoffel H., Wilson W.~E., 1982, A\&AS, 47, 1 

\bibitem[\protect\citeauthoryear{Hazelton, Morales \& Sullivan}{2013}]{2013ApJ...770..156H} Hazelton B.~J., Morales M.~F., Sullivan I.~S., 2013, ApJ, 770, 156 

\bibitem[\protect\citeauthoryear{Hogg}{1999}]{1999astro.ph..5116H} Hogg D.~W., 1999, preprint (astro-ph/9905116)

\bibitem[\protect\citeauthoryear{Ibar et al.}{2009}]{2009MNRAS.397..281I} Ibar E., Ivison R.~J., Biggs A.~D., Lal D.~V., Best P.~N., Green D.~A., 2009, MNRAS, 397, 281 

\bibitem[\protect\citeauthoryear{Intema et al.}{2009}]{2009A&A...501.1185I} Intema H.~T., van der Tol S., Cotton W.~D., Cohen A.~S., van Bemmel I.~M., R{\"o}ttgering H.~J.~A., 2009, A\&A, 501, 1185 

\bibitem[\protect\citeauthoryear{Jeli{\'c} et al.}{2008}]{2008MNRAS.389.1319J} Jeli{\'c} V. et al., 2008, MNRAS, 389, 1319 

\bibitem[\protect\citeauthoryear{Jeli{\'c} et al.}{2010}]{2010MNRAS.409.1647J} Jeli{\'c} V., Zaroubi S., Labropoulos P., Bernardi G., de Bruyn A.~G., Koopmans L.~V.~E., 2010, MNRAS, 409, 1647 

\bibitem[\protect\citeauthoryear{Kogut et al.}{1996}]{1996ApJ...460....1K} Kogut A., Banday A.~J., Bennett C.~L., Gorski K.~M., Hinshaw G., Reach W.~T., 1996, ApJ, 460, 1 

\bibitem[\protect\citeauthoryear{Kohn et al.}{2016}]{2016ApJ...823...88K} Kohn S.~A., et al., 2016, ApJ, 823, 88 

\bibitem[\protect\citeauthoryear{Komatsu et al.}{2011}]{2011ApJS..192...18K} Komatsu E. et al., 2011, ApJS, 192, 18 

\bibitem[\protect\citeauthoryear{La Porta et al.}{2008}]{2008A&A...479..641L} La Porta L., Burigana C., Reich W., Reich P., 2008, A\&A, 479, 641 

\bibitem[\protect\citeauthoryear{Lane et al.}{2014}]{2014MNRAS.440..327L} Lane W.~M., Cotton W.~D., van Velzen S., Clarke T.~E., Kassim N.~E., Helmboldt J.~F., Lazio T.~J.~W., Cohen A.~S., 2014, MNRAS, 440, 327 

\bibitem[\protect\citeauthoryear{Lidz et al.}{2008}]{2008ApJ...680..962L} Lidz A., Zahn O., McQuinn M., Zaldarriaga M., Hernquist L., 2008, ApJ, 680, 962 

\bibitem[\protect\citeauthoryear{Liu et al.}{2009}]{2009MNRAS.398..401L} Liu A., Tegmark M., Bowman J., Hewitt J., Zaldarriaga M., 2009, MNRAS, 398, 401 

\bibitem[\protect\citeauthoryear{Liu \& Tegmark}{2011}]{2011PhRvD..83j3006L} Liu A., Tegmark M., 2011, Phys. Rev. D, 83, 103006 

\bibitem[\protect\citeauthoryear{Liu \& Tegmark}{2012}]{2012MNRAS.419.3491L} Liu A., Tegmark M., 2012, MNRAS, 419, 3491 

\bibitem[\protect\citeauthoryear{Liu, Parsons \& Trott}{2014a}]{2014PhRvD..90b3018L} Liu A., Parsons A.~R., Trott C.~M., 2014a, Phys. Rev. D, 90, 023018 

\bibitem[\protect\citeauthoryear{Longair}{2011}]{2011hea..book.....L} Longair M.~S., 2011, High Energy Astrophysics. Cambridge Univ. Press, Cambridge, UK  

\bibitem[\protect\citeauthoryear{Mao et al.}{2008}]{2008PhRvD..78b3529M} Mao Y., Tegmark M., McQuinn M., Zaldarriaga M., Zahn O., 2008, Phys. Rev. D, 78, 023529 

\bibitem[\protect\citeauthoryear{Mao}{2012}]{2012ApJ...744...29M} Mao X.-C., 2012, ApJ, 744, 29 

\bibitem[\protect\citeauthoryear{McGreer, Mesinger, \& D'Odorico}{2015}]{2015MNRAS.447..499M} McGreer I.~D., Mesinger A., D'Odorico V., 2015, MNRAS, 447, 499 

\bibitem[\protect\citeauthoryear{McQuinn et al.}{2006}]{2006ApJ...653..815M} McQuinn M., Zahn O., Zaldarriaga M., Hernquist L., Furlanetto S.~R., 2006, ApJ, 653, 815 

\bibitem[\protect\citeauthoryear{Mellema et al.}{2013}]{2013ExA....36..235M} Mellema G. et al., 2013, Exp. Astron., 36, 235 

\bibitem[\protect\citeauthoryear{Mesinger \& Furlanetto}{2007}]{2007ApJ...669..663M} Mesinger A., Furlanetto S., 2007, ApJ, 669, 663 

\bibitem[\protect\citeauthoryear{Mesinger, Furlanetto \& Cen}{2011}]{2011MNRAS.411..955M} Mesinger A., Furlanetto S., Cen R., 2011, MNRAS, 411, 955 

\bibitem[\protect\citeauthoryear{Mesinger, Ferrara \& Spiegel}{2013}]{2013MNRAS.431..621M} Mesinger A., Ferrara A., Spiegel D.~S., 2013, MNRAS, 431, 621 

\bibitem[\protect\citeauthoryear{Mesinger, Ewall-Wice \& Hewitt}{2014}]{2014MNRAS.439.3262M} Mesinger A., Ewall-Wice A., Hewitt J., 2014, MNRAS, 439, 3262 

\bibitem[\protect\citeauthoryear{Moore et al.}{2013}]{2013ApJ...769..154M} Moore D.~F., Aguirre J.~E., Parsons A.~R., Jacobs D.~C., Pober J.~C., 2013, ApJ, 769, 154 

\bibitem[\protect\citeauthoryear{Morales \& Hewitt}{2004}]{2004ApJ...615....7M} Morales M.~F., Hewitt J., 2004, ApJ, 615, 7 

\bibitem[\protect\citeauthoryear{Morales, Bowman \& Hewitt}{2006}]{2006ApJ...648..767M} Morales M.~F., Bowman J.~D., Hewitt J.~N., 2006, ApJ, 648, 767 

\bibitem[\protect\citeauthoryear{Nyquist}{1928}]{1928TAIEE..47..617N} Nyquist H., 1928, Trans. Am. Inst. Electri. Eng., 47, 617 

\bibitem[\protect\citeauthoryear{O'Dea}{1998}]{1998PASP..110..493O} O'Dea C.~P., 1998, PASP, 110, 493 

\bibitem[\protect\citeauthoryear{Oh \& Mack}{2003}]{2003MNRAS.346..871O} Oh S.~P., Mack K.~J., 2003, MNRAS, 346, 871 

\bibitem[\protect\citeauthoryear{Owen \& Morrison}{2008}]{2008AJ....136.1889O} Owen F.~N., Morrison G.~E., 2008, AJ, 136, 1889 1644 

\bibitem[\protect\citeauthoryear{Paciga et al.}{2013}]{2013MNRAS.433..639P} Paciga G. et al., 2013, MNRAS, 433, 639 

\bibitem[\protect\citeauthoryear{Parsons \& Backer}{2009}]{2009AJ....138..219P} Parsons A.~R., Backer D.~C., 2009, AJ, 138, 219 

\bibitem[\protect\citeauthoryear{Parsons et al.}{2010}]{2010AJ....139.1468P} Parsons A.~R. et al., 2010, AJ, 139, 1468 

\bibitem[\protect\citeauthoryear{Parsons et al.}{2012a}]{2012ApJ...756..165P} Parsons A.~R., Pober J.~C., Aguirre J.~E., Carilli C.~L., Jacobs D.~C., Moore D.~F., 2012a, ApJ, 756, 165 

\bibitem[\protect\citeauthoryear{Parsons et al.}{2012b}]{2012ApJ...753...81P} Parsons A., Pober J., McQuinn M., Jacobs D., Aguirre J., 2012b, ApJ, 753, 81 

\bibitem[\protect\citeauthoryear{Petrovic \& Oh}{2011}]{2011MNRAS.413.2103P} Petrovic N., Oh S.~P., 2011, MNRAS, 413, 2103 

\bibitem[\protect\citeauthoryear{Pober et al.}{2013}]{2013ApJ...768L..36P} Pober J.~C. et al., 2013, ApJ, 768, L36 

\bibitem[\protect\citeauthoryear{Pober et al.}{2014}]{2014ApJ...782...66P} Pober J.~C. et al., 2014, ApJ, 782, 66 

\bibitem[\protect\citeauthoryear{Pritchard et al.}{2015}]{2015aska.confE..12P} Pritchard J. et al., 2015, Proc. Adv. Astrophys. SKA, 12 

\bibitem[\protect\citeauthoryear{Randall et al.}{2012}]{2012MNRAS.421.1644R} Randall K.~E., Hopkins A.~M., Norris R.~P., Zinn P.-C., Middelberg E., Mao M.~Y., Sharp R.~G., 2012, MNRAS, 421, 

\bibitem[\protect\citeauthoryear{Reich \& Reich}{1988}]{1988A&AS...74....7R} Reich P., Reich W., 1988, A\&AS, 74, 7 

\bibitem[\protect\citeauthoryear{Remazeilles et al.}{2015}]{2015MNRAS.451.4311R} Remazeilles M., Dickinson C., Banday A.~J., Bigot-Sazy M.-A., Ghosh T., 2015, MNRAS, 451, 4311 

\bibitem[\protect\citeauthoryear{Robert, Chopin \& Rousseau}{2008}]{2008arXiv0804.3173R} Robert C.~P., Chopin N., Rousseau J., 2008, preprint (arXiv:0804.3173)

\bibitem[\protect\citeauthoryear{Robert, Chopin \& Rousseau}{2009}]{2009arXiv0909.1008R} Robert C.~P., Chopin N., Rousseau J., 2009, preprint (arXiv:0909.1008) 

\bibitem[\protect\citeauthoryear{Robertson et al.}{2015}]{2015ApJ...802L..19R} Robertson B.~E., Ellis R.~S., Furlanetto S.~R., Dunlop J.~S., 2015, ApJ, 802, L19

\bibitem[\protect\citeauthoryear{Rogers \& Bowman}{2008}]{2008AJ....136..641R} Rogers A.~E.~E., Bowman J.~D., 2008, AJ, 136, 641 

\bibitem[\protect\citeauthoryear{Santos, Cooray, \& Knox}{2005}]{2005ApJ...625..575S} Santos M.~G., Cooray A., Knox L., 2005, ApJ, 625, 575 

\bibitem[\protect\citeauthoryear{Scott \& Rees}{1990}]{1990MNRAS.247..510S} Scott D., Rees M.~J., 1990, MNRAS, 247, 510 

\bibitem[\protect\citeauthoryear{Shaver et al.}{1999}]{1999A&A...345..380S} Shaver P.~A., Windhorst R.~A., Madau P., de Bruyn A.~G., 1999, A\&A, 345, 380 

\bibitem[\protect\citeauthoryear{Steppe et al.}{1995}]{1995A&AS..113..409S} Steppe H., Jeyakumar S., Saikia D.~J., Salter C.~J., 1995, A\&AS, 113, 409 

\bibitem[\protect\citeauthoryear{Sun \& Reich}{2009}]{2009A&A...507.1087S} Sun X.~H., Reich W., 2009, A\&A, 507, 1087 

\bibitem[\protect\citeauthoryear{Taylor, Carilli \& Perley}{1999}]{1999ASPC..180.....T} Taylor G.~B., Carilli C.~L., Perley R.~A., eds, 1999, ASP Conf. Ser. Vol. 180, Synthesis Imaging in Radio Astronomy II. Astron. Soc. Pac., San Francisco 

\bibitem[\protect\citeauthoryear{Tegmark}{1997}]{1997ApJ...480L..87T} Tegmark M., 1997, ApJ, 480, L87 

\bibitem[\protect\citeauthoryear{Tegmark et al.}{2000}]{2000ApJ...530..133T} Tegmark M., Eisenstein D.~J., Hu W., de Oliveira-Costa A., 2000, ApJ, 530, 133 

\bibitem[\protect\citeauthoryear{Thyagarajan et al.}{2015}]{2015ApJ...804...14T} Thyagarajan N. et al., 2015, ApJ, 804, 14 

\bibitem[\protect\citeauthoryear{Tingay et al.}{2013}]{2013PASA...30....7T} Tingay S.~J. et al., 2013, Publ. Astron. Soc. Aust., 30, 1 

\bibitem[\protect\citeauthoryear{Trott et al.}{2016}]{2016ApJ...818..139T} Trott C.~M., et al., 2016, ApJ, 818, 139 

\bibitem[\protect\citeauthoryear{van Haarlem et al.}{2013}]{2013A&A...556A...2V} van Haarlem M.~P. et al., 2013, A\&A, 556, A2 

\bibitem[\protect\citeauthoryear{Wang et al.}{2013}]{2013ApJ...763...90W} Wang J. et al., 2013, ApJ, 763, 90 

\bibitem[\protect\citeauthoryear{Wilman et al.}{2008}]{2008MNRAS.388.1335W} Wilman R.~J. et al., 2008, MNRAS, 388, 1335 

\bibitem[\protect\citeauthoryear{Zheng et al.}{2012}]{2012MNRAS.424.2562Z} Zheng Q., Wu X.-P., Gu J.-H., Wang J., Xu H., 2012, MNRAS, 424, 2562   

\end{thebibliography}
\end{document}